\documentclass[journal]{IEEEtran}

\usepackage{xcolor,soul,framed} 

\colorlet{shadecolor}{yellow}
\usepackage[pdftex]{graphicx}
\graphicspath{{../pdf/}{../jpeg/}}
\DeclareGraphicsExtensions{.pdf,.jpeg,.png}

\usepackage[cmex10]{amsmath}
\usepackage{amssymb}
\usepackage{amsfonts}
\usepackage{url}
\usepackage{booktabs,makecell,tabularx,threeparttable}
\usepackage[bookmarks=true,hidelinks,pdfpagelabels=true]{hyperref} 
\hypersetup{
    pdftitle = {A tutorial on 5G positioning},
    pdfauthor = {L. Italiano, B. Camajori Tedeschini, M. Brambilla, H. Huang, M. Nicoli, H. Wymeersch}
}
\usepackage{newtxtext,newtxmath}
\usepackage{diagbox}

\usepackage[printonlyused,nohyperlinks]{acronym}

\newcommand{\acapo}[1]{%
  \begin{tabular}{@{}c@{}}\strut#1\strut\end{tabular}%
}

\usepackage{bm}
\usepackage{multirow}
\usepackage{calrsfs}
\usepackage{pgfplots}
\usepackage{tikz}  
\usetikzlibrary[topaths]
\usetikzlibrary{positioning}
\usetikzlibrary{arrows.meta}
\usetikzlibrary{shapes.misc}
\usetikzlibrary{shapes.arrows}
\usetikzlibrary{plotmarks}
\usetikzlibrary{decorations}
\usetikzlibrary{decorations.markings}
\usetikzlibrary{backgrounds,shapes,arrows,positioning,calc,fit}
\usepgflibrary{decorations.markings}
\usetikzlibrary{mindmap,trees}
\tikzset{cross/.style={cross out, draw=black, minimum size=2*(#1-\pgflinewidth), inner sep=0pt, outer sep=0pt},
cross/.default={1pt}}
\usepgfplotslibrary{patchplots}

\usepackage[caption=false,font=footnotesize]{subfig}
\usepackage[export]{adjustbox}

\usepackage{cite}  

\let\mathcal\undefined
\DeclareMathAlphabet{\mathcal}{OMS}{cmsy}{m}{n}
\DeclareMathAlphabet\mathbfcal{OMS}{cmsy}{b}{n}

\usepackage{pifont}
\newcommand{\cmark}{\text{\ding{51}}}
\newcommand{\xmark}{\text{\ding{55}}}

\definecolor{lightgrey}{rgb}{0.85, 0.85, 0.85}
\definecolor{morelightgrey}{rgb}{0.95, 0.95, 0.95}
\definecolor{amber}{rgb}{1.0, 0.75, 0.0}
\definecolor{airforceblue}{rgb}{0.36, 0.54, 0.66}
\definecolor{aquamarine}{rgb}{0.5, 1.0, 0.83}
\definecolor{azure}{rgb}{0.0, 0.5, 1.0}
\definecolor{cadmiumorange}{rgb}{0.93, 0.53, 0.18}
\definecolor{babypink}{rgb}{0.96, 0.76, 0.76}
\definecolor{bubblegum}{rgb}{0.99, 0.76, 0.8}
\definecolor{emerald}{rgb}{0.18039, 0.8, 0.443137}
\definecolor{amaranth}{rgb}{0.9, 0.17, 0.31}
\definecolor{azure}{rgb}{0.0, 0.5, 1.0}
\definecolor{darkred}{rgb}{0.63529,0.07843,0.18431}
\definecolor{smoothgreen}{rgb}{0,0.597,0}
\definecolor{lightviolet}{rgb}{0.8, 0.6, 1}
\definecolor{forestgreen}{rgb}{0.13, 0.55, 0.13}

\usepackage{xcolor}
\definecolor{wildstrawberry}{rgb}{1.0, 0.26, 0.64}

\definecolor{caribbeangreen}{rgb}{0.0, 0.8, 0.6}

\newcommand{\T}{\text{T}}

\newlength\fheight
\newlength\fwidth
\pgfplotsset{compat=1.18}

\begin{document}

\bstctlcite{BSTcontrol}

\title{A Tutorial on 5G Positioning}
\author{Lorenzo~Italiano,~\IEEEmembership{Graduate Student Member,~IEEE}, 
  Bernardo Camajori Tedeschini,~\IEEEmembership{Graduate Student Member,~IEEE}, 
  Mattia Brambilla,~\IEEEmembership{Member,~IEEE},
  Huiping Huang,~\IEEEmembership{Member,~IEEE}, \\
  Monica Nicoli,~\IEEEmembership{Senior Member,~IEEE},
  Henk Wymeersch,~\IEEEmembership{Fellow,~IEEE}
  \thanks{This work was partially funded by the European Union—NextGenerationEU under the National Sustainable Mobility Center CN00000023, Italian Ministry of University and Research (MUR) Decree n. 1033—17/06/2022 (Spokes 6 and 9), and under the MUR Decree n. 352—09/04/2022, by the Vinnova B5GPOS Project under Grant 2022-01640, and by the Swedish Research Council (VR grant 2022-03007).}
\thanks{L.\ Italiano, B.\ Camajori Tedeschini, and M.\ Brambilla are with Dipartimento di Elettronica, Informazione e Bioingegneria (DEIB), Politecnico di Milano, Milan, Italy.
(e-mail: lorenzo.italiano@polimi.it, bernardo.camajori@polimi.it, mattia.brambilla@polimi.it).

M. Nicoli is with Dipartimento di Ingegneria Gestionale (DIG), Politecnico di Milano, Milan, Italy. 
(e-mail: monica.nicoli@polimi.it). 

H. Huang and H. Wymeersch are with the Department of Electrical Engineering, Chalmers University of Technology, Göteborg, Sweden. (e-mail: huiping@chalmers.se, henkw@chalmers.se). 

  }  
}

\markboth{IEEE COMMUNICATIONS SURVEYS \& TUTORIALS, VOL.~V, NO.~N, MONTH~YEAR
}{}

\maketitle


\begin{abstract}
\textcolor{black}{The widespread adoption of the fifth generation (5G) of cellular networks has brought new opportunities for the development of localization-based services. 
High-accuracy positioning use cases and functionalities defined by the standards are drawing the interest of vertical industries. In the transition towards the deployment, this paper aims to provide an in-depth tutorial on 5G positioning, summarizing the evolutionary path that led to the standardization of cellular-based positioning, describing the localization elements in current and forthcoming releases of the Third Generation Partnership Project (3GPP) standard, and the major research trends.}
By providing fundamental notions on wireless localization, comprehensive definitions of measurements and architectures, examples of algorithms, and details on simulation approaches, this paper is intended to represent an exhaustive guide for researchers and practitioners. 
Our approach aims to merge practical aspects of enabled use cases and related requirements with theoretical methodologies and fundamental bounds, allowing to understand the trade-off between system complexity and achievable, i.e., tangible, benefits of 5G positioning services.
\textcolor{black}{We analyze the performance of 3GPP Rel-16 positioning by standard-compliant simulations in realistic outdoor and indoor propagation environments, investigating the impact of the system configuration and the limitations to be resolved for delivering accurate positioning solutions.}
\end{abstract}


%
\IEEEpeerreviewmaketitle


\section{Introduction}


The recent enhancement of the \ac{5G} of cellular communications unveiled an era of unprecedented connectivity, embracing altogether the \ac{eMBB}, \ac{URLLC} and \ac{mMTC} scenarios~\cite{Wunder_j17}.

\textcolor{black}{In this new era of connectivity, \ac{5G} has not only accelerated data transmission to unprecedented speeds~\cite{Saxena_j16}, it has also catalyzed innovation across various sectors~\cite{saeed_state---art_2019}, promising groundbreaking possibilities and redefining the way we interact with technology and the world around us~\cite{GhoMaeBakCha:J19}.
A main application area that is benefiting from the adoption of the \ac{5G} technology is the \ac{IOT}~\cite{Abu-Mahfouz_j18, LI20181}, where the high density of connected devices calls for the design of enhanced radio access methodologies for mutual coordination~\cite{7499809}. In the \ac{IOT}, \ac{5G} connectivity enables real-time data analytics~\cite{verma_survey_2017}, representing a game changer for industries~\cite{Bera_j20} and redesigning the business models of vendors~\cite{Ladid:j16}.
Visions on the \ac{IOT} ecosystem expect a growing impact from \ac{B5G} communication technologies~\cite{malik_energy-efficient_2022, guo_enabling_2021}.
The empowered \ac{5G} connectivity will bring major enhancements in mobility, including road vehicles~\cite{eskandarian_research_2021}, trains~\cite{ko_high-speed_2022}, and drones~\cite{Cuevas_j19}, with
\ac{5G} \ac{V2X} communications~\cite{alalewi_5g-v2x_2021, GarCasKou:J21, Zeadally_j18} are fostering the rollout of enhanced automotive services demanding for high-speed data transfer.
Major impact is also expected in healthcare services~\cite{Kok-Lim_j19, 9305243} and large-scale network automation~\cite{chi_survey_2023, coronado_zero_2022, arzo_theoretical_2021}.}

Within such an evolution for the telecommunication market, \ac{5G} positioning stands out as a key fundamental enabler that promises to unlock and revolutionize location-based services\textcolor{black}{~\cite{Bartoletti_book_2023, Melazzi2021}}.
Positioning has been a desired feature of cellular communications since the \ac{2G}~\cite{RosRauGon:J18}; however, with the deployment of \ac{5G} networks, it has undergone a paradigm shift, leveraging the unique capabilities of this new wireless technology in providing unprecedented location accuracy~\cite{Popovski_j14, 10101780}\textcolor{black}{, navigation augmentation capabilities and competitiveness against other technologies~\cite{laoudias_survey_2018}}.

The popularity of positioning is remarked by the significant efforts in technological research frontiers about \textcolor{black}{\ac{UWB}~\cite{UWB_1, UWB_logistics}}, \ac{mmWave}~\cite{Shahmansoori_j19, Aryanfar_j14,Rappaport_mmwave}, \ac{THz}~\cite{wang_key_2021, sarieddeen_overview_2021, rikkinen_thz_2020, polese_toward_2020} 
and wireless optical networks~\cite{sun_review_2020, keskin_localization_2018} 
that allow improving positioning services by exploring larger signal bandwidths. 
\textcolor{black}{Improvements in positioning are also being investigated by developing new technologies that allow to control of the interaction of the radio signal with the propagation environment by \acp{RIS}~\cite{liu_reconfigurable_2021, elmossallamy_reconfigurable_2020}.}

The ongoing research works encompass the integration of pervasive \ac{AI}~\cite{shen_holistic_2022,baccour_pervasive_2022,narayanan_key_2020}, the implementation of all-spectrum reconfigurable front-end technologies facilitating dynamic spectrum access~\cite{jabbari_dynamic_2010,akyildiz_next_2006,leaves_dynamic_2004}, the exploration of quantum communications~\cite{ali2023quantum, Rahman_j22}, as well as blockchain mechanisms~\cite{kalla_survey_2022,9277899,9508931,9430905}\textcolor{black}{, and energy-efficient communication methodologies~\cite{zhuang_exploiting_2022, yao_backscatter_2020 , huang_survey_2019}, such as ambient back-scattering communications~\cite{kishore_opportunistic_2019, van_huynh_ambient_2018}}. These emerging paradigms signify a notable transformation in the landscape of communication technologies, offering the potential for enhanced efficiency, security, and sustainability~\cite{DiTaranto_m14}.

Furthermore, this research path is underpinned by a shifting architectural framework, wherein the transition towards a \ac{3D} network architecture becomes increasingly prominent~\cite{giordani_non-terrestrial_2021, Abu_j18, cheng_air-ground_2018}, presenting novel possibilities for extending network coverage, improving connectivity, and addressing the evolving demands of precise and ubiquitous positioning~\cite{shit_ubiquitous_2019} for autonomous driving vehicles~\cite{ghorai_state_2022, turay_toward_2022,  feng_deep_2021, grigorescu_survey_2020, yurtsever_survey_2020} 
or \acp{UAV}~\cite{fotouhi_survey_2019, mozaffari_tutorial_2019, zeng_accessing_2019, li_uav_2019, khuwaja_survey_2018}
, in contexts such as augmented and \ac{VR}~\cite{ruan_networked_2021,chakareski_viewport-adaptive_2020,perfecto_taming_2020,xu_energy_2019,sukhmani_edge_2019}, industry 4.0~\cite{oztemel_literature_2020, beier_industry_2020, ghobakhloo_determinants_2020} 
and robotics~\cite{karoly_deep_2021,alattas_evolutionary_2019}. 
In the following, we delve deeper into this topic, providing an overview of the main ongoing research on 5G positioning, including standardization and experimental activities. 

\subsection{Related work on 5G positioning}
\label{sec:relatedWorks5Gpos}


\textcolor{black}{A first investigation of the potentials of \ac{5G} positioning is in~\cite{shahmansoori_survey_2017}, where the authors highlight how \ac{mmWave} and massive \ac{MIMO} technologies represent key enablers for localization. They discuss general concepts of location-aware communications and use path-loss models to motivate the need for beamforming to counteract the high propagation losses at \ac{mmWave}. 
The performed simulations, using \ac{AOD}, \ac{AOA}, and \ac{TOA} measurements extracted from large bandwidth (600 MHz) signals at \ac{mmWave} (60~GHz), prove an achievable cm-level positioning accuracy.}

\textcolor{black}{More recent studies addressed the topic of \ac{5G} positioning, focusing on cellular positioning architectures, algorithms
and envisioned applications~\cite{liyanage_5g_2018,8000805,wen_survey_2019,fischer,mogyorosi_positioning_2022,jia_link-level_2023,Blefari21,WymGraDest2017,7583720,10485369,Ciaramitaro2023,Tagliaferri2021,MedComNet2024,Seco2024,Deng2019,Aboutanios_j17, El-Sheimy_j21, surveymmWave, Riva_2014, Brink_c22, Brink_c23, Witrisal_assistedliving, dwivedi_positioning_2021, morselli_soft_2023}.
The work in~\cite{liyanage_5g_2018} provides a concise and thorough analysis of how cellular systems have changed from the \ac{1G} to the \ac{4G}, also offering a basic introduction to the architecture and security protocols employed in each generation. 
A more detailed review of the architecture evolution and the positioning technologies is in~\cite{8000805}. Key enablers are discussed in~\cite{wen_survey_2019}, where the authors give an overview of \ac{5G} massive \ac{MIMO} localization, with a main focus on \ac{mmWave} frequencies. They discuss channel modeling and localization algorithms, outlining possible research directions. 
A comprehensive explanation of the \ac{5G} positioning signals and methodologies, with some insights into the architectures, is provided in~\cite{fischer}.
Non-standardized, e.g., \ac{ML}-based algorithms, are discussed in~\cite{mogyorosi_positioning_2022} and compared (from a theoretical perspective) with conventional (i.e., non \ac{ML}-based) algorithms. 
Given the lack of a unified 
platform to support the research on \ac{5G} localization algorithms, authors in~\cite{jia_link-level_2023} introduce a link-level simulator for \ac{CSI}-based localization in \ac{5G}  networks, which can realistically depict physical behaviors of the system.}

\textcolor{black}{Moving to application-oriented works, the main interest is in the potential of \ac{5G} positioning, especially in terms of accuracy and latency in vehicular networks. Therein, the \ac{5G} hardware can act as an additional sensor of the vehicular onboard sensor suite, providing communication, positioning, and sensing functionalities~\cite{Blefari21}. 
In the vehicular context,  \ac{5G} \ac{mmWave} positioning was shown to provide high-accuracy localization, thanks to the large bandwidth~\cite{WymGraDest2017,7583720}, provided that the communication beams are correctly steered\textcolor{black}{~\cite{10485369}}. This can be achieved with the assistance of onboard navigation sensors~\cite{Ciaramitaro2023,Tagliaferri2021}.}
\textcolor{black}{The \ac{5G} technology has also been used for pedestrian positioning
\cite{MedComNet2024}, also complementing \ac{GNSS} 
\cite{Seco2024,Deng2019} in outdoor positioning and navigation. }

\textcolor{black}{Another main context for research is indoor positioning, whose evolution and applications are studied in~\cite{Aboutanios_j17} and further investigated in the fields of \ac{IOT} and device-free localization~\cite{El-Sheimy_j21,surveymmWave, Riva_2014} where deep shadowing and dense multipath represent severe impairments for positioning. 
Authors in~\cite{Brink_c22, Brink_c23} have proposed techniques to efficiently remove outliers for \ac{5G} indoor positioning in smart factories. Multipath is being exploited as a friend instead of a foe~\cite{Witrisal_assistedliving} by gaining insightful information for positioning from wall reflections.
\Ac{3GPP} standard-compliant simulations are carried out in~\cite{dwivedi_positioning_2021, morselli_soft_2023}, where the positioning capabilities of \ac{3GPP} Rel-16 have been investigated in the \ac{UMi}, \ac{UMa}, and \ac{IOO} scenarios, considering multi-cell \ac{RTT}, \ac{DL}-\ac{TDOA}, and \ac{UL}-\ac{AOA} positioning.
\textcolor{black}{Lastly, \ac{5G}, WiFi and their fusion are compared in~\cite{Moreno2024} for fingerprinting with incomplete maps.}
}

\textcolor{black}{Concerning experimental validation, at present, most of the experiments have been performed using \ac{SDR} with \ac{LTE}~\cite{shamaei_exploiting_2018} or \ac{5G}~\cite{abdallah_opportunistic_2022, sci_rep_5G, IMPos2024}. 
\acp{SDR} have been used for positioning purposes by extracting  \ac{CSI}~\cite{Ruan2024,RuaCheLia:J22} or \ac{CIR} parameters~\cite{Blefari2023,PalBarBiaMelBle:c22}, resulting into time-domain techniques.
\ac{SDR} hardware such as \ac{USRP} can also be used for phase tracking, reaching a sub-meter positioning accuracy in indoor environments~\cite{chen_carrier_2022}.} 


\textcolor{black}{A main topic of research is positioning augmentation in harsh environments with low \acp{BS} visibility and multipath exploitation.
Authors in~\cite{guo_performance_2022} combine \ac{AOD} with multi-\ac{RTT} to cope with a limiting number of visible \acp{BS}, still neglecting reflections and scattering due to the absence of \ac{RT} simulations. In an urban environment, authors in~\cite{cui_novel_2023} exploit the \ac{DRSS} to avoid dealing with synchronization issues.
Further studies on \ac{5G} positioning in harsh environments can be found in~\cite{Mendrzik_j19,torsoli_blockage_2023,wen_5g_2021, xhafa_evaluation_2022, Dardari_j22}.
The work in~\cite{Mendrzik_j19} provides a theoretical analysis of the position and orientation accuracy 
achieved by harnessing \ac{NLOS} components.
In~\cite{torsoli_blockage_2023}, the concept of blockage intelligence is introduced, showing that a probabilistic description of the propagation environment (especially indoors, such as factories) can be profitably embedded into positioning algorithms.
Authors of~\cite{xhafa_evaluation_2022} demonstrate that joint synchronization, positioning,
and mapping are possible even when the \ac{LOS} path is
blocked, and the reflecting surfaces are only characterized by diffuse scattering.
Lastly, in~\cite{Dardari_j22}, the feasibility of localizing a \ac{UE} with one \ac{BS} under \ac{NLOS} conditions is shown by exploiting the reflections from a
\ac{RIS} in near-field propagation regime.}

\textcolor{black}{Most of the other existing surveys and tutorials currently available in the literature are not fully focused on \ac{5G} positioning; still, they cover a variety of related topics.
\begin{table*}[!ht]
    \centering
\caption{\textcolor{black}{Comparison of existing surveys and tutorials on cellular positioning} } 
    \label{tab:contributions}
    \begin{tabular}{ l | c | c | c | c | c | c | c | c | c | c | c}
        \toprule
        \multirow{2}{4em}{\textbf{Ref.}} & \multirow{2}{*}{\textbf{Year}} & \multicolumn{3}{c|}{\textbf{Cellular Evolution}} & \multirow{2}{*}{\textbf{Use Cases}} & \textbf{Positioning} & \textbf{Positioning} & \multicolumn{4}{c}{\textbf{Simulations}} \\
          &  & \textbf{1G $\rightarrow$ 4G} & \textbf{5G} & \textbf{B5G} &  & \textbf{Architecture} & \textbf{Method} & \textbf{Analytic} & \textbf{Ray Tracing} & \textbf{Outdoor} & \textbf{Indoor} \\
         \midrule
         \cite{Wunder_j17} & 2017 & \xmark & \textcolor{lightgray}{\cmark} & \xmark  & \cmark & \xmark & \cmark & \xmark & \xmark  & \xmark & \xmark \\
         \cite{RosRauGon:J18} & 2017 & \cmark & \textcolor{lightgray}{\cmark} &  \xmark &  \xmark & \cmark &  \textcolor{lightgray}{\cmark} &  \xmark &  \xmark &  \xmark & \xmark \\
         \cite{shahmansoori_survey_2017} & 2017 & \xmark & \textcolor{lightgray}{\cmark} &  \xmark &  \xmark & \textcolor{lightgray}{\cmark} &  \xmark &  \cmark &  \xmark &  \xmark & \xmark \\
         \cite{Aboutanios_j17} & 2017 & \textcolor{lightgray}{\cmark} & \textcolor{lightgray}{\cmark} & \xmark  & \xmark & \xmark & \textcolor{lightgray}{\cmark} & \xmark & \xmark  & \xmark & \xmark \\
         \cite{8000805} & 2017 & \cmark & \textcolor{lightgray}{\cmark} & \xmark & \xmark & \cmark & \textcolor{lightgray}{\cmark} & \xmark & \xmark & \xmark & \xmark \\
         \cite{GioPolZor:J19} & 2018 &  \xmark & \textcolor{lightgray}{\cmark} &  \xmark &  \xmark &  \xmark & \textcolor{lightgray}{\cmark} & \cmark &  \xmark & \cmark & \xmark \\
         \cite{laoudias_survey_2018} & 2018 & \textcolor{lightgray}{\cmark}  & \textcolor{lightgray}{\cmark} & \xmark  & \xmark & \textcolor{lightgray}{\cmark} & \xmark & \xmark & \xmark & \xmark & \xmark\\
         \cite{wen_survey_2019} & 2019 & \xmark & \xmark  & \xmark & \xmark & \xmark & \textcolor{lightgray}{\cmark} & \xmark & \xmark & \xmark & \xmark\\
         \cite{navarro-ortiz_survey_2020} & 2020 & \xmark & \textcolor{lightgray}{\cmark} & \xmark & \cmark & \xmark & \xmark & \xmark & \xmark & \xmark & \xmark\\
         \cite{JiaHanHabSch:J21} & 2021 & \cmark & \xmark  & \cmark & \textcolor{lightgray}{\cmark} & \xmark & \xmark & \xmark & \xmark & \xmark & \xmark \\
         \cite{alalewi_5g-v2x_2021} & 2021 & \xmark & \textcolor{lightgray}{\cmark} & \textcolor{lightgray}{\cmark} & \cmark & \xmark & \xmark & \xmark & \xmark & \xmark & \xmark\\
         \cite{mogyorosi_positioning_2022} & 2022 & \xmark & \textcolor{lightgray}{\cmark} & \cmark & \cmark & \textcolor{lightgray}{\cmark} & \xmark  & \xmark & \xmark & \xmark & \xmark\\
         \cite{surveymmWave} & 2022 & \xmark & \textcolor{lightgray}{\cmark} & \cmark & \textcolor{lightgray}{\cmark} & \textcolor{lightgray}{\cmark} & \cmark  & \xmark & \xmark & \xmark & \xmark\\
         \cite{Vaezi_cellular_2022} & 2022 & \xmark & \textcolor{lightgray}{\cmark} & \cmark & \textcolor{lightgray}{\cmark} & \xmark & \xmark & \xmark & \xmark & \xmark & \xmark\\
         \cite{MolSopEduSam:J22} & 2022 & \xmark & \textcolor{lightgray}{\cmark} & \textcolor{lightgray}{\cmark} &  \textcolor{lightgray}{\cmark} & \xmark & \xmark & \xmark & \xmark & \xmark & \xmark\\
         \cite{chen2022tutorial} & 2022 &  \xmark &  \xmark & \cmark &  \xmark &  \xmark &  \xmark & \cmark & \xmark & \xmark & \xmark \\
         \midrule
         \textbf{This Work} & 2024 & \cmark & \cmark & \cmark & \cmark & \cmark & \cmark & \xmark & \cmark & \cmark & \cmark \\
        \bottomrule
    \end{tabular}
    \begin{tablenotes}
    \item[] Symbol {\cmark} indicates that the work fully covers the topic, while \textcolor{lightgray}{$\cmark$} indicates a partial coverage of the topic. Symbol {\xmark} specifies the topic is not addressed. 
    \end{tablenotes}
\end{table*}
The tutorial in~\cite{GioPolZor:J19} focuses on beam management procedures for \ac{mmWave} cellular networks.
Mobile traffic and its characterization according to the application are discussed in~\cite{navarro-ortiz_survey_2020}.
The visions on \ac{B5G} drivers, use cases,  requirements, \acp{KPI}, architectures, enabling technologies, and algorithms given in~\cite{JiaHanHabSch:J21,Vaezi_cellular_2022,iliadis_road_2022} attempt to shape the forthcoming revolution brought by \ac{6G} technology. 
Specifically, authors in~\cite{JiaHanHabSch:J21} provide a general view by explaining the motivation for the advent of \ac{6G}; the work in~\cite{Vaezi_cellular_2022} is dedicated to the application of \ac{IOT} in the contexts of cellular, wide-area, and \acp{NTN}; while~\cite{iliadis_road_2022} is focused on \ac{DNN} application for cell-free massive \ac{MIMO}. 
Looking towards \ac{6G}, tutorials on \ac{mmWave} and \ac{THz} communication and localization have been proposed~\cite{MolSopEduSam:J22,chen2022tutorial}; 
the former work is focused on mathematical modeling, while the latter is shaped with an application-oriented perspective and compares \ac{mmWave} and \ac{THz} technologies on the achievable localization performances.}

\textcolor{black}{Previous works highlight the necessity for a comprehensive guideline on \ac{5G} positioning, guiding the reader from the fundamentals of positioning to the latest literature enhancements, complemented by a side vision of the evolution of the standards and applications. 
We acknowledge a gap in developing realistic environment-dependent simulations through \ac{RT} tools, which are essential for accurately accounting for the presence of obstacles impacting the \ac{UE}-\ac{BS} visibility.
Most of the prior art is typically focused on a single scenario; thus, the findings have poor generalization. Here, we exhaustively analyze several combinations of environments, mobility conditions, visibility, and 5G signal configurations, offering a thorough set of outcomes and conclusions encompassing a complete vision of the potential of 5G positioning.} 

\textcolor{black}{A comparison of this work with respect to existing surveys and tutorials available in the literature is summarized in Table~\ref{tab:contributions}, where we highlight the contents of each reference in terms of the cellular technology addressed, use case descriptions and requirements, discussion of the positioning architecture and methods, and types of simulation analyses.}

\tikzset{every node/.append style={scale=0.99}}  
\definecolor{lightseagreen}{rgb}{0.13, 0.7, 0.67}
\definecolor{Color1}{HTML}{F4D6B0}
\definecolor{Color2}{HTML}{EEB8C6}
\definecolor{Color3}{HTML}{B1A6CE}
\definecolor{Color4}{HTML}{A7D4E7}
\definecolor{Color5}{HTML}{ABD4CE}

\begin{figure*}[!ht]
\centering
    \begin{tikzpicture}[scale=0.99]
\path[mindmap,concept color=lightgrey!50,grow cyclic]
    node[concept, text=black] {A Tutorial on 5G Positioning}
    [counterclockwise from=126, level 1 concept/.append style={sibling angle=72}]
    child[concept color=Color1!70] {
      node[concept] {Historical Overview {\footnotesize (Sec. \ref{sec:State of the art})}}
      [counterclockwise from=57]
      child [concept color=Color1!30]{ node[concept]  {Use Cases} }
      child [concept color=Color1!30]{ node[concept] {From 1G to 4G} }
      child [concept color=Color1!30]{ node[concept]  {5G from Rel-15 to Rel-19} }
      child [concept color=Color1!30]{ node[concept]  {Beyond 5G} }
    }
    child[concept color=Color2!70] {
      node[concept] {Fundamentals {\footnotesize (Sec. \ref{sec:System Model})}}
      [counterclockwise from=135]
      child [concept color=Color2!30]{ node[concept]  {Wireless Channel} }
      child [concept color=Color2!30]{ node[concept]  {Measurements} }
      child [concept color=Color2!30]{ node[concept]  {Algorithms} }
      child [concept color=Color2!30]{ node[concept]  {Bayesian Tracking} }
    }  
    child[concept color=Color3!70] { 
      node[concept] {5G Positioning Technology {\footnotesize (Sec. \ref{sec:5G Positioning Technology (Rel-16)})}} [counterclockwise from=205]
      child [concept color=Color3!30]{ node[concept] {Architecture} }
      child [concept color=Color3!30]{ node[concept] {Signals} }
      child [concept color=Color3!30]{ node[concept] {Methods} }
      child [concept color=Color3!30]{ node[concept] {Measurement Extraction} }
    }
    child[concept color=Color4!70] { 
    node[concept] {Simulations {\footnotesize (Sec. \ref{sec:Simulation Analysis})}} [clockwise from=45]
      child [concept color=Color4!30]{ node[concept]  {Environments} }
      child [concept color=Color4!30]{ node[concept]  {Settings} }
      child [concept color=Color4!30]{ node[concept]  {Results} }
    }   
    child[concept color=Color5!70] { 
    node[concept] {Challenges {\footnotesize (Sec. \ref{sec:lessons_learned_limitations})}} [clockwise from=180]
      child [concept color=Color5!30]{ node[concept]  {Lessons Learned}}
      child [concept color=Color5!30]{ node[concept]  {Synch.} }
      child [concept color=Color5!30]{ node[concept]  {Deployment} }
      child [concept color=Color5!30]{ node[concept]  {Hardware} }
      child [concept color=Color5!30]{ node[concept]  {Private Networks} }
    };   
    \end{tikzpicture}
    \caption{
    \textcolor{black}{Mind map visualizing the contents of this manuscript and the associated sections.}}
    \label{fig:flowchart}
\end{figure*}
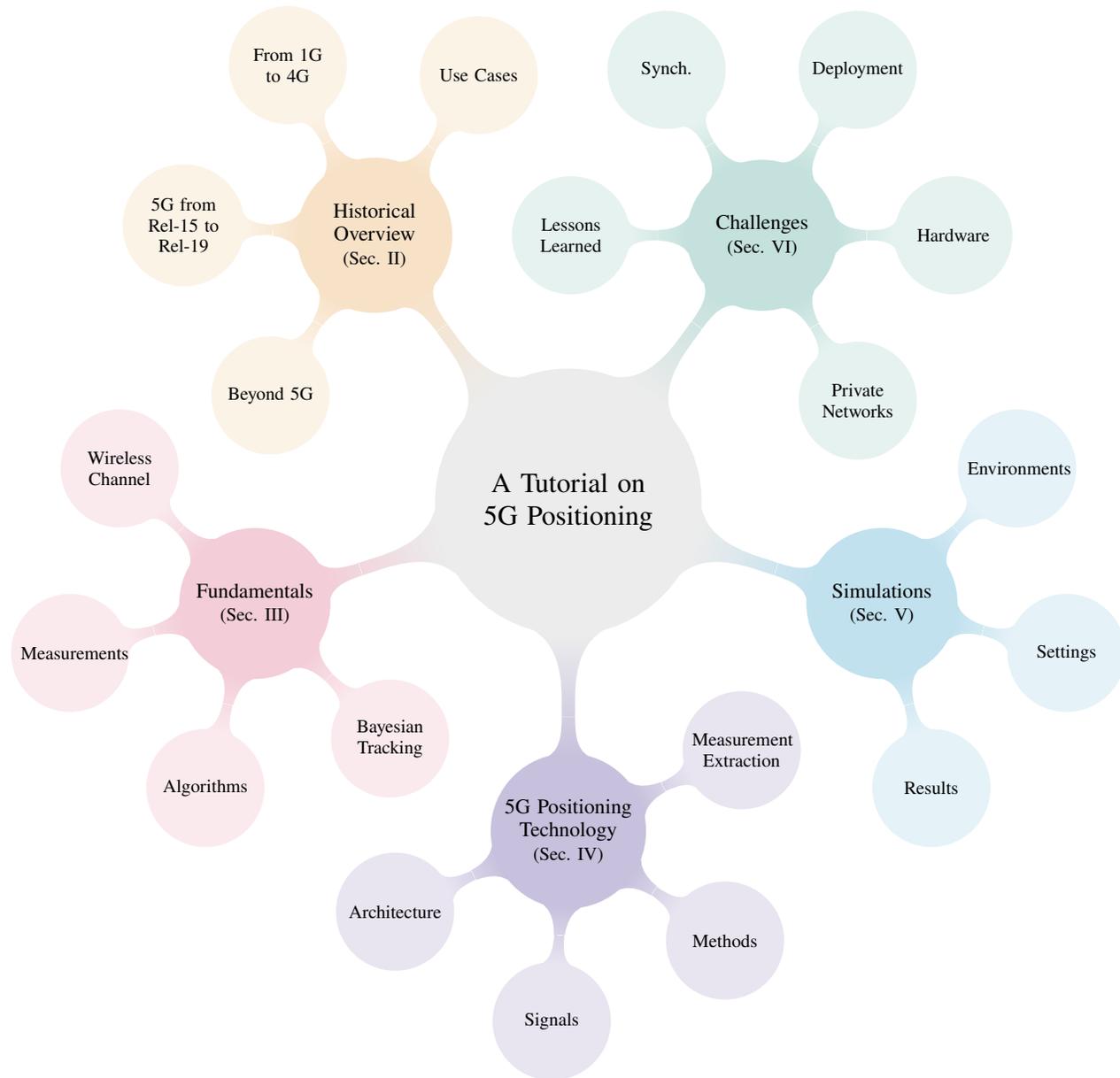


\subsection{Contribution}
\label{sec:contribution}
By proceeding over the survey in~\cite{RosRauGon:J18}, which provides an historical overview of cellular positioning from \ac{1G} to \ac{3GPP} Rel-15, this tutorial paper aims to provide 
the reader a comprehensive and accessible reference guideline to the convoluted world of \ac{5G} positioning, by offering a short summary of historical developments, contextualization of the current state of research, and an outlook over future developments. It is designed to cater to a diverse audience, ranging from researchers and engineers seeking an in-depth understanding of the subject to practitioners looking for practical insights into harnessing \ac{5G} positioning for real-world applications.
With this approach, we characterize the maturity level of the technology and analyze the enabled use cases. We also discuss the main industrial and technological trends, as well as research advances inherited by previous generations of cellular networks. By providing an overview of standardization activities and highlighting fundamental research, we define potential directions of forthcoming \ac{B5G} systems and their associated breakthrough applications.
We also review experimental positioning activities by analyzing state-of-the-art solutions and algorithms. At the same time, this work presents a thorough assessment of \ac{5G} positioning capabilities under different system configurations that are useful to understand the achievable performance by varying the settings.

The main contributions are the following:
\begin{itemize}
\item We provide an overview of the evolution of cellular positioning, from the first development until the current \ac{5G} version, with an overlook over the forthcoming releases, analyzing the enhancements introduced over the generations and the current innovation trends;
\item We provide a detailed description of the standardized \ac{5G} positioning signals as foreseen by the \ac{3GPP} standard, specifying their configuration parameters and usability. This involves an exploration of the specific features of these signals 
and their role in enabling accurate and efficient positioning;
\item We conduct a thorough examination of \ac{5G} positioning architectures and methods by discussing the various solutions that can be employed to achieve precise positioning;
\item We carry out extensive \ac{5G} positioning simulations in outdoor and indoor scenarios that are relevant for challenging use cases such as automotive or industrial automation. We consider both static and mobile \ac{UE} positioning, analyzing different system parameters and configurations such as numerology, positioning methodology, and antenna array configuration; 
\item We discuss the current limitations of \ac{5G} positioning by providing the reader an easy understanding of the main challenges that research and industry are addressing for releasing cellular-based location services. Lastly, we delineate potential avenues for future research in cellular positioning.
\end{itemize}


\subsection{Tutorial organization}
\textcolor{black}{As highlighted in the mind map in Fig.~\ref{fig:flowchart}, this tutorial is organized as follows:
Section~\ref{sec:State of the art} starts by motivating why \ac{5G} positioning is useful in exemplary use cases taken from industrial and automotive domains, and then presents the evolution of cellular positioning from a historical perspective from \ac{1G} to the latest releases, diving into the future of \ac{B5G} trends. 
In Section~\ref{sec:System Model}, we first review the fundamentals of wireless localization, describing the different classes of positioning measurements and positioning/tracking algorithms.
Section~\ref{sec:5G Positioning Technology (Rel-16)} is devoted to the description of the \ac{5G}  positioning architecture, the associated reference signals, as well as the \ac{5G} positioning methods. 
Section~\ref{sec:Simulation Analysis} focuses on simulation analyses, with a description of performance metrics, the simulation environment, and parameters, and achieved results for a number of different system configurations.
\textcolor{black}{Section~\ref{sec:lessons_learned_limitations} analyzes the results, highlighting the lessons learned in the previous sections and delineating current limitations impairing cellular positioning.}
Concluding remarks and future directions are discussed in Section~\ref{sec:Conclusions}.}

\subsection{Notation}

\textcolor{black}{Vectors are denoted by boldface lower-case letters (e.g., $\boldsymbol{a}$) and matrices by boldface upper-case letters (e.g., $\boldsymbol{A}$).
The number of elements of an array, i.e., the cardinality, is indicated by $|\boldsymbol{a}|$, while $\lVert \boldsymbol{a} \rVert$ denotes the l2-norm of $\boldsymbol{a}$.
The transpose of a matrix $\boldsymbol{A}$ is written as $\boldsymbol{A}^{\T}$, its Hermitian as $\boldsymbol{A}^{\mathrm{H}}$, while $\boldsymbol{A}^{-1}$ denotes the inverse operation. 
The notation $\mathrm{diag}(\boldsymbol{a})$ is used to denote a diagonal matrix with vector $\boldsymbol{a}$ as its main diagonal, $\mathrm{tr}(\boldsymbol{A})$ is the trace of matrix $\boldsymbol{A}$. 
$[\boldsymbol{A}]_{i,j}$ indicates the $i$-th row and $j$-th column of the matrix $\boldsymbol{A}$, and $[\boldsymbol{A}]_{i_1:i_2,j_1:j_2}^{}$ indicates the  selection of the matrix rows between indices $i_1^{}$ and $i_2^{}$ and matrix columns between indices $j_1^{}$ and $j_2^{}$.
$\mathrm{Cov}(\cdot)$ denotes the covariance and $\mathrm{E}[\cdot]$ the expected value.
When vector $\boldsymbol{a}$ follows a Gaussian distribution, it is referred to as
$\boldsymbol{a} \sim \mathcal{N}(\mathrm{E}[\boldsymbol{a}],\mathrm{Cov}(\boldsymbol{a}))$.
$\mathbb{R}$ and $\mathbb{C}$ indicate the sets of real and complex numbers, respectively.}


\section{5G Positioning: History, Present, and Future}
\label{sec:State of the art}

In this section, we provide an overview of cellular positioning, starting from the targeted use cases to the technological evolution put in place to satisfy the performance requirements of such use cases, with a closer look at the latest \ac{5G} releases and future trends. Section~\ref{sec:use_cases} investigates the positioning use case requirements; Section~\ref{sec:The Evolution of Cellular Positioning} summarizes the evolution of the technology from the early days of analog cellular networks to the modern era of \ac{5G}  positioning; Section~\ref{sec:5G Cellular Positioning (from Rel-15 to Rel-19)} discusses the specific features of \ac{5G} positioning, from the first release of \ac{5G} (3GPP Rel-15) up to the forthcoming Rel-19.
By the end of this section, the reader should have a better understanding of the evolution of cellular positioning and the advancements conceived in the design of \ac{5G} positioning.

\subsection{Cellular positioning use cases}
\label{sec:use_cases}

\ac{5G} positioning targets a wide range of use cases with highly different performance requirements. Main positioning \acp{KPI} includes accuracy,  availability, latency, coverage, energy consumption, and update rate, which contribute to determining the feasibility (or not) of a specific service. 
To this extent, the document~\cite{ts122261} specifies seven service levels to be guaranteed by \ac{5G} positioning systems.
Regarding the association between positioning accuracy and the standard releases, we report that Rel-16 for commercial use cases aims to guarantee 3~m for horizontal accuracy~\cite{tr122872}, while in Rel-17 it is set to 20~cm. 
Other safety-critical metrics to be taken into account are reliability and integrity, which are related to the degradation of the positioning accuracy and the trustworthiness of the positioning system~\cite{Blefari21}.

Among the verticals that would benefit from \ac{5G}  positioning, a critical one is the automotive sector, where the enhancements on automated (and autonomous) services call for highly accurate positioning with ultra-low latency and high reliability~\cite{ts122186,Kuutti:j18}.
A description of the envisioned automotive use cases as prescribed by the \ac{5GAA}~\cite{5GAA_UC2, 5GAA_UC3} with associated positioning accuracy is reported in
Table~\ref{tab:UC5GAA}. These requirements were already envisioned in~\cite{WymGraDest2017}, where \ac{5G}  is indicated as the most promising technology able to meet all of them.

Another major class of use cases refers to indoor positioning, which has been widely studied and discussed due to the necessity to guarantee safety for clients and workers such as in hospital~\cite{lu_localization_2013, konecny_real-time_2018, buettner_rfid_2020} or workspace~\cite{buffi_rfid-based_2021, zhao_enhancing_2022}. 
In particular, we can distinguish between consumer applications and industrial services. The former can tolerate relatively low positioning accuracy (3~m) and high latency (1~s), while the latter has stricter requirements. Specifically, most of the industrial needs are related to asset tracking~\cite{5GACIA_UC}, where positioning accuracy in the order of centimeters and latency in the order of milliseconds is requested~\cite{indoor, ts122261}. Table~\ref{tab:UC5GACIA} reports some indoor use cases, specifying horizontal accuracy, maximum \ac{UE} speed, and latency.

\begin{table*}[ht]
    \centering
    \caption{5G positioning: C-V2X enhanced services and requirements~\cite{5GAA_UC2, 5GAA_UC3} } 
    \label{tab:UC5GAA}
    \begin{tabular}{p{5cm}  c  c  c }
        \toprule
         \textbf{Use case} & \textbf{Positioning accuracy [cm]} & \textbf{Latency [ms]} &
         \textbf{Max UE speed [m/s]}
         \\
         \midrule 
              & & &  \\[-9pt]
        Cooperative lateral parking & 20 & 
        10 & 1.38 \\
        \midrule
        Automated intersection crossing & 15 & 10 & 
        33.3  \\
        \midrule
         \shortstack[l]{Cooperative maneuvers of autonomous \\ vehicles for emergency situations} & 20 & 10 & 
        69.4\\
        \midrule
        \shortstack[l]{Infrastructure assisted environment \\ perception} & 10 & 100 & 69.4\\
        \midrule
        Vehicles platooning in steady state & 50 & 
        50 & 27.8 \\
        \midrule
        Vehicle decision assist & 150 & 50 & 27.8\\
        \midrule
        Cooperative adaptive cruise control  & 50 & 
        10 & 60 \\
        \bottomrule
    \end{tabular}
\end{table*}
\begin{table*}[ht]
    \centering
    \caption{5G positioning: indoor services and requirements~\cite{5GACIA_UC, tr122872, ts122261}}
    \label{tab:UC5GACIA}
    \begin{tabular}{p{5cm}  c  c  c }
        
        \toprule
        \textbf{Use case} & \textbf{Positioning accuracy [cm]} & \textbf{Latency [ms]} & \textbf{Max \ac{UE} speed [m/s]} \\
        \midrule
         &   &    &  \\[-9pt]
        Augmented reality in smart factories &  100 &  15  &  
        2.8\\
        \midrule
         \shortstack[l]{Mobile control panels with safety functions \\ within factory danger zones} &  100 &  1000 & -\\
        \midrule
         \shortstack[l]{Inbound logistics for manufacturing \\ (goods storage)} &  20 &  1000 & 
        8.3\\
        \midrule
        Trolley location in factories & 50 & 20 & 
        13.9\\
        \midrule
        eHealth: patient tracking & 100-300 & - & 
        5.6\\
        \bottomrule
        
    \end{tabular}
\end{table*}


\textcolor{black}{The reported use cases for \ac{C-V2X} and indoor services are recognized as benchmarks and contain valuable information for the research and industries. 
Notice that a critical aspect of the specification of requirements (especially for safety-related constraints) is also attributable to the speed of involved terminals, which affects positioning accuracy, latency, and integrity. Guaranteeing the same level of positioning accuracy requirement at higher speeds poses a greater challenge compared to nearly-static mobility conditions.}

\subsection{Evolution of cellular positioning technology from 1G to 4G}

\label{sec:The Evolution of Cellular Positioning}
Localization functionalities were introduced for the first time in cellular networks in the mid-1990s due to the specific requirements issued by enhanced emergency call services in the \ac{US}~\cite{RosRauGon:J18}. 
\textcolor{black}{Even if localization procedures were not mentioned in the early cellular standards, localization solutions had been adopted since \ac{1G} to target the \ac{UE} position, particularly for vehicles. In the beginning, only methods based on signal strength were used, although the idea of exploiting a coarse  \ac{AOA}  estimation by directive antennas had been raised~\cite{Ott1977}.}

The \ac{e991} requirements approved by the \ac{FCC}~\cite{e911} encouraged the study for more accurate localization methods in \ac{2G}  cellular systems, introduced with the \ac{GSM} standard. 
In \ac{2G} systems, while the primary focus was on \ac{UL}-\ac{TDOA} for localization, the framework also acknowledged the potential of \ac{AOA}, fingerprinting, and other methods. 
Indeed, further studies demonstrated the feasibility of \ac{AOA} estimation with \ac{GSM} network by using \ac{DRSS}~\cite{DebVan2007}.

With the introduction of the \ac{3G} and the globalization of cellular communications driven by the \ac{3GPP}, cellular localization methods initiated a standardization process. 
The goal of \ac{3GPP} was to support emergency services and foster location-based applications.
With the advent of \ac{3G}, the following network-based localization solutions have been introduced: \ac{TOA}, \ac{TDOA}, \ac{AOA}, \ac{CID}, fingerprinting, and hybrid methods~\cite{zhao_standardization_2002}. Moreover, \ac{3G} was used to augment \ac{GPS} with differential corrections, providing a navigation message to reduce the \ac{TTFF} and facilitate tracking. This method was already standardized in \ac{2G} under the name of \ac{A-GPS}.
The \ac{UMTS}, as the successor of \ac{GSM}, was one of the candidate technologies to define an international standard for \ac{3G} networks. \ac{UMTS} was delineated by \ac{3GPP} and its main air interface was called \ac{UTRA}.

\textcolor{black}{Transitioning from \ac{3G} to \ac{4G}, the \ac{LTE} standard marked the progression from \ac{GSM} and \ac{UMTS}, introducing the \ac{E-UTRA} air interface.}
\ac{E-UTRA} is based on \ac{OFDMA} in \ac{DL} and \ac{SC-FDMA} in \ac{UL}. One of the objectives of \ac{LTE} localization was to act as a backup to the \ac{A-GPS} when satellite visibility is not ensured. Therefore, a \ac{PRS} was designed for \ac{DL} purposes. With Rel-9 in 2009, \ac{LTE} positioning had a major breakthrough. Multiple positioning methods were defined, such as \ac{eCID} and \ac{OTDOA}, adopting the newly designed \ac{PRS}. Moreover, the \ac{LPP} was defined in \ac{3GPP} \ac{TS} 36.355~\cite{ts136355}, and \ac{A-GNSS} was included in \ac{3GPP} \ac{TS}  36.305~\cite{ts136305}.

From Rel-10, the standardization of \ac{LTE-A} starts to include the \ac{UL}-\ac{TDOA} method based on \acp{SRS} to complement \ac{A-GNSS}. Furthermore, an improvement of \acp{PRS} was proposed to increase the hearability. The hearability problem arises when a user needs to communicate with multiple \acp{BS} and differentiate the communication systems from positioning systems. In Rel-13, a further enhancement has been made with the \ac{LTE-A} Pro, mainly addressed for strict indoor environments.
Two of the main improvements referred to \ac{OTDOA} enhancement (new \ac{PRS} patterns and bandwidth extension) and \ac{MIMO} introduction (multi-antenna arrays for beamforming).
The introduction of \ac{3GPP} Rel-14, as well as continuing the \ac{LTE} evolution, also sets the starting point for \ac{5G}~\cite{TR121914}.

\subsection{5G positioning from Rel-15 to Rel-19}
\label{sec:5G Cellular Positioning (from Rel-15 to Rel-19)}
Between 2017 and 2018, Rel-15 established the \ac{5G} technology foundation~\cite{tr121915}, which includes a range of features and capabilities designed to improve the performance and functionality of cellular networks. Rel-15, also known as \textit{5G Phase 1}, supports the use of both sub-6~GHz and millimeter-wave bands for \ac{5G}  communications and defines the following main use cases:
\begin{itemize}
    \item \textit{\ac{eMBB}}: designed to support data rates of up to several gigabits per second and to enable the use of high-bandwidth applications;
    \item \textit{\ac{mMTC}}: designed to support a large number of connected devices and to enable low-power, low-cost communication for these devices;
    \item \textit{\ac{URLLC}}: designed to support latency of less than 1 ms and reliability of up to 99.999\%.
\end{itemize}
Rel-15 mainly focuses on the first use case, also thanks to the introduction of network slicing, which allows different parts of a \ac{5G}  network to be configured and optimized for specific use cases, allowing for higher flexibility and supporting a wider range of services.
Moreover, the adoption of mobile edge computing is able to improve the performance of \ac{5G} networks and reduce latency~\cite{yifan2016}. Lastly, it includes enhanced \ac{V2X} communications, enabling vehicles to communicate with each other and with infrastructure elements, such as \acp{RSU}. 
Since Rel-15 primarily lays the foundations for the \ac{5G} \ac{NR} technology, no further positioning enhancements have been developed with respect to \ac{LTE}.

\textit{5G Phase 2} starts with Rel-16 at the end of 2018, which is built on the characteristics of Rel-15 and includes additional features and enhancements. In particular, it focuses on \ac{URLLC} and \ac{mMTC} use cases and includes support for the 6~GHz bands~\cite{parkvall_2020}. 
From a positioning point of view, Rel-16 is one of the most valuable releases.
\textcolor{black}{First of all, 3GPP Rel-16 sets the basis for the 5G \acp{LCS} in the \ac{TS} 23.273~\cite{ts123273}.}
Then, using older signals as a basis, Rel-16 defines \ac{DL}-\ac{PRS} and \ac{UL}-\ac{SRS} signals, i.e., the enhanced versions of \ac{PRS} in LTE and \ac{SRS} of Rel-15, respectively. For this reason, throughout this tutorial, they will be referred to as \ac{PRS} and \ac{SRS}.
These new reference signals improve the positioning accuracy and lower the communication overhead. In fact, \acp{PRS} have the capability to report \acp{TOA} from multiple \acp{gNB} simultaneously, and, together, they can be employed to compute \ac{RTT}. Furthermore, Rel-16 supports operations in the \ac{FR}1 and \ac{FR}2, covering the ranges of 410~MHz~–~7.125~GHz and 24.25~–~52.6~GHz, respectively, where larger bandwidths are available, thus enhancing the ranging accuracy. \textcolor{black}{In Rel-16, 3GPP also mentions the possibility of introducing a new \ac{FR} (unofficially referred to as \ac{FR}3) to enable cellular communication in the range between 7 and 24 GHz~\cite{tr138802}. Its standardization is expected to be included in future releases.}

At the end of 2020, \ac{3GPP} published Rel-17 based on the features proposed in the previous release. Key contributions for \ac{5G} positioning are the introduction of the support for 2.5~GHz and 4.5~GHz bands, the increased \acp{gNB}' coverage, and the improvements related to edge computing, network slicing, and \ac{V2X} communications. Moreover, \ac{FR}2 is extended up to 71~GHz. 
The main positioning improvements include~\cite{ren2021progress}:  
\begin{itemize}
    \item \textit{Timing delay correction at \ac{Tx} and \ac{Rx} sides}: \ac{Tx}/\ac{Rx} timing delay is a problem affecting ranging measurements, and it involves the generation, transmission, and reception of \ac{PRS} and \ac{SRS}. This error persists even after the internal calibration of \ac{UE} and \ac{TRP}, and the accuracy of timing-related positioning methods may be significantly affected, as reported in \ac{3GPP} \ac{TR} 38.857~\cite{tr138857}. Rel-17 introduces \acp{TEG} in order to mitigate this phenomenon~\cite{ts138214}. When multiple signals are sent from the same \ac{TRP}, they are expected to have a similar \ac{Tx} error; therefore, they are associated with the same group. Instead, signals from different \acp{TRP} should have a different \ac{Tx} error and may belong to different groups. Therefore, associating the \ac{TEG} identifier to the signal could be helpful for reducing  \ac{Tx}/\ac{Rx} timing delay error~\cite{ts138214,tr138857}.
    \item \textit{\ac{UL}-\ac{AOA} and \ac{DL}-{AOD} enhancements}: \ac{UL}-\ac{AOA} enhancements include additional assistance data, such as expected \ac{AOA} and its uncertainty through a search window, and multi-angle reporting.
    In particular, this last feature permits to discern the \ac{LOS} within a group of multipath components that exhibit similar delay profiles.
    Rel-17 also introduces the \ac{UL}-\ac{SRS} \ac{RSRPP}, which indicates the power of the received \ac{SRS} for a given path. On the other hand, \ac{DL}-{AOD} is based on \ac{DL}-\ac{PRS} \ac{RSRP}, which is the measurement used to select the best \ac{AOD}. However, this measurement also takes into account multipath components, which are undesirable. Therefore, as for its \ac{UL} counterpart, Rel-17 introduces the \ac{DL}-\ac{PRS} \ac{RSRPP}, which is a measurement associated with the path and not with the entire channel, as well as the search window for \ac{DL}-\ac{AOD}.
    \item \textit{Multipath mitigation}: it consists of reporting not only a single path but also additional paths (up to 8) as a part of timing estimation.
    \item \textit{\ac{LOS}/\ac{NLOS} identification}: it is provided using additional information, such as \ac{LOS}/\ac{NLOS} indicators, which could be a boolean value (i.e., 0 or 1) or a likelihood (between 0 and 1 with a step of 0.1)~\cite{ts137355}.
\end{itemize}
\textcolor{black}{Moreover, the concept of position integrity is improved over Rel-15, and the positioning integrity monitoring, already supported by \ac{GNSS}, is included in Rel-17 \cite{tr138857}. The following \acp{KPI} are defined:
\begin{itemize}
    \item \textit{\ac{AL}}: The maximum positioning error allowed for the specific use case;
    \item \textit{\ac{TTA}}: The maximum elapsed time to provide an alert when the positioning error exceeds the \ac{AL};
    \item \textit{\ac{TIR}}: The probability that the positioning error exceeds the \ac{AL} without warnings within the \ac{TTA}.
\end{itemize}}

In June 2021, at the \ac{3GPP} \ac{RAN} Rel-18 Workshop, the concept of \textit{5G Advanced} was proposed with the aim of paving the way for \ac{6G}. Rel-18 is expected to bring further enhancements over the previous releases and introduce more intelligence into the wireless cellular network, with pervasive \ac{AI} solutions spread over different network layers~\cite{burghal_comprehensive_2020}. 
The main focus of Rel-18 is to enhance network energy savings, coverage, mobility support, \ac{MIMO} evolution, multicast and broadcast service, and positioning~\cite{lin_overview_2022}. Related to positioning, it should accommodate for \ac{CPP}, a \ac{GNSS}-native technology capable of reaching cm-level accuracy~\cite{chen_carrier_2022, 9977723} but limited to outdoor applications, adapting the already standardized signals. \textcolor{black}{Open challenges and potential solutions for indoor scenarios are provided in~\cite{10475845}.}
At the same time, Rel-18 will support \ac{LPHAP} requirements and positioning functionalities for \ac{RedCap} \acp{UE}. 
\textcolor{black}{Moreover, the enhanced support for \ac{AI} and \ac{ML} solutions is driving researchers to revolutionize beam management through spatial and temporal prediction, as well as to improve positioning directly (e.g., fingerprinting) or by using \ac{ML} models to infer and refine measurements \cite{tr138843,alawieh20235g}.}
Lastly, Rel-18 reports the requirements for \ac{SL} positioning and the implementation of ad-hoc \ac{SL} signals based on \ac{PRS} and \ac{SRS}, called \ac{SL}-\ac{PRS}~\cite{tr138859}.

The timeline of standardization bodies will periodically foresee new releases, starting with Rel-19  (work activities opened since mid-2021~\cite{ts122261}) and proceeding over advanced standards defining the evolution of cellular networks.
\textcolor{black}{The new studies involving Rel-19 address the industrial needs not considered in the previous releases. Examples include metaverse services and energy harvesting for \ac{IOT}-enabled factories. Both topics are strongly related to positioning: the estimate of user position and orientation is essential for the representation and interaction of the avatars~\cite{tr122856}, and energy-harvesting tags are a cost-effective way for asset tracking~\cite{tr122882}. 
To better support the applications of \ac{AI}/\ac{ML}, 
future cellular releases will aim to decentralize intelligence across devices rather than confining it solely to the network infrastructure. 
Therefore, data and models will be shared directly between devices without traversing the 5G core network~\cite{tr122876}. Consequently, objectives involve researching potential service and performance requirements necessary to facilitate efficient AI/ML operations via direct device connections.}
\textcolor{black}{During a recent 3GPP meeting held in May 2024, the primary objective was to enhance positioning using AI/ML. Building on the Rel-18 baseline, the discussions focused on assisted and direct AI/ML positioning, improved beam management, and \ac{CSI} feedback enhancements~\cite{R1-2405385}.}

\begin{figure*}
    \centering
    \begin{tikzpicture}

\draw [-{Triangle[width=30pt,length=15pt]}, line width=15pt, azure!20] (-3.95,1.0)to node[black]{\small \textsc{PHASE 1}} (-1.65,1.0);
\draw [-{Triangle[width=30pt,length=15pt]}, line width=15pt, azure!20] (-1.65,1.0)to node[black]{\small \textsc{PHASE 2}} (4,1.0);
\draw [-{Triangle[width=30pt,length=15pt]}, line width=15pt, azure!20] (4,1.0)to node[black]{\small \textsc{5G ADVANCED}} (10,1.0);

\draw[line width = 1pt, lightgrey] (-3.95,-1) -- (-3.95,0.5);
\node[] (Rel15_name) at(-3.4,-0.8){Rel-15};
\draw[line width = 1pt, lightgrey] (-1.65,-1) -- (-1.65,0.5);
\node[] (Rel16_name) at(-1.1,-0.8){Rel-16};
\draw[line width = 1pt, lightgrey] (0.825,-1) -- (0.825,0.5);
\node[] (Rel17_name) at(1.37,-0.8){Rel-17};
\draw[line width = 1pt, lightgrey] (4,-1) -- (4,0.5);
\node[] (Rel18_name) at(4.55,-0.8){Rel-18};
\draw[line width = 1pt,lightgrey] (6.35,-1) -- (6.35,0.5);
\node[] (Rel19_name)at(6.9,-0.8){Rel-19};
\draw[line width = 1pt, dashed,lightgrey] (9,0.5) -- (9,-1);
\node[lightgrey] (Rel20_name)at(9.55,-0.8){Rel-20};
    
    \draw [-stealth,line width=1.5pt,azure](-6,0) -- (11,0);
    
        \foreach \i in {1,2,3,4,5,6,7,8,9,10}
{
    \draw[fill=azure] (-6+1.5*\i, 0) circle (3pt);
}
\node[azure] at(-4.5,0.35){2017};
\node[azure] at(-3,0.35){2018};
\node[azure] at(-1.5,0.35){2019};
\node[azure] at(-0,0.35){2020};
\node[azure] at(1.5,0.35){2021};
\node[azure] at(3,0.35){2022};
\node[azure] at(4.5,0.35){2023};
\node[azure] at(6,0.35){2024};
\node[azure] at(7.5,0.35){2025};
\node[azure] at(9,0.35){2026};

\node[rectangle, draw, rounded corners,  text width=1.9cm, color = azure!35, text=black, below=0.6cm of Rel15_name, xshift=-2.15cm] (Rel15_text)  {\footnotesize 5G foundations: \\
\,\,- eMBB\\
\,\,- URLLC\\
\,\,- mMTC\\
};
\draw[-*,color=azure!35] (Rel15_text.north)--(Rel15_name.south);

\node[rectangle, draw, rounded corners,  text width=3.5cm, color = azure!35, text=black, below=0.6cm of Rel16_name, xshift=-1.3cm] (Rel16_text)  {\footnotesize New positioning signals with increased hearability: \\
\,\,- DL-PRS (downlink)\\
\,\,- UL-SRS (uplink)\\
};
\draw[-*,color=azure!35] (Rel16_text.north)--(Rel16_name.south);

\node[rectangle, draw, rounded corners,  text width=3.1cm, color = azure!35, text=black, below=0.6cm of Rel17_name, xshift=-0cm] (Rel17_text)  {\footnotesize Positioning enhancements:\\
    \,\,- Delay correction\\
    \,\,- Angle enhancement\\
    \,\,- Multipath mitigation\\
    \,\,- NLOS detection \\
    };
\draw[-*,color=azure!35] (Rel17_text.north)--(Rel17_name.south);

\node[rectangle, draw, rounded corners,  text width=3.8cm, color = azure!35, text=black,below=0.6cm of Rel18_name, xshift=0.75cm] (Rel18_text)  {\footnotesize Advanced positioning:\\
\,\,- AI/ML integration\\
\,\,- New methods (CPP, LPHAP) \\
\,\,- New signals (SL-PRS)\\
};
\draw[-*,color=azure!35] (Rel18_text.north)--(Rel18_name.south);

\node[rectangle, draw, rounded corners,  text width=3.6cm, color = azure!35, text=black,below=0.6cm of Rel19_name, xshift=2.55cm] (Rel19_text)  {\footnotesize Full 5G positioning potentials:\\
\,\,- Improved AI/ML\\
\,\,- Energy efficient network\\
};
\draw[-*,color=azure!35] (Rel19_text.north)--(Rel19_name.south);

    \end{tikzpicture}
    \caption{
    \textcolor{black}{Timeline of cellular communication reporting the phases of 5G evolution, the associated 3GPP releases, and the main positioning enhancements.}
    }
    \label{fig:5G_evolution}
\end{figure*}
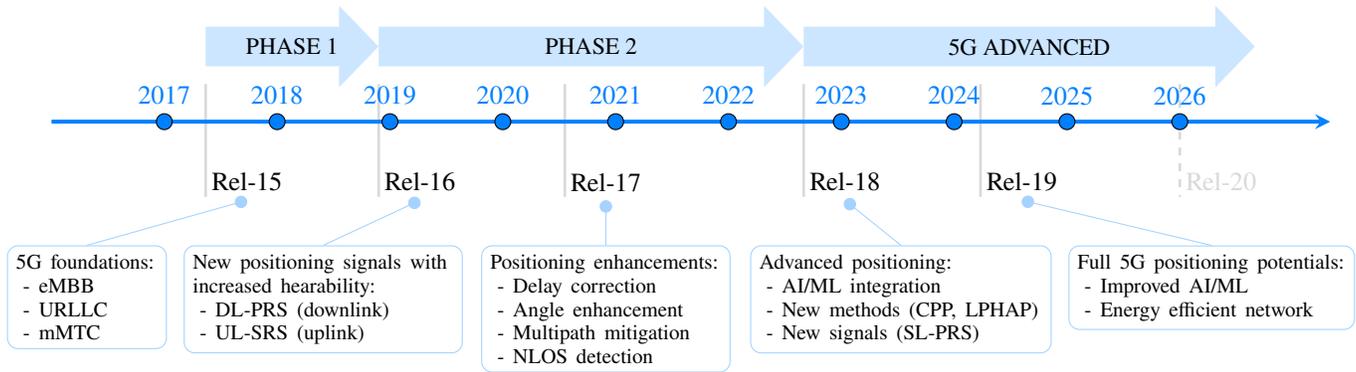

Fig. \ref{fig:5G_evolution} shows the 5G evolution timeline, with a recap of the main positioning enhancements.

\subsection{Positioning trends beyond 5G}
\label{sec:5G positioning trends}

\begin{figure}[!tb]
    \centering
    \includegraphics[width=0.99\columnwidth]{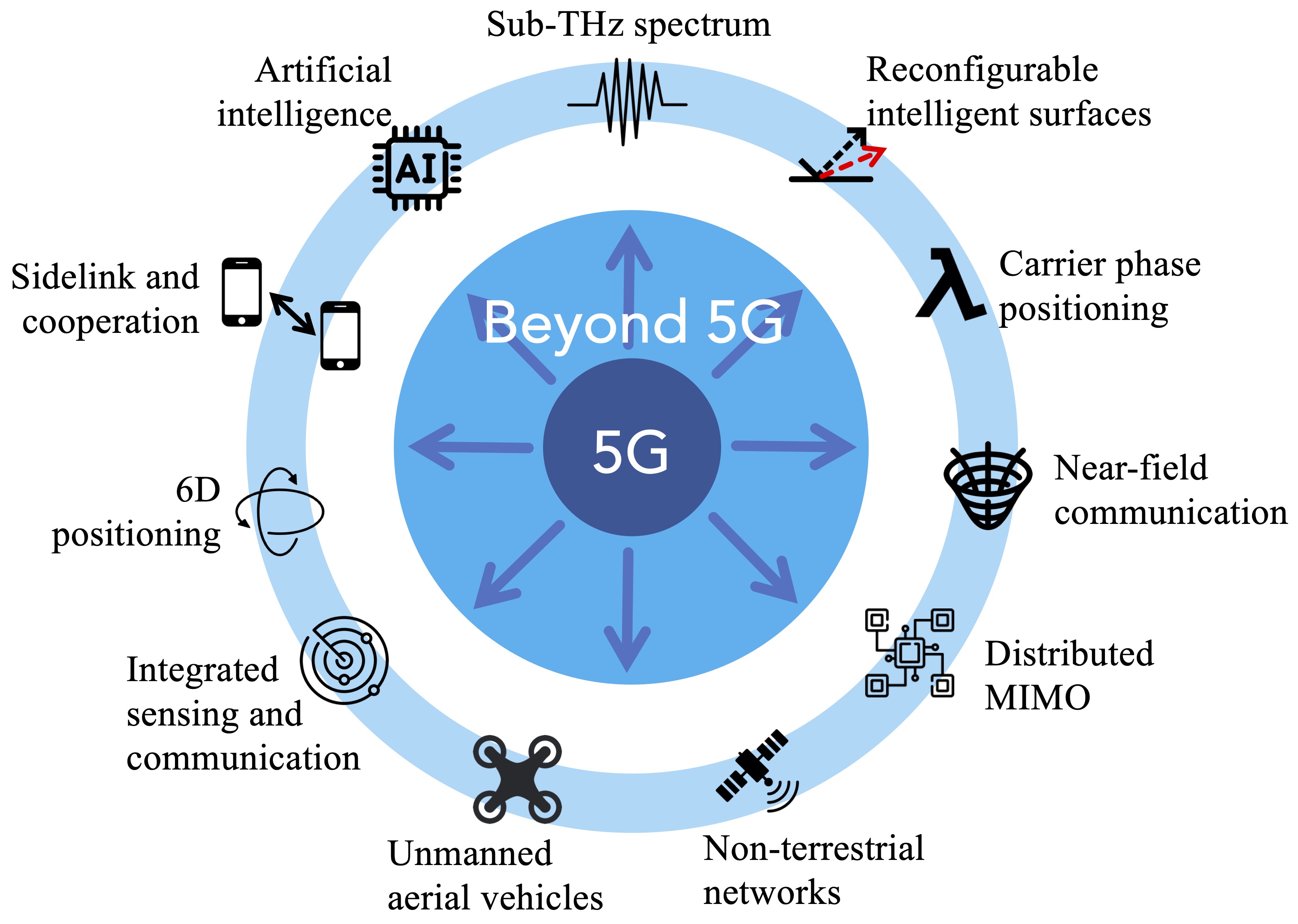}
    \caption{Overview of the positioning trends beyond 5G.}
    \label{fig:Beyond5GEnablers}
\end{figure}

The advent of \ac{B5G} will represent a significant transformation for wireless communications~\cite{Giuliano2024}. With the potential to revolutionize location-based services, the forthcoming cellular technology will ensure unprecedented positioning accuracy and high-speed connectivity. In this subsection, we briefly discuss the foreseen innovations related to technological and methodological aspects, covering topics such as the use of \ac{THz} bands, \ac{RIS}, \ac{CPP},  \ac{NFC}, \ac{D-MIMO}, \ac{NTN}, \ac{UAV}, \ac{ISAC}, \ac{6D} positioning and orientation, \ac{SL} and \ac{CP}, and lastly \ac{AI}. These aspects are summarized in Fig.~\ref{fig:Beyond5GEnablers} and described in the following.

\subsubsection{THz bands}
Even though the challenges of \ac{5G} are still to be resolved, research on \ac{B5G}  systems has already started~\cite{Bi:J19}. In particular, the next-generation of cellular networks taps into the \ac{THz} spectrum, a frequency band with the availability of larger bandwidths, enabling higher data rates, lower latency, and enhanced positioning accuracy~\cite{GioPolMezRanZor:J20}. The unique propagation characteristics of the \ac{THz} band allow for an improved ability to determine the precise location of devices and users. This is thanks to the two-fold effect of \textit{(i)} larger available bandwidth at such frequencies, providing improved delay resolution, and \textit{(ii)} miniaturization possibilities, allowing packing of more antennas in a small area, improving angular resolution~\cite{chen2022tutorial}. 
\textcolor{black}{Moreover, leveraging \ac{THz} imaging and high-frequency \ac{SLAM},
a high-accuracy positioning is expected in the coming decades, also leveraging \ac{NLOS} scenario involving multipath reflections~\cite{thz_review}.}
On the other hand, the use of \ac{THz} also comes with major challenges, such as high path loss (limiting the coverage) and sensitivity to atmospheric conditions~\cite{Trichopoulos_j19} that call for enhanced precoding strategies~\cite{Saad_j23} to avoid loss of connection.

\subsubsection{RIS}
\ac{B5G} systems are expected to standardize and introduce in the market the concept of \ac{RIS}~\cite{Zhang_j19} (also referred to as \ac{RIM}~\cite{loscri_joint_2021,loscri_best-rim:_2023}), which leverages the deployment of programmable surfaces with electromagnetic properties that can be controlled by software~\cite{Al-Dhahir_j21}. These surfaces can manipulate the wireless signal environment~\cite{he_beyond_2022}, facilitating better signal quality and enabling precise positioning also when \ac{LOS} path is not guaranteed~\cite{tang_wireless_2021}.
The adoption of \ac{RIS} will improve \ac{UE} positioning as it will behave as a multipath controller~\cite{wymeersch_radio_2020}, which may provide both new location references and new measurements (e.g., angles, delays).  Every single antenna of the surface can be treated as a local emitter, which makes the \ac{BS}-\ac{UE} link more robust even in poor propagation conditions~\cite{strinati_reconfigurable_2021, auto_RIS_2023}. 
\textcolor{black}{Further advances on smart surfaces include  \acp{TIS} (which support both outdoor and indoor positioning by adopting semi-transparent antennas)~\cite{TIS6G},
space-time modulated metasurfaces~\cite{MizTagSpa:j24}, 
and fully‑passive, flexible, and chipless smart skins~\cite{lynch20245g}.}
The installation of \ac{RIS} can be constrained by the physical properties of the objects: conformal metasurfaces can aid the installation over curved surfaces, such as over vehicles~\cite{Mizmizi_23}.
The research on \ac{RIS} suggests an ever-increasing interest in controlling electromagnetic waves, allowing to shape the environment according to the desired purposes. As a result, full control and exploitation of the wireless link enables \ac{HL}, where \acp{RIS} or \acp{LIS}~\cite{alghamdi_intelligent_2020,hu_beyond_2018} together with \ac{NFC} provide a great opportunity to move towards the ultimate capacity limit of the wireless channel~\cite{dardari_communicating_2020} and enhance positioning capabilities~\cite{Elzanaty2021Towards6H} \textcolor{black}{even in NLOS conditions~\cite{RIS_NLOS}}.

\subsubsection{CPP}
 The absolute phase of a signal, which relates to the distance between a transmitter and receiver, is used in \ac{CPP}. In~\cite{wymeersch2023fundamental}, \Ac{CPP} signals have been used for highly accurate positioning, with the potential for orders-of-magnitude performance improvements compared to standard \ac{TDOA} positioning. Recent studies have explored \ac{CPP} in cellular positioning, both integrated with \ac{GNSS} and as a stand-alone solution, examining its application in different frequency ranges, its challenges, and its potential in various configurations like massive MIMO~\cite{chen_carrier_2022}.

\subsubsection{NTN}
An \ac{NTN} refers to a novel communication infrastructure that extends beyond Earth's surface, encompassing communication links established through satellites, drones, and other space-based platforms~\cite{Federica_non_2020}. These networks have gained prominence as a potential solution to address connectivity gaps in remote and underserved regions, offering improved global coverage and high-speed data transmission~\cite{Volk_field_2022}. The \ac{NTN} technology leverages advancements in satellite technology, inter-satellite links, and emerging concepts like constellations of low Earth orbit satellites to create a seamless and interconnected network that can support various applications, from broadband internet access to \ac{IOT} connectivity and emergency communication services~\cite{Vaezi_cellular_2022}. From the positioning perspective, \ac{NTN} has been investigated in~\cite{dureppagari2023ntn}, and it was shown to have the potential to improve positioning accuracy by using the \ac{CRB} analysis.

\subsubsection{UAV}
\ac{UAV} \ac{5G} positioning leverages the capabilities of \ac{5G}  networks to enhance the accuracy and reliability of \ac{UAV} navigation and location tracking. By utilizing the high data rates, the low latency, and the extensive coverage of \ac{5G} networks, a joint design of passive beamforming, blocklength, and \ac{UAV} positioning has been developed in~\cite{Ranjha_URLLC_2021}, which has excellent positioning precision. This technology enables \acp{UAV} to perform tasks that demand cm-level accuracy, such as aerial mapping, surveying, and critical infrastructure inspection. \ac{UAV} was studied in~\cite{Afifi_autonomous_2021} for autonomous positioning based on supervised \ac{DNN} and reinforcement learning approaches. The integration of \ac{5G} positioning not only improves the \ac{UAV}'s ability to maintain its intended flight path but also enhances the safety and efficiency of operations, making it a crucial advancement in the realm of \ac{UAV}-based applications~\cite{Shehzad_backhaul_2021}.

\subsubsection{NFC}
The effect of near-field communications should be taken into account in situations where extremely large antenna arrays, \acp{RIS} and/or \ac{D-MIMO} are adopted~\cite{XLMIMO,Eldar_j23,Ozdenizci_b12}. \ac{NFC} mainly contains three features: spherical wavefront, spatial non-stationarity, and beam squint effect. Enhanced positioning capabilities can be achieved by incorporating these features and using specialized signal processing methods~\cite{10529957}. \textcolor{black}{For example, the authors in~\cite{Djurić_j21} derived the \ac{PCRB} and discussed how the loss of positioning information outside the Fresnel region results from an increase of the ranging error rather than from inaccuracies of angular estimation.} This provides means to position devices using very limited bandwidth, though often at a high complexity cost.

\subsubsection{D-MIMO}
\ac{D-MIMO} is another key technology shaping \ac{B5G} positioning. Unlike conventional \ac{MIMO}, where multiple antennas are placed close together on a single device, in the \ac{D-MIMO} paradigm, antennas are placed on separate phase-coherent devices distributed over a geographical area~\cite{IEEEproc_DMIMO}. 
\textcolor{black}{A substantial body of literature on \ac{D-MIMO} in \ac{B5G} has been introduced in the community.
For example, \cite{Keskin22} demonstrated the potential of integrating fiber technologies with \ac{D-MIMO} for precise localization, while  \cite{demirhan2023} explored \ac{D-MIMO} systems for joint radar and communication functionalities, proposing a strategy that optimizes both sensing and communication. The challenge of deploying \ac{D-MIMO} in underwater environments was addressed in~\cite{Wang_b14}. Surveys in~\cite{elhoushy_cell-free_2022, chen_survey_2022, ammar_user-centric_2022, he_cell-free_2021} discussed the scalability, performance improvements, and future outlook of cell-free massive \ac{MIMO} systems, emphasizing their role in enhancing user experience, network efficiency, and meeting the ambitious goals of future wireless communications. \cite{zhang_cell-free_2019} highlighted the paradigm shift towards cell-free massive \ac{MIMO}, underlining its transformative potential for next-generation networks.
Note that in some literature, such as \cite{Hadaschik2017} and \cite{Kakkavas2018}, multi-array positioning has been considered, where multiple antenna arrays (placed in different locations) were used as \ac{Tx} and/or \ac{Rx} for radio positioning\textcolor{black}{, revealing increased positioning accuracy, with respect to the \ac{3GPP} studies, along with improved robustness and multipath mitigation.}}
\textcolor{black}{The same concept is also referred to as \ac{DAS}, especially in the vehicular context~\cite{5GAA_DAS}, indicating the installation of multiple antenna panels at different locations (e.g., one for each side of the vehicle). Compared to a single antenna, the redundancy of panels and their spatial distribution increase the quality of communication links by minimizing blockage conditions. Moreover, with two or more antennas, a single \ac{BS} is sufficient for \ac{TDOA} positioning.
Although a new paradigm is required, the use of \ac{DAS} is expected to improve positioning performance~\cite{5GAA_HAP_CV2X} and spectral efficiency~\cite{5G_UHD-DAS}.}
A distributed arrangement of arrays enhances spatial diversity and provides a better channel matrix, leading to improved signal quality, enhanced network capacity, and more accurate positioning~\cite{Lemey_j22}.
\textcolor{black}{Many methods have been proposed to achieve this advantage, including graph-based methods, linear \ac{MMSE}, sequential \ac{MMSE}, \ac{ZF}, among others~\cite{ammar_user-centric_2022,he_cell-free_2021, Lemey_j22}.} \ac{D-MIMO} is especially useful in high-density environments, such as urban settings and large public venues, where accurate positioning is critical~\cite{Armada_j22}. While \ac{D-MIMO} is often operated in phase-coherent mode, at higher frequencies, frequency-coherent \ac{D-MIMO} is more practical to implement, leading to separate channels per antenna~\cite{haliloglu2023distributed}. Phase-coherent and frequency-coherent \ac{D-MIMO} are both attractive for positioning, though with different benefits.

\subsubsection{ISAC}
\label{sec:ISCA}
\ac{ISAC} involves merging sensor networks and communication systems to gather real-time data and facilitate seamless information exchange. This integration greatly benefits \ac{B5G}  positioning by enabling multi-sensor fusion for more accurate positioning, providing redundancy for reliability, and supporting adaptive algorithms that respond to changing conditions~\cite{ISAC_Proc2024,Liu_integrated_2022}. \ac{ISAC} will not only provide new sensing functions (both radar-like and spectroscopy-like), but integrated sensing enhances existing positioning and localization techniques, contributing to highly accurate and resilient positioning solutions in various scenarios and environments~\cite{Luettel_autonomous_2012, Chen_deep_2021, Yang_hybrid_2022, Blefari21}. \textcolor{black}{The authors in~\cite{Yang_hybrid_2022} extended the classic probabilistic data association \ac{SLAM} mechanism to achieve \ac{UE} localization, using \ac{ISAC} systems and showing better performance without any prior information. Besides, in~\cite{Blefari21}, a case study for \ac{ISAC} using experimental data showcased the potential of the new enablers that are paving the way toward enhanced road safety in \ac{B5G} scenarios.} Finally, the \ac{ISAC} paradigm also provides enhancements for communication itself, as time-consuming beam training and handover can be avoided.

\subsubsection{6D positioning}
The significance of joint \ac{3D} position and \ac{3D} orientation estimation, commonly referred to as \ac{6D} localization, cannot be overstated~\cite{nazari2023mmwave}. While \ac{5G} \ac{mmWave} primarily focused on \ac{UE} position estimation, the demands of \ac{B5G} necessitate comprehensive \ac{6D} information. This encompasses both \ac{3D} positioning and \ac{3D} orientation, often termed pose in robotics. For instance, \ac{C-ITS} require vehicle position and heading for advanced features like driving assistance and platooning. In assisted living environments, a resident's pose can offer insights into their health. Similarly, \acp{UAV} in search-and-rescue missions rely on precise pose data for effective operations. Typical 6G applications such as \ac{VR}, augmented reality, robot interactions, and digital twins will further underscore the need for \ac{6D} localization~\cite{Liu2024,behravan2022positioning}. While external systems, like the fusion of \ac{GNSS} (for positioning) and \ac{IMU} (for orientation), offer solutions, they have limitations like indoor inefficiencies or error accumulation. A more integrated approach would harness existing cellular infrastructure for \ac{6D} localization, utilizing multiple \acp{BS} for accurate \ac{UE} orientation and position estimation.

\subsubsection{SL and CP}
\label{sec:Cooperative and sidelink positioning}

In \ac{B5G}  systems, the development of direct device-to-device communication is fundamental to lower latency and guarantee the service even in out-of-coverage conditions (i.e., areas without cellular \acp{BS})~\cite{Araniti_j23}.
This is facilitated through \ac{SL} communications (e.g., \ac{V2V} communications~\cite{noor-a-rahim_6g_2022}), which allow to bypass the traditional routing through a \acp{BS} and core network~\cite{Wymeersch_j23}, enhancing the reliability of positioning service, reducing latency, and enabling accurate relative positioning in proximity~\cite{Decarli2023}. 
Sidelink communications can also benefit from a-priori knowledge of digital maps or channel information for a more efficient link~\cite{MIZMIZI2022100402}.
%
The evolution of \ac{3GPP} standards looks towards the development of a unique technology jointly guaranteeing \ac{SL} communications and positioning, like for uplink and downlink, complying with the convergence of communication, localization, and sensing in forthcoming \ac{6G} networks~\cite{6GFlagship:J21}. These features are inherently suited for the rise of \ac{CP} solutions~\cite{wymeersch_cooperative_2009,Win:J18,Win:Proc18,Buehrer:18, 10227084}. 
In \ac{CP}, signal processing techniques operate on either centralized or distributed network architectures and typical application domains include \ac{IOT}~\cite{Yang:J23,Bi:j22,saucan_information-seeking_2020,Jian:J19,soatti_consensus_2017}, \ac{C-ITS}~\cite{brambilla_augmenting_2020,soatti_implicit_2018,Kuutti:j18,barbieri_decentralized_2022,Fas_coop_18,10215362}, maritime surveillance~\cite{Bra:J16,ferri2017cooperative}, collaborative robotics~\cite{Savazzi:m21}, drones or \acp{UAV}~\cite{wu_cooperative_2018,zhang_cooperative_2022,Guerra_M22}. 
These systems critically necessitate sensing agents perceiving the environment in proximity and making informed decisions based on the data received from both individual sensors and communication links.
The collaboration among distributed agents also enhances situational awareness, allowing for improved localization resolution of both agents and potential obstacles or targets~\cite{brambilla_cooperative_2022,teague_network_2022,meyer_scalable_2020,10472719,barbieri_implicit_2023}. 
\textcolor{black}{In this framework, the use of \ac{RIS} working as anchor nodes with known positions has been recently proposed~\cite{RACLN2024}.}


\subsubsection{AI}
\label{sec:Artificial intelligence}


\textcolor{black}{The role of \ac{AI} is already emerging to a certain extent for Rel-18, but its pervasive realization will rise only with the advent of \ac{6G}~\cite{AkyKakNie:J20}.
The first expected \ac{AI} applications  within next \ac{3GPP} releases refer to resource block allocation and mobility management~\cite{prado_enabling_2023}, channel estimation~\cite{guo_robust_2023}, scheduling policies~\cite{min_meta-scheduling_2023}, and beam management~\cite{li_machine_2023}. 
Regarding positioning, \ac{ML} algorithms can be divided into \ac{AI}/\ac{ML}-assisted positioning and direct \ac{AI}/\ac{ML} positioning~\cite{lin_overview_2022}. 
The former category includes the methods to improve conventional geometric-based algorithms. Examples are the geometric measurements estimation and corrections~\cite{10274628, liu_machine_2023, wymeersch_machine_2012}, the improvement of Bayesian tracking filters~\cite{ko_high-speed_2022}, \ac{CSI} prediction and compression~\cite{8395053}.
The latter category focuses on the design of algorithms that learn the relation between the channel characteristics (i.e., fingerprint) and the \ac{UE} position~\cite{lv_deep_2022, malmstrom_5g_2019}. By directly addressing the positioning problem with \ac{AI}, the focus is given to the generalization capabilities~\cite{10274764} and the type of input features~\cite{10274110}.
}

\textcolor{black}{Regarding the adopted \ac{AI} algorithms, a variety of methods are present in the literature, ranging from conventional \ac{ML}~\cite{wymeersch_machine_2012, van_nguyen_machine_2015} to \acp{DNN}~\cite{tedeschini_latent_2023_2, wu_learning_2021}, \acp{GNN}~\cite{10274765}, \ac{FL}~\cite{10274418,10.48550/ARXIV.1602.05629, 10.1109/ACCESS.2024.3446577}\textcolor{black}{, and \acp{BNN}~\cite{10.1109/LSP.2021.3130504, tedeschini_latent_2024}}.
In~\cite{wymeersch_machine_2012} and~\cite{van_nguyen_machine_2015}, \ac{SVM} and \ac{RVM} are employed for \ac{NLOS} identification and correction with \ac{CSI} features, such as \ac{TOF}, energy and kurtosis.
To avoid limiting the performances with hand-crafted features, \ac{DNN} methods, such as \acp{CNN} or \ac{AE}~\cite{zeng_nlos_2019,stahlke_estimating_2022, tedeschini_latent_2023}, can be used to directly estimate the position from the full \ac{CIR}. Examples can be found in both indoor~\cite{chen2017confi,Nessa_ML_2020,amiri_indoor_2023} and outdoor~\cite{tedeschini_latent_2023_2, li_machine-learning-based_2019} environments.}
\textcolor{black}{
Regarding the \ac{FL} paradigm to improve the location estimate while maintaining the privacy of locally stored data, authors in \cite{10274418} introduce a framework for map matching, enabling multiple data sources to train a shared model collaboratively without exchanging raw data.}
\textcolor{black}{When dealing with out-of-distribution areas, it is important to have a reliability measure of the model's output. To this concern, recently \acp{BNN} have been adopted for producing static point estimates with related uncertainties in \ac{mmWave} \ac{MIMO} scenarios~\cite{10.1109/LSP.2021.3130504}. \acp{BNN} have also been integrated into tracking filters to provide mobile positioning under \ac{NLOS} conditions~\cite{tedeschini_latent_2024}.}

\textcolor{black}{For a more in-depth analysis of these topics, we refer to the surveys in~\cite{li_machine-learning-based_2019,Nessa_ML_2020,sonny2024survey,nugroho_federated_2024} which provide comprehensive insights on the role of \ac{AI}, \ac{ML}, and \ac{FL} in enhancing positioning accuracy and improving localization techniques, also outlining key challenges and open issues.}





\section{Fundamentals of Wireless Positioning}
\label{sec:System Model}

\textcolor{black}{In this section, we provide the fundamentals of network positioning, starting from the model of the wireless channel (Section~\ref{sec:Channel_Model}) and the location-related measurements that can be extracted from it for localization purposes (Section~\ref{sec:Fundamentals of wireless localization}). Then, we discuss techniques allowing the estimation of the \ac{UE} position from such measurements, with a focus on snapshot algorithms (Section~\ref{sec:Positioning algorithms}) and tracking filters (Section~\ref{sec:EKF_tracking}). }

\subsection{Wireless channel model}
\label{sec:Channel_Model}

\textcolor{black}{We consider a time-slotted \ac{UL} wireless \ac{MIMO} \ac{OFDM} communication system, as the one used in \ac{5G}, with $N_{\mathrm{tx}}^{}$ transmit and $N_{\mathrm{rx}}^{}$ receiving antenna elements. We assume a block-fading time-invariant channel response, i.e., constant within an \ac{OFDM} symbol, 
with maximum delay contained within the cyclic prefix $T_{\mathrm{cp}}^{}$. Let 
the matrix $\boldsymbol{\mathbfcal{H}}_{\tau}^{} \in \mathbb{C}^{N_{\mathrm{rx}} \times N_{\mathrm{tx}}}$ represent the  $\tau$-th tap of the equivalent base-band \ac{MIMO} channel response, the signal received at discrete time $t=1, 2, \ldots, T$ (sampled at symbol time $T_s$), denoted as  $\boldsymbol{z}_t^{} \in \mathbb{C}^{N_{\mathrm{rx}}\times 1}$, is modeled as
\begin{equation}
    \boldsymbol{z}_t^{} = \sum_{\tau=0}^{T_{\mathrm{cp}}} \mathbfcal{H}_{\tau}^{} \, \boldsymbol{y}_{t-\tau}^{} + \boldsymbol{\xi}_t^{}  ,
    \label{eq:rx_signal}
\end{equation}
where  $\boldsymbol{y}_t^{} \in \mathbb{C}^{N_{\mathrm{tx}}\times 1}$ is the transmitted signal and $\boldsymbol{\xi}_t^{} \in \mathbb{C}^{N_{\mathrm{rx}}\times 1}$ the background noise.
It is common in the literature to assume the noise as spatially and temporally uncorrelated zero-mean complex Gaussian. Non-diagonal covariance can be considered to model the presence of directional interference. }

\textcolor{black}{The \ac{MIMO} channel within a generic \ac{OFDM} symbol time can be modeled as a combination of $P$  paths as follows:
\begin{equation}
    \mathbfcal{H}_{\tau}^{} = \sum^P_{p=1} 
    \alpha_{p}^{} \,
    \boldsymbol{a}^{}_{\mathrm{rx}}(\phi_{\mathrm{rx},p}^{}, \psi_{\mathrm{rx},p}^{}) \,
    \boldsymbol{a}^{\T}_{\mathrm{tx}}(\phi_{\mathrm{tx},p}^{}, \psi_{\mathrm{tx},p}^{}) \,
    \mathrm{g}(\tau - \tau_{p}^{}),
    \label{eq:channel}
\end{equation}
where path $p$ is characterized by the complex fading amplitude $\alpha_{p}^{}$, 
the transmitting antenna array response $\boldsymbol{a}_{\mathrm{tx}}^{}(\cdot) \in \mathbb{C}^{N_{\mathrm{tx}}\times 1}$ to the azimuth ($\phi_{\mathrm{tx},p}^{}$) and elevation ($\psi_{\mathrm{tx},p}^{}$) \acp{AOD}, 
the receiving antenna array response  $\boldsymbol{a}_{\mathrm{rx}}^{}(\cdot)\in \mathbb{C}^{N_{\mathrm{rx}}\times 1}$ to the azimuth ($\phi_{\mathrm{rx},p}^{}$) and the elevation ($\psi_{\mathrm{rx},p}^{}$) \acp{AOA}, 
and the pulse waveform $\mathrm{g}(\cdot)$ delayed by the path delay $\tau_{p}$, with ${\mathrm{max}_p}(\tau_{p}^{}) \leq T_{\mathrm{cp}^{}}$. 
We consider the fading amplitudes $\alpha_{p}^{}$ as \ac{OFDM}-block-fading, 
while delays and angles are assumed to be constant over a number of \ac{OFDM} symbol transmissions. Characterization of the antenna array responses depends on the antenna configuration geometry and design method~\cite{OrfanidisBook}.}

\begin{figure}
    \centering
    \subfloat[\label{fig:LOS_channel_profile}]
    {
        \begin{tikzpicture}  
        \node[]at(0,0){\includegraphics[width=1\columnwidth]{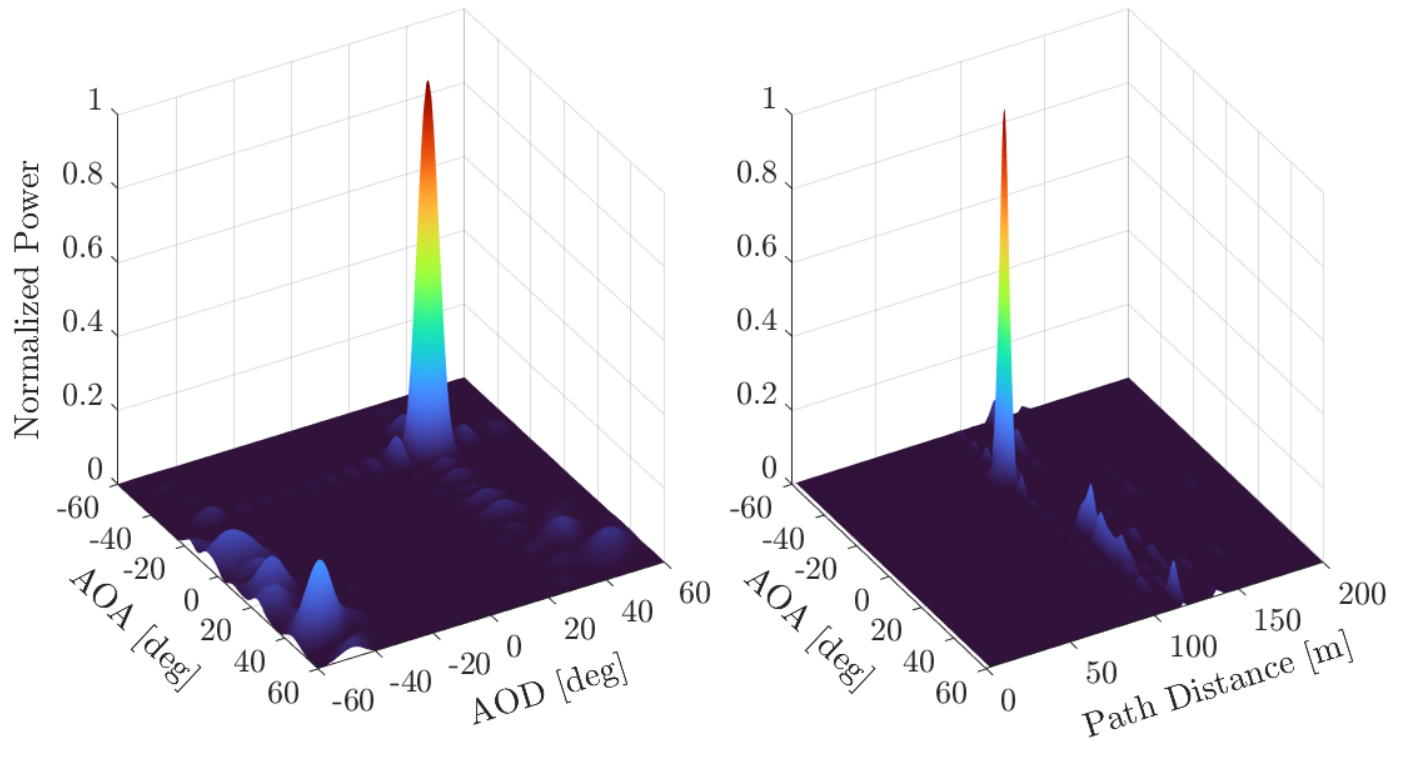}};
        \end{tikzpicture}
    }
    \\ \vspace{-15pt}
    \subfloat[\label{fig:NLOS_channel_profile}]
    {
        \begin{tikzpicture}  
        \node[]at(0,0){\includegraphics[width=1\columnwidth]{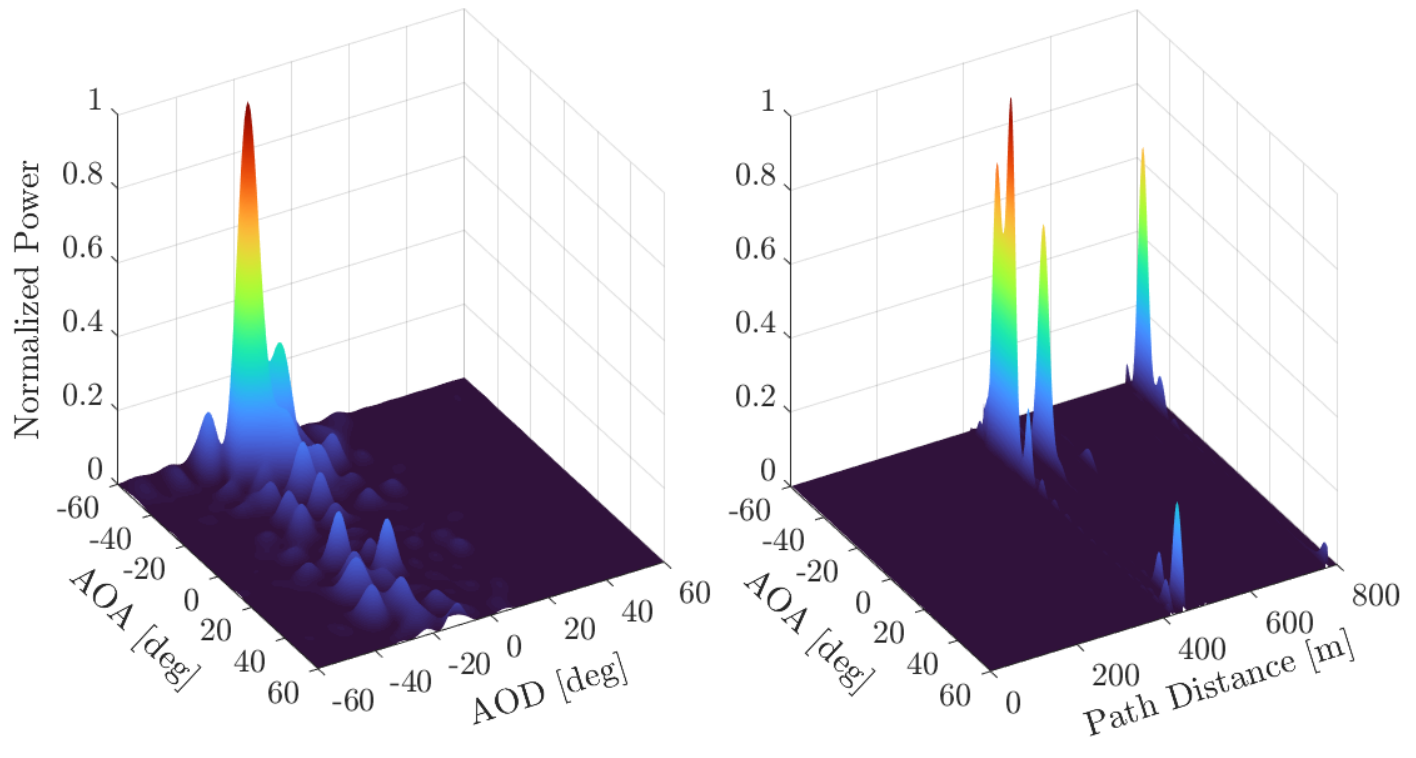}};
        \end{tikzpicture}
    }
    \caption{\textcolor{black}{Beam space representation of a MIMO channel. (a) LOS channel; (b) NLOS channel. On the left, spatial representation of the normalized received power versus the azimuth AOAs and AODs. On the right, power-angle-delay profile of the received signal, with path delay converted into distance for an easier interpretation.}}    \label{fig:channel_profile}
\end{figure}

\textcolor{black}{Fig.~\ref{fig:channel_profile} shows two examples of \ac{MIMO} channels, represented in terms of power-angle (left) and power-angle-delay  (right) profiles for \ac{LOS} (Fig.~\ref{fig:LOS_channel_profile}) and \ac{NLOS} (Fig.~\ref{fig:NLOS_channel_profile}) propagation conditions. 
The communication system considers a  $16\times 16$  planar antenna array at both the \ac{Tx} and \ac{Rx} devices. 
Fig.~\ref{fig:LOS_channel_profile} illustrates a scenario where the \ac{Tx} and the \ac{Rx} are in \ac{LOS}, separated by a distance of $100$\,m, with azimuth \ac{AOA} and \ac{AOD} of \mbox{-30} deg and 30 deg, respectively. 
Fig.~\ref{fig:NLOS_channel_profile}) provides a more complex scenario characterized by the absence of a direct path between the \ac{Tx} and the \ac{Rx}.
The figures display the different multipath components of the channel, facilitating the identification of the dominant channel paths, their power, \ac{AOA}, \ac{AOD}, and delay. 
We can observe that the channel carries relevant information for positioning: in \ac{LOS} condition, the first received signal peak corresponds to the direct path, which, besides carrying power information, allows the estimate of the angle and distance with respect to the \ac{Tx}, enabling localization. In \ac{NLOS} conditions, instead, gathering position measurements is more intricate, and the usage of advanced algorithms is necessary (see Section~\ref{sec:NLOS_detection}). 
The following section delves into the modeling of the positioning measurements extracted from the received signal \eqref{eq:rx_signal} exploiting the location features embedded in the wireless channel.}

\subsection{Location measurements from cellular signals}
\label{sec:Fundamentals of wireless localization}
Let us consider a \ac{UE}, connected to a number of cellular \acp{BS}.
The \ac{UE} location can be estimated by extracting different types of measurements from the radio signals \eqref{eq:rx_signal}, either in \ac{UL} (at the \ac{BS}) or in \ac{DL} (at the \ac{UE}). 
Typical measurements include distance, angle, or power.  

The distance can be measured by computing the delay or the power loss experienced by the signal during the propagation from the \ac{BS} and the \ac{UE} (or viceversa). The delay, referred to as \ac{TOF}, is the time difference between the \ac{TOA} and the transmission time.
The difference between two \acp{TOA}, instead, is the \ac{TDOA}, while the \ac{RTT} is a \textit{two-way} TOA obtained as detailed later in this section.
The power measurement is obtained by reading the \ac{RSS} at the \ac{Rx} side.

The angle measurement refers to the main direction from which the signal~\eqref{eq:rx_signal} is received or transmitted, and it is denoted as \ac{AOA} or \ac{AOD}, accordingly. It is obtained by employing directional or \ac{MIMO} antenna systems. A typical condition in cellular networks involves \acp{BS} with many antennas and \acp{UE} with only one (or limited, e.g., $2 \times 2$ \ac{MIMO}) antenna. It follows that the \ac{AOD} coincides with the direction of beam pointing, i.e., where the \ac{BS} emits most of its radiation beam pattern.

We denote with $\boldsymbol{u}=[u_x^{} \, u_y^{} \, u_z^{}]$  the unknown \ac{3D} \ac{UE}  location, and with  $\boldsymbol{s}_i^{}=[s_{x,i}^{} \, s_{y,i}^{} \, s_{z,i}^{}]$ the \ac{3D} coordinate of the $i$-th \ac{BS}, with $i = 1, ..., N_{\text{BS}}^{}$, defined in a convenient spatial reference system (e.g., a Cartesian, ellipsoidal or geographic coordinate system).
We indicate with ${\rho}_{i}^{}$ the single measurement generated or collected by \ac{BS} $i$, defined as
\begin{equation}
    \rho_{i}^{} = h_{i}^{}(\boldsymbol{s}_{i}^{},\boldsymbol{u}^{}_{})+{n}_{i}^{},
\label{eq:measurement_single}  
\end{equation}
where $h_{i}^{}(\cdot)$ is a known non-linear function that deterministically relates the measured parameter to the positions of the \ac{BS} ($\boldsymbol{s}_i^{}$) and the \ac{UE} ($\boldsymbol{u}$); $n_{i}^{}$ is an additive term describing the measurement error.
Vector 
$
 \boldsymbol{\rho}_{i}^{} = \boldsymbol{h}_{i}^{}(\boldsymbol{s}_{i}^{},\boldsymbol{u})+\boldsymbol{n}_{i}^{}
$ aggregates all the measurements (e.g., \ac{TOA}, \ac{AOA}, \ac{TDOA}, \ac{RSS}) generated by the $(\boldsymbol{s}_{i}^{},\boldsymbol{u})$ pair. 
The overall vector of measurements for all the $N_{\text{BS}}^{}$ \acp{BS} is indicated with
$    \boldsymbol{\rho} = \left[\boldsymbol{\rho}^{\T}_{1}  \cdots \boldsymbol{\rho}^{\T}_{N_{\text{BS}}^{}} \right]^{\T} =  \boldsymbol{h}(\boldsymbol{s},\boldsymbol{u})+\boldsymbol{n},
$
where $\boldsymbol{s} = \left[\boldsymbol{s}^{\T}_{1}  \cdots \boldsymbol{s}^{\T}_{N_{\text{BS}}^{}} \right]^{\T}$  and $\boldsymbol{n} = \left[\boldsymbol{n}^{\T}_{1}  \cdots \boldsymbol{n}^{\T}_{N_{\text{BS}}^{}} \right]^{\T}$  collect all the \ac{BS} locations and measurement noises, respectively. The overall number of measurements is $M=|\boldsymbol{\rho}|$.

\input{Figure_tex/Fig_Positioning_Measurements}

Depending on the available hardware technology, protocol, or signal, different definitions hold for the model~\eqref{eq:measurement_single}~\cite{ZafGkeLeu2019}. In the following, we introduce the models used for the cases of interest in cellular localization systems, whereas the methods for extracting such measurements are detailed later in Section~\ref{sec:5G measurement extraction}, with specific reference to \ac{5G} radio systems.
An illustrative example of \ac{UE} localization with four \acp{BS} is reported in Fig.~\ref{fig:positioning_meas}, where we represent the spatial information carried by the main types of measurements.



\subsubsection{TOF measurement}
A radio signal can be used to estimate the distance between a \ac{Tx} and a \ac{Rx}, knowing the propagation speed of the radio wave and measuring the travel time. 
In order to obtain the \ac{TOA} (which identifies a circular set of candidate \ac{UE} locations, see Fig.~\ref{fig:TOFmeas}) 
at the \ac{Rx} side, a replica of the (known) transmitted signal is needed to compute the cross-correlation with the received signal. 
In ideal \ac{LOS} \ac{AWGN} channels, the optimal \ac{TOA} estimate is obtained by searching the peak of the cross-correlation output~\cite{fischer}. 


Assuming a \ac{DL} measurement (i.e., the signal is sent by the \ac{BS} and received by the \ac{UE}) and indicating with $t_{\text{rx},i}^{}$ the \ac{TOA} at the \ac{UE} of a signal transmitted by \ac{BS} $i$ at time $t_{\text{tx},i}^{}$, the measured \ac{TOF} is:
\begin{equation}
    \tau_i^{} =  t_{\text{rx},i}^{}-t_{\text{tx},i}^{} = 
    \frac{d_i^{}}{c},
    \label{eq:toa_LOS}
\end{equation}
where $d_i^{}$ is the length of the propagation path traveled by the signal at speed  $c$. 

The resulting \ac{TOF} measurement relating the \ac{UE} and \ac{BS} $i$ is:
\begin{align}
    \rho_{i}^{\text{TOF}} & = \tau_i^{} +n_i^{\text{TOF}} .
\label{eq:toameasurement} 
\end{align}
Note that an analogous disclosure is also applicable in \ac{UL} (i.e., the \ac{BS} measures the \ac{TOA} of a signal transmitted by the \ac{UE}) and for \ac{RTT}.

A major problem for \ac{TOF}-based localization is that a precise measurement of $t_{\text{tx},i}$ must be available at the \ac{Rx} side, and the internal clocks of \ac{Tx} and \ac{Rx} must be synchronized~\cite{Serpedin:M11}. Typically, the clock of the \ac{UE} has a poorer quality compared to the one of the \ac{BS}; thus, it can introduce large errors in the \ac{TOF} measurement. 
To bypass the low quality of \ac{UE} hardware, \ac{TDOA} measurements can be used.



\subsubsection{TDOA measurement}
\ac{DL}-\ac{TDOA} is the measurement of the difference between the arrival times of the signals transmitted simultaneously by two distinct \acp{BS} and received by the \ac{UE}, i.e., the \ac{TDOA} is the difference between two \ac{TOA} measurements. 
It results that \ac{TDOA} measurements draw a hyperbolic line in space (see Fig.~\ref{fig:TDOAmeas}).
Unlike \ac{TOA} measurements, transmitted signals are not requested to carry any time stamp, and the \ac{Rx} does not need to be synchronized with the \acp{Tx}~\cite{fischer}. On the other hand, the involved \acp{BS} need a precise synchronization. 
This feature allows overcoming the errors due to the clock offset at the \ac{UE} (which typically has lower quality hardware compared to the \acp{BS}). 
For the computation of \ac{TDOA} measurements, a \ac{BS} has to be selected as a reference (e.g., in Fig.~\ref{fig:TDOAmeas} the \ac{BS} on the left is chosen as reference), and thereby the number of available \ac{TDOA} measurements reduces to $N_{\text{BS}}^{}-1$. A possible choice for the selection of the reference \ac{BS} is to take the \ac{BS} with the highest \ac{SNR} after the cross-correlation, although different selection criteria exist~\cite{deng2020base,tsumachi2021base,torsoli_selection_2022}.

Indicating  the reference \ac{BS} with index $i=1$, the \ac{TDOA} for \ac{BS} $i \neq 1$  is computed as
\begin{align}
    \Delta \tau_{i,1}^{} & = \tau_i^{} - \tau_1^{} \nonumber
    \\ & =  \left(t_{\text{rx},i}^{}-t_{\text{rx},1}^{}\right) - \left(t_{\text{tx},i}^{}-t_{\text{tx},1}^{}\right) \nonumber
    \\ & = \frac{d_i^{}-d_1^{}}{c}, \quad\quad\quad\quad\quad i = 2,...,N_{\text{BS}}^{}, 
    \label{eq:tdoa}
\end{align}
and the \ac{TDOA} measurement $\rho_{i}^{\text{TDOA}}$ as
\begin{align}
    \rho_{i}^{\text{TDOA}} & = \Delta \tau_{i,1}^{} + \left(n_i^{\text{TOF}}-n_1^{\text{TOF}}\right)  \nonumber
    \\ & =\frac{d_i^{}-d_1^{}}{c} + n_i^{\text{TDOA}}, \quad i = 2,...,N_{\text{BS}}^{}.
\label{eq:tdoameasurement} 
\end{align}
For an accurate measurement,  the synchronization offset between the \acp{BS}, i.e., $t_{\text{tx},i}^{}-t_{\text{tx},1}^{}$, has to be negligible or known.

\subsubsection{RTT measurement}
\ac{RTT} is a ranging technique which involves both \ac{UL} and \ac{DL} measurements. It is also known as  \textit{two-way} \ac{TOA} because the \ac{TOA} measurement is provided by both the initiating device and the responding device.


The initiating device (either a \ac{BS} $i$ or the \ac{UE}) transmits a signal at time $t_0^{}$, which is received by the responding device (\ac{UE} or \ac{BS}) at time $t_1^{}=t_0^{}+\tau_i$. After a time interval $\tau_{i,\text{reply}}^{}$ due to internal processing and switch from transmission to reception, the responding device sends another signal at time $t_2$, which arrives at time $t_3^{} = t_2^{} + \tau_i^{}$ at the initiating device. 
The overall $\text{RTT}$ over link $i$ is computed at the initiating device as the difference between its own transmit and receive times as
\begin{equation}
    \text{RTT}_i^{} = t_3^{}-t_0^{} .
\label{eq:rttmeasurement} 
\end{equation}
Assuming perfect knowledge of the reply time (computed at the responding device as $\tau_{i,\text{reply}}^{}=t_2^{}-t_1^{}$ and included in the payload, or known a priori) the \ac{TOF} $\tau_{i}$ can be then extracted as
\begin{equation}
    \tau_i^{} = \frac{\text{RTT}_i^{}-\tau^{}_{i,\text{reply}}}{2} .
\label{eq:rtt_tao} 
\end{equation}

The resulting \ac{RTT} measurement $\rho_{i}^{\text{RTT}}$ can be modeled similar to \eqref{eq:toameasurement}.
Different from \ac{TDOA} measurements, the \ac{RTT} measurement does not require synchronized \acp{BS}, as the time differences involve only the local clock of the devices.

\subsubsection{AOA/AOD measurement}
The \ac{AOA} indicates the spatial direction of the \ac{UL} signal sent by the \ac{UE} and received by the \ac{BS}. It can be estimated using directional antennas, such as phased arrays, which allow steering the radio signal over confined spatial directions called beams~\cite{wiley_2002}.  
Conventional methods estimate the \ac{AOA} by performing beamforming over various directions and selecting the beam with the highest received power. Higher resolution can be obtained by maximum-likelihood or subspace-based
algorithms (e.g., \ac{ESPRIT}, \ac{MUSIC}~\cite{wiley_2002, Cerutti_j20}). The main drawback is the high hardware-software complexity (and cost) required to get precise angular information (i.e.,  small beamwidth or equivalently large number of antennas), the high sensitivity to multipath, as well as the increasing location uncertainty with the distance (see Fig.~\ref{fig:AOAmeas}). 
On the other hand, synchronization among \acp{BS} is not required, and high-precision localization can be achieved when large arrays are available.

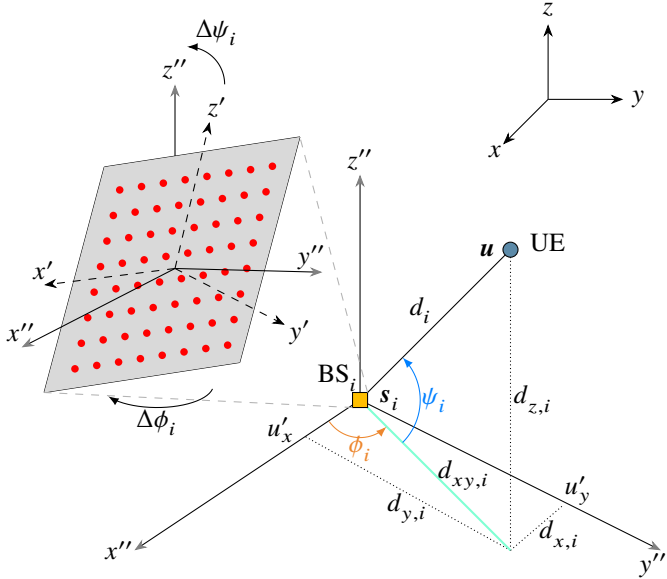
\begin{figure}[tb]
    \centering

    \begin{tikzpicture}
   
        \draw[-{Stealth[gray]}] (0,0) -- (4,-2) node[ yshift = -2mm , xshift=-3pt] {$y''$}; 
        \draw[-{Stealth[gray]}] (0,0) -- (0,3) node[yshift = 2mm ] {$z''$};
        \draw[-{Stealth[gray]}] (0,0) -- (-3,-2) node[xshift = -2mm ] {$x''$};
        \draw[] (0,0) -- (2,2) node[midway, yshift = 2mm , xshift = -2mm  ] {$d_i^{}$}; 
        \draw[densely dotted] (2,2) -- (2,-2) node[midway , xshift = 3mm ] {$d_{z,i}^{}$};
        \draw[densely dotted] (2,-2) -- (2.7,-1.4) node[midway , xshift = 3mm , yshift=-1mm] {$d_{x,i}^{}$};
        \draw[densely dotted] (2,-2) -- (-0.75,-0.475) node[midway , yshift = -2mm  ] {$d_{y,i}^{}$};
        \draw[thick,aquamarine,text=black] (0,0) -- (2,-2) node[midway , xshift = 3.8mm  ] {$d_{xy,i}^{}$};

        \draw [azure,-{latex},domain=-45:45] plot ({0.8*cos(\x)}, {0.8*sin(\x)}) node[midway, xshift = 10mm]{$\psi_i^{}$};
        \draw [cadmiumorange,-{latex},domain=-145:-45] plot ({0.5*cos(\x)}, {0.5*sin(\x)}) node[midway, yshift = -6.5mm]{$\phi_i^{}$};
        
        \draw[fill=amber] (-0.1,-0.1) rectangle (0.1,0.1) node[xshift = -4mm ,yshift=2mm] {BS$_i^{}$};
        \node[xshift=4mm] at (0,0) {$\boldsymbol{s}_i^{}$};
        \draw[fill=airforceblue] (2,2) circle (1mm) node[xshift = 5mm, yshift= 1mm ] {UE};
        \node[xshift=-3mm] at (2,2) {$\boldsymbol{u}$};




        \node[]at(2.9,-1.2){$u'_{y}$};
        \node[]at(-1.05,-0.4){$u'_{x}$};

        \coordinate (O') at (-2.46,1.75); 
        \coordinate (NW) at (-3.4, 3.1);
        \coordinate (NE) at (-0.8, 3.5);
        \coordinate (SW) at (-4.2, 0.1);
        \coordinate (SE) at (-1.6, 0.5);
        \draw[-{Stealth[gray]}] (O')--(-2.46,4.2) node [yshift = 2mm] {$z''$}; 
        \draw[] (NE)--(SE) node {}; 
        \draw[] (NE)--(NW) node {}; 
        \draw[] (SE)--(SW) node {}; 
        \draw[] (NW)--(SW) node {}; 

        \draw [fill=gray!30!, draw=none] (NE)--(NW)--(SW)--(SE);

        \foreach \y in {1,...,8}
        {
        \foreach \x in {1,...,8}
            {
                \draw[fill=red, draw=none] (-3.4+\x*0.29-\y*0.085,3.09+\x*0.045-\y*0.34) circle (0.5mm) node {};
            }
        }
        
        \draw[-{Stealth[gray]}] (O')--(-0.5,1.7) node [xshift=-1.5mm, yshift = 2mm] {$y''$}; 
        \draw[-{Stealth[gray]}] (O')--(-4.5,0.7) node [yshift=2mm] {$x''$}; 

        \draw[-{Stealth[black]}, dashed] (O')--(-2,3.7) node [yshift = 2mm, xshift = 1mm] {$z'$}; 
        \draw[-{Stealth[black]}, dashed] (O')--(-1, 1) node [yshift = -1mm, xshift = 2mm] {$y'$}; 
        \draw[-{Stealth[black]}, dashed] (O')--(-4.2,1.53) node [yshift = 2mm] {$x'$}; 

        \draw [black,{latex}-,domain=90:0,xshift=-2.3cm, yshift=4.2cm] plot ({0.5*cos(\x)}, {0.5*sin(\x)}) node[midway, xshift = 4mm, yshift=7mm]{$\Delta\psi_i^{}$};
        \draw [black,-{latex},domain=-10:-150,xshift=-2.7cm, yshift=2mm] plot ({0.75*cos(\x)}, {0.3*sin(\x)}) node[midway, yshift=-5mm]{$\Delta\phi_i^{}$};
        
        \draw[dashed,color=black!30] (-0.1,-0.1)--(SW) node {};
        \draw[dashed,color=black!30] (0.1,0.1)--(NE) node {};
        
        \coordinate (O) at (2.5,4); 
        \draw[-{Stealth[black]}] (O)--(2.5,5) node [yshift = 2mm] {$z$}; 
        \draw[-{Stealth[black]}] (O)--(3.5, 4) node [xshift = 2mm] {$y$}; 
        \draw[-{Stealth[black]}] (O)--(1.9,3.4) node [yshift = -1mm, xshift=-1mm] {$x$}; 
    \end{tikzpicture}

    \caption{UE and BS LOS geometry in a 3D Cartesian coordinate system with a focus on the BS array orientation.}
    \label{fig:cartesian_reference_system}
\end{figure}

The \ac{AOA} is defined as the \ac{3D} direction of the \ac{LOS} link to the $i$-th \ac{BS}, 
which includes the azimuth $\phi_i^{}$  and the elevation $\psi_i^{}$. This is estimated by the BS in a local reference system ($x'$, $y'$, $z'$) referred to the antenna array (see Fig.~\ref{fig:cartesian_reference_system}) and then converted into the global one for \ac{UE} positioning. We denote  with $\left( \Delta\phi_i^{}, \Delta\chi_i, \Delta\psi_i^{} \right)$ the orientation of the array,  where $\Delta\phi_i^{}, \Delta\chi_i^{}$ and $\Delta\psi_i^{}$ are respectively the rotation over the axis $z, y$ and $x$ and known as yaw, pitch and roll. Assuming a null pitch ($\Delta\chi_i^{}=0$), the \ac{AOA} measurement $\angle{(\boldsymbol{u'}-\boldsymbol{s'}_{i})}$ extracted by the antenna array  is rotated through a rotation matrix $\boldsymbol{R}_{xz}$ that combines the rotations around the $x'$ and $z'$ axes as follows~\cite{Blanco2012ATO}: 
\begin{equation}
    \boldsymbol{R}_{xz} = 
    \begin{bmatrix}
        \cos{\Delta\phi_i^{}} & -\sin{\Delta\phi_i^{}}\cos{\Delta\psi_i^{}} & \sin{\Delta\phi_i^{}}\cos{\Delta\psi_i^{}} \\
        \sin{\Delta\phi_i^{}} & \cos{\Delta\phi_i^{}}\cos{\Delta\psi_i^{}} & -\cos{\Delta\phi_i^{}}\sin{\Delta\psi_i^{}} \\
        0 & \sin{\Delta\psi_i^{}} & \cos{\Delta\phi_i^{}}
    \end{bmatrix} ,
\label{eq:rotmatrix} 
\end{equation}
and the \ac{AOA} is obtained as  $\angle{\boldsymbol{R}_{xz}^{}(\boldsymbol{u'}-\boldsymbol{s'}_{i})}$. 
The resulting azimuth ($\phi_i^{}$) and elevation ($\psi_i^{}$) angles are:
\begin{align}
    \phi_i^{} &= \
    \phi'_{i} + \Delta\phi_{i}^{} =
    \text{tan}^{-1}\left(\frac{u_y^{}-s_{y,i}^{}}{u_x^{}-s_{x,i}^{}}\right),
\\
    \psi_i^{} &= 
    \psi'_{i} + \Delta\psi_{i}^{} = 
    \text{tan}^{-1}\left(\frac{u_z^{}-s^{}_{z,i}}{d^{}_{xy,i}}\right),
\end{align}
with  $d_{xy,i}^{}=\sqrt{\left(u_x^{}-s_{x,i}^{}\right)^2 +\left(u_y^{}-s_{y,i}^{}\right)^2}$. Note that this is true only for $\Delta\chi_i^{} = 0$, otherwise additional algebraic transformations are requested.


The \ac{AOA} measurement vector is finally modeled as
\begin{equation}
    \boldsymbol{\rho}_{i}^{\text{AOA}} = 
    \begin{bmatrix}
         \phi_i^{}\\
         \psi_i^{}\end{bmatrix}+\boldsymbol{n}_i^{\text{AOA}},
\label{eq:aoameasurement} 
\end{equation}
which includes the measurement error $\boldsymbol{n}_i^{\text{AOA}}$.

On the other hand, \ac{AOD} measurements use \ac{DL} signals, which are sent by the \ac{BS} and received by the \ac{UE}. Still, the resulting angle is with respect to the \ac{BS} array; therefore, the \ac{AOD} measurement vector is modeled similarly to the \ac{AOA}.

\subsubsection{RSS measurements}
Distance information can also be obtained from power-based measurements, which are easy to extract, both in \ac{DL} and \ac{UL}.  
According to the  path-loss model~\cite{Sarkar_channel_model,tr138901, Hata_channel_model, Rappaport_channel_model_1}, the average power received over link $i$ (expressed in logarithmic scale) can be related to the distance as
\begin{equation}
    P_{\text{rx},i}^{}={P}_\text{0}^{} -10\,\beta \, \log_{10}\left(\frac{d_i^{}}{d_0^{}}\right),
\end{equation}
where $P_0^{}$ is the power received at a reference distance  $d_0^{}$, while $\beta$ is the path-loss index that depends on the propagation environment.  
The \ac{RSS} measurement is then defined as
\begin{equation}
    \rho_{i}^{\text{RSS}} = P^{}_{\text{rx},i}+ n_{i}^{\text{RSS}}={P}^{}_\text{0}-10\,\beta \, \log_{10}\left(\frac{d_i^{}}{d_0^{}}\right) + n_{i}^{\text{RSS}},
\end{equation}
where $n_{i}^{\text{RSS}}$ accounts for shadowing fluctuations and measurement errors.  

Unfortunately, power-based measurements reveal reasonable distance indicators only if the \acp{BS} is near to the \ac{UE}, as shadowing and multipath fading significantly affect the power values, and the propagation environment needs to be accurately modeled. The latter aspect can be really complex to achieve, as calibration procedures have to be performed and repeated anytime the environment changes. Overall, analytical modeling tends to be unrealistic in environments with severe multipath and obstructions. It results that \ac{RSS}-based positioning method is more suited, and generally used, for proximity detection and fingerprinting~\cite{ostling_performance_1996, kjaergaard_efficient_2010, savic_fingerprinting-based_2015, soltanieh_review_2020}.

\subsubsection{Digital maps and AI-based fingerprints}
\textcolor{black}{Fingerprinting localization is employed in complex multipath environments where analytical models are not able to describe the location-measurement relation. The analytical function  $h_{i}(\boldsymbol{s}_{i},\boldsymbol{u})$ is thus replaced by a digital map built ad-hoc during a training phase. A  database ${\mathcal{D}_{i} = \big\{ \boldsymbol{\rho}_i^{(m)}, \boldsymbol{u}_{}^{(m)} \big\}_{m=1}^M}$ is created by collecting channel fingerprints $\boldsymbol{\rho}_i^{(m)}$ over $M$ locations $\boldsymbol{u}^{(m)}$
in the area of interest, for each \ac{BS} $i$. 
The channel measurements can be derived from the \ac{CIR} (e.g., \ac{TOA}, \ac{AOA}, \ac{TDOA}, \ac{RSS}) or can be represented by the whole \ac{CIR}. Examples in this direction are the \ac{CFRM}~\cite{gao_toward_2022, hoang2020cnn} or \ac{ADCPM}~\cite{tedeschini_latent_2023_2, wu_learning_2021, jing2019learning}, which encode all the essential information of the environment, i.e., \ac{TOF}, \ac{AOA}, and \ac{RSS} for each path.}

\textcolor{black}{Once the position-referenced dataset is available, according to the type of channel measurement, different algorithms can be adopted for real-time localization. In the case of \ac{RSS} measurements, algorithms like HORUS~\cite{10.1145/1067170.1067193} or RADAR~\cite{832252}, based on probabilistic methods and \ac{KNN}, respectively, have been proposed in the past. 
With the advent of \ac{AI}, \ac{AE}-based structures, which are already foreseen in future \ac{3GPP} releases~\cite{38.857_rel_18}, allow to encode the input channel measurements into compressed versions, called latent features. This permits the reduction of the input dimensionality and performs feature extraction for subsequent position estimation through \ac{DNN} algorithms~\cite{tedeschini_latent_2023_2}. In case the database is incomplete, spectrum cartography techniques for estimating missing values and reconstructing the whole \ac{RSS} map can be used~\cite{Hong:j22,Roger:j23}.
For incomplete full-\ac{CIR} measurements, semi-supervised learning methods~\cite{tedeschini_latent_2023, stahlke_estimating_2022} or \ac{GAN}~\cite{8663987} can be adopted to limit the necessary labels information or generate new data, respectively.
}






\begin{table*}[!ht]
\centering
\caption{Measurement models and entries of the Jacobian matrix for 3D localization algorithms. Angles are referred to the \ac{UE}.} 
    \label{tab:Hel}
    \begin{tabularx}{0.95\linewidth}{ m{2.2cm} c c } 
    \toprule
    \textbf{Method} & $h_i^{}(\boldsymbol{s},\boldsymbol{u})$ & $
     \left[\boldsymbol{H}\right]_i 
     \triangleq
     \left[\boldsymbol{H}(\boldsymbol{u})\right]_i
     =
     \dfrac{\partial h_i^{}(\boldsymbol{s},\boldsymbol{u})}{\partial\boldsymbol{u}}$ \\[5pt]
    \midrule
    & & \\[-9pt]
    \ac{TOA} & $d_i^{} = \lVert\boldsymbol{s}_i - \boldsymbol{u} \rVert$ & $\left[\dfrac{u_x^{}-s_{x,i}^{}}{d_i^{}} \ \ \ \ \ \dfrac{u_y^{}-s_{y,i}^{}}{d_i^{}} \ \ \ \ \ \dfrac{u_z^{}-s_{z,i}^{}}{d_i^{}}\right]$\\[5pt]
    \ac{TDOA} & $
    d_i^{} - d_1^{}= \lVert\boldsymbol{s}_i^{} - \boldsymbol{u} \rVert-\lVert\boldsymbol{s}_1^{} - \boldsymbol{u}  \rVert$ & $\left[\dfrac{u_x^{}-s_{x,i}^{}}{d_i^{}}-\dfrac{u_x^{}-s_{x,1}^{}}{d_1^{}} \ \ \ \ \dfrac{u_y^{}-s_{y,i}^{}}{d_i^{}}-\dfrac{u_y^{}-s_{y,1}^{}}{d_1^{}} \ \ \ \ \dfrac{u_z^{}-s_{z,i}^{}}{d_i^{}}-\dfrac{u_z^{}-s_{z,1}^{}}{d_1^{}}\right]$\\[5pt]
    AOA (az.) & $\phi_i^{} = \text{tan}^{-1}\left(\dfrac{s_{y,i}-u_y^{}}{s_{x,i}-u_x^{}}\right)$ & $\left[\dfrac{d_{i,y}^{}}{d^2_{i,xy}} \ \ \ \ \ -\dfrac{d_{i,x}^{}}{d^2_{i,xy}} \ \ \ \ \  0\right]$ \\[5pt]
    AOA (el.) & $\psi_i^{}=\text{tan}^{-1}\left(\dfrac{s_{z,i}^{}-u_z^{}}{\sqrt{\left(s_{x,i}^{}-u_x^{}\right)^2 +\left(s_{y,i}-u_y^{}\right)^2}}\right)$ & $\left[\dfrac{d_{i,z}^{} \cdot d_{i,x}^{}}{d_i^2 \cdot d_{i,xy}^{}} \ \ \ \ \  \dfrac{d_{i,z}^{} \cdot d_{i,y}^{}}{d_i^2 \cdot d_{i,xy}} \ \ \ \ \ -\dfrac{d_{i,xy}}{d_i^2}\right]$ \\[5pt]

    RSS &  $P_{\text{rx},i}^{}={P}_\text{0}^{} -10\alpha \, \log_{10}\left(\dfrac{\lVert\boldsymbol{s}_i - \boldsymbol{u} \rVert}{d_0^{}}\right)$  & $\left[ -\dfrac{10\alpha}{\ln{10}} \dfrac{u_x^{}-s_{x,i}}{d_i^2} \ \ \ \ \ -\dfrac{10\alpha}{\ln{10}} \dfrac{u_y^{}-s_{y,i}}{d_i^2} \ \ \ \ \ -\dfrac{10\alpha}{\ln{10}} \dfrac{u_z^{}-s_{z,i}}{d_i^2}\right]$ \\[5pt]
    \bottomrule
    \end{tabularx}
\end{table*}

\subsection{Positioning algorithms}
\label{sec:Positioning algorithms}

Estimation of the \ac{UE} position $\boldsymbol{u}$  from the collected measurements $\boldsymbol{\rho}$  (delay, angle, power parameters, or any combination of them) can be obtained by conventional inference algorithms~\cite{gustafsson_positioning_2003, Bellusci_j13}.
The estimation problem amounts to solving a system of non-linear equations in the unknown location $\boldsymbol{u}$  by minimizing a cost function embedding the difference between the available measurements and the related analytical models. Different cost functions are used according to the selected optimization criteria~\cite{Gustafsson_m05}.

A popular approach in positioning systems is the \ac{NLS}~\cite{lu_non-linear_2020, LiLanJia2022},  a non-probabilistic method 
minimizing the square difference between the  measurements and the corresponding models as
\begin{equation}
    \boldsymbol{\hat{u}} = \underset {\boldsymbol{u}} {\text{arg min}} \,
    \lVert\boldsymbol{\rho}-\boldsymbol{h}(\boldsymbol{s},\boldsymbol{u})\rVert^{2}.
    \label{eq:nls}
\end{equation}
An extension of \ac{NLS} is the \ac{WNLS}\cite{jazaeri_weighted_2015}, which takes into account the different degrees of reliability of the measurements (i.e., different statistics) by weighting the error terms as follows:
\begin{align}
    \boldsymbol{\hat{u}} &= \underset {\boldsymbol{u}} {\text{arg min}} 
    \,
    \left\lVert\boldsymbol{\rho}-\boldsymbol{h}(\boldsymbol{s},\boldsymbol{u})\right\rVert^{2}_{\boldsymbol{R}^{-1}}  ,
    \label{eq:wnls}
\end{align}
where 
$\boldsymbol{R}= \mathrm{Cov}(\boldsymbol{\rho})$. 
Under the assumption of uncorrelated measurements, the measurement covariance  matrix $\boldsymbol{R}$  reduces to
a diagonal matrix.

In general, there is no closed-form solution to the non-linear optimization, and thereby, numerical search methods are used. Iterative \ac{NLS} estimation is obtained by initializing the location with a starting guess $\boldsymbol{\hat{u}}_0$ and refining the estimate over the iterations by local linearization and linear resolution. Indicating with $k$ the single iteration, the update is in the form of 
$\boldsymbol{\hat{u}}_{k+1}^{} = \boldsymbol{\hat{u}}_{k}^{}+\Delta\boldsymbol{\hat{u}}_k^{}$,
where $k=0,1,...,K$, with $K$ the maximum number of iterations, and $\Delta\boldsymbol{\hat{u}}_k^{}$ the correction. 
Within the iterative \ac{NLS} category, different implementations exist, such as the Gauss-Newton and Levenberg–Marquardt algorithms~\cite{guvenc_fundamental_2012, more_levenberg-marquardt_1978, gratton_approximate_2007}.

Linearization involves the computation of the Jacobian matrix $\boldsymbol{H}_k^{}$ $\triangleq \boldsymbol{H}_k^{}(\boldsymbol{u}_k^{})$ to be performed at each $k$-th iteration as follows:
 \begin{equation}
     \boldsymbol{H}_k^{} = 
     \left. \frac{\partial\boldsymbol{h}(\boldsymbol{s},\boldsymbol{u})}
     {\partial\boldsymbol{u}} \right|_{\substack{\boldsymbol{u}=\boldsymbol{\hat{u}}_k^{}}}.
     \label{eq:jacob}
 \end{equation}
The element of the Jacobian matrix $\boldsymbol{H}_k^{}$ for each type of measurement considered in this tutorial are reported in Table~\ref{tab:Hel} (Fig.~\ref{fig:cartesian_reference_system} is taken as a reference for notation). 

Depending on the algorithm implementation, the update function of \ac{UE} estimate can slightly differ. As an example, considering the Gauss-Newton algorithm, the update rule for the iterative \ac{NLS} is the following:
\begin{equation}
    \boldsymbol{\hat{u}}_{k+1}^{} = \boldsymbol{\hat{u}}_{k}^{}+\eta\left(\boldsymbol{H}_k^{\T}\boldsymbol{H}_k^{}\right)^{-1}\boldsymbol{H}^{\T}_k \, \Delta\boldsymbol{\rho},
    \label{eq:updateNLS}
\end{equation}
where $\eta$ is a step-size scaling parameter and $\Delta\boldsymbol{\rho} = \boldsymbol{\rho}-\boldsymbol{h}(\boldsymbol{s},\boldsymbol{\hat{u}}_k^{})$ the residual error. Similarly, the update for the iterative \ac{WNLS} with Gauss-Newton implementation becomes:
\begin{equation}
    \boldsymbol{\hat{u}}_{k+1}^{} = \boldsymbol{\hat{u}}_{k}^{}+\eta\left(\boldsymbol{H}_k^{\T}\boldsymbol{R}^{-1}\boldsymbol{H}_k^{}\right)^{-1}\boldsymbol{H}_k^{\T}\boldsymbol{R}^{-1} \, \Delta\boldsymbol{\rho}.
    \label{eq:updateWNLS}
\end{equation}

An alternative implementation of iterative \ac{NLS} is by the Levenberg-Marquardt algorithm, which uses the Hessian matrix instead of the Jacobian one, i.e., considering the second-order derivative of the measurement model $\boldsymbol{h}(\boldsymbol{s},\boldsymbol{u})$~\cite{wilamowski_improved_2010}.

\textcolor{black}{The accuracy of any unbiased positioning algorithm is lower bounded by the \ac{CRB} \cite{Kay}. Denoting the covariance of the location estimate as $\boldsymbol{C} = \mathrm{Cov}(\boldsymbol{u}) =\mathrm{E}[(\hat{\boldsymbol{u}}-\boldsymbol{u})(\hat{\boldsymbol{u}}-\boldsymbol{u})^{\T}]$, the \ac{CRB} specifies that  $\boldsymbol{C}\succeq\boldsymbol{C}_{\text{CRB}}=\boldsymbol{J}^{-1}(\boldsymbol{u})$, where $\boldsymbol{J}(\boldsymbol{u})$ is the \ac{FIM}. 
For Gaussian measurements,  the \ac{FIM} can be expressed in closed form as  $\boldsymbol{J}(\boldsymbol{u}) = \boldsymbol{H}^{\T} \boldsymbol{R}^{-1} \boldsymbol{H}$, with $\boldsymbol{H}$ defined as in Table~\ref{tab:Hel} \cite{Kay}.
The \ac{CRB} represents a useful benchmark for designing localization algorithms and provides a practical tool for optimizing the \ac{BS} deployment.  Furthermore, it is the performance reached asymptotically (i.e., for a large number of measurements or large \ac{SNR}) when the maximum likelihood estimation algorithm is adopted. Indeed, in this specific case, the location estimate is $\hat{\boldsymbol{u}} \sim \mathcal{N}(\boldsymbol{u},\boldsymbol{J}(\boldsymbol{u})^{-1})$~\cite{lehmann1983}. }

\subsection{Bayesian tracking filters}
\label{sec:EKF_tracking}

As an alternative to \ac{NLS} solutions which do not include a-priori knowledge of the \ac{UE} dynamics, Bayesian tracking methods can be implemented to improve positioning accuracy over a trajectory, as well as to embed tracking of higher order kinematic quantities (such as velocity and acceleration). 
In addition to the measurement model (see Section~\ref{sec:Fundamentals of wireless localization}), Bayesian tracking also requires a dynamic system model describing the evolution of the \ac{UE} location over the time $t$. Overall, the two following models are considered: 
\begin{align}
    \boldsymbol{x}_t^{} &= \boldsymbol{f}_t^{} (\boldsymbol{x}_{t-1}^{}) + \boldsymbol{\upsilon}_t,
    \label{eq:system_model_EKF}
\\
    \boldsymbol{\rho}_t^{} &= \boldsymbol{h}_t^{}(\boldsymbol{x}_t^{}) + \boldsymbol{n}_t,
    \label{eq:meas_model_EKF}
\end{align}
where $\boldsymbol{x}_t^{}$ and $\boldsymbol{\rho}_t^{}$ are the vectors of the state (collecting all the relevant kinematic parameters) and the observation vectors at time $t$, respectively, $\boldsymbol{\upsilon}_t^{}$ is the driving process accounting for model uncertainties,  $\boldsymbol{n}_t$ is the measurement error,  $\boldsymbol{f}_t^{}(\cdot)$ and $\boldsymbol{h}_t^{}(\cdot)$ are non-linear functions describing the state evolution in time and mapping the state to the measurement, respectively. The definition of the function $\boldsymbol{h}_{t}^{}(\cdot)$ depends on the type of available measurement (see Table~\ref{tab:Hel}).

One of the most widely-used algorithms in mobile positioning is the \ac{EKF}. The basic principle of \ac{EKF} is to convert a non-linear system into a system of linear equations by focusing on the first-order Taylor expansion of the estimate~\cite{ribeiro2004kalman}.
Other Bayesian solutions include the Unscented Kalman filter~\cite{UKF_04}, the cubature Kalman filter~\cite{Haykin_CKF}, the particle filter~\cite{Arulampalam_j02,Gustafsson_j02}, and the belief propagation~\cite{Loeliger_01}.

Starting from an initialization of the estimated state mean $\hat{\boldsymbol{x}}_0^{}$ and covariance $\boldsymbol{\Sigma}_0^{}$, at the successive time instants $t>0$ the \ac{EKF} performs a prediction and update steps for tracking the \ac{UE} state $\boldsymbol{x}_t^{}$.
The prediction step uses the state transition model \eqref{eq:system_model_EKF} to predict the next state mean $\boldsymbol{x}^-_t$ and covariance $\boldsymbol{\Sigma}^-_t$, as follows:
\begin{equation}
    \boldsymbol{x}^-_t = \boldsymbol{F}_{t}^{} \boldsymbol{\hat{x}}_{t-1}^{},
    \label{eq:state_prediction_mean}
\end{equation}
\begin{equation}
     \boldsymbol{\Sigma}^-_t = \boldsymbol{F}_{t}^{\T}\boldsymbol{\hat{\Sigma}}_{t-1}\boldsymbol{F}_{t}+\boldsymbol{Q}_t,
    \label{eq:state_prediction_cov}
\end{equation}
where
\begin{align}
    \boldsymbol{F}_{t}^{} = \left. \dfrac{\partial \boldsymbol{f}_t^{} (\boldsymbol{x}) }{\partial \boldsymbol{x}} \right|_{\boldsymbol{x} = \boldsymbol{\hat{x}}_{t-1}^{}},
\end{align}
and $\boldsymbol{Q}_t^{} = \mathrm{Cov} (\boldsymbol{\upsilon}_t^{})$.
The update step first requires the computation of the so-called Kalman gain defined as
\begin{equation}
    \boldsymbol{G}_t^{} = \boldsymbol{\Sigma}^-_t\boldsymbol{H}_{t}^{\T} \left(\boldsymbol{H}_{t}^{}\boldsymbol{\Sigma}^-_t\boldsymbol{H}_{t}^{\T} + \boldsymbol{R}_t^{}\right)^{-1},
\end{equation}
where
\begin{align}
    \boldsymbol{H}_{t}^{} = \left. \dfrac{\partial \boldsymbol{h}_t^{} (\boldsymbol{x}) }{\partial \boldsymbol{x}} \right|_{\boldsymbol{x} = \boldsymbol{x}^-_t} ,
\end{align}
followed by the update of state mean and covariance estimates as
\begin{align}
    \boldsymbol{\hat{x}}_t^{} &= \boldsymbol{x}^-_t +\boldsymbol{G}_t^{} \left(\boldsymbol{\rho}_t^{}-\boldsymbol{h}_t^{}(\boldsymbol{x}^-_t)\right),
\\
    \boldsymbol{\hat{\Sigma}}_t &= \boldsymbol{\Sigma}^-_t - \boldsymbol{G}_t^{}\boldsymbol{H}_{t}^{} \boldsymbol{\Sigma}^-_t.
\end{align}

\textcolor{black}{As for the stationary case, fundamental performance bounds can be computed by deriving the \ac{CRB} for mobile positioning employing Bayesian tracking. This holds true as the \ac{CRB} considers asymptotic information and is, therefore, also conservative in filtering. 
The \ac{CRB} for the dynamic case, also known as Bayesian or \ac{PCRB}, can be derived as in \cite{Gustafsson_m05} and varies according to the motion model used in~\eqref{eq:system_model_EKF}. 
In the case of random walk, the lower bound at time $t$ is $\boldsymbol{C}_t^{}=\mathrm{Cov}(\boldsymbol{\hat{x}}_t^{})\succeq \boldsymbol{P}_t^{}$
with $\boldsymbol{P}_t$ 
given by~\cite{Gustafsson_m05}:
\begin{equation}
    \boldsymbol{P}_{t}^{} = \left( (\boldsymbol{P}_{t-1}^{} + \boldsymbol{T}_s^{} \boldsymbol{Q}_{t-1})^{-1} + \boldsymbol{J}(\boldsymbol{x}_{t-1}^{})\right)^{-1}.
\end{equation} 
}

The selection and calibration of the most suitable model of dynamics depend on the considered problem, which might require (or not) the tracking of position, velocity, acceleration, or other kinematic parameters. Examples of motion models are given in~\cite{Gustafsson_m05}. Note that it is also possible to merge more than one model for a quicker reaction to unpredictable motion or to better adhere to highly predictable conditions, such as by \ac{IMM} filtering~\cite{IMM}.


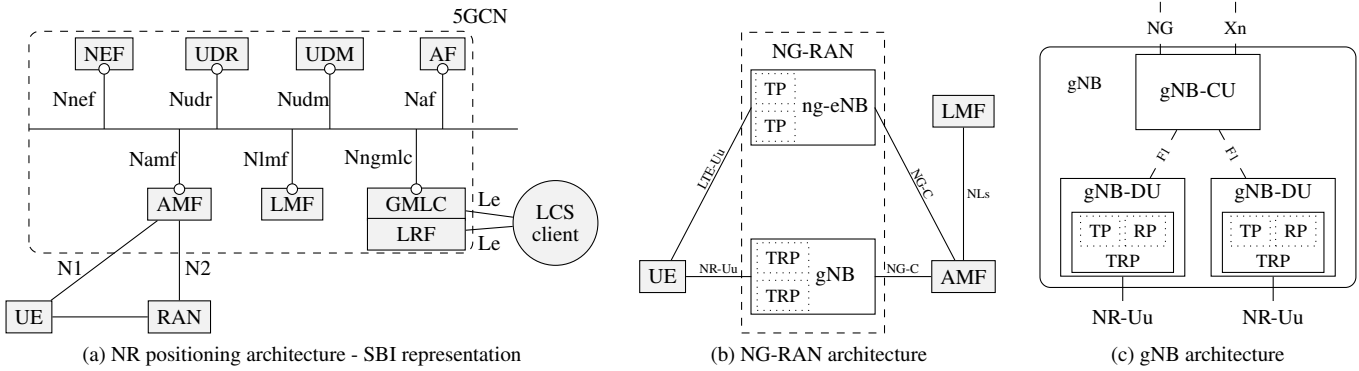
\begin{figure*}[!ht]
    \centering
        
\subfloat[NR positioning architecture - SBI representation]
{
\label{fig:architecture}
\begin{tikzpicture}
    \tikzstyle{every node}=[font=\footnotesize]

\node[rectangle, draw, fill=morelightgrey] (UE) at (-5,-3) {UE}; 
\node[rectangle, draw, fill=morelightgrey] (RAN) at (-3,-3) {RAN};
\draw  (UE)--(RAN);
\node[rectangle, draw, fill=morelightgrey] (AMF) at (-3,-1.5) {AMF};
\draw  (AMF)--(RAN) node[midway, xshift=2.5mm , yshift=-1mm]{N2};
\draw  (AMF)--(UE) node[midway, xshift=-4.5mm,yshift=-1mm]{N1};
\node[rectangle, draw, fill=morelightgrey] (LMF) at (-1.5,-1.5) {LMF};
\node[rectangle, draw, fill=morelightgrey, minimum width = 1.3 cm] (GMLC) at (0.15,-1.5) {GMLC};
\node[rectangle, draw, fill=morelightgrey, minimum width = 1.3 cm] (LRF) at (0.15,-1.9) {LRF};
\node[circle, draw, fill=morelightgrey] (LCS) at (2,-1.75) {\shortstack{LCS \\ client}};
\draw  (GMLC)--(LCS) node[midway, yshift=2mm]{Le};
\draw  (LRF)--(LCS) node[midway, yshift=-2mm]{Le};

\node[rectangle, draw, fill=morelightgrey] (NEF) at (-4,0.5) {NEF};
\node[rectangle, draw, fill=morelightgrey] (UDR) at (-2.5,0.5) {UDR};
\node[rectangle, draw, fill=morelightgrey] (UDM) at (-1,0.5) {UDM};
\node[rectangle, draw, fill=morelightgrey] (AF) at (0.5,0.5) {AF};

\draw  (-5,-0.5)--(1.5,-0.5) ;

\draw   (-4,0.3)--(-4,-0.5) node[midway, xshift=-4mm]{Nnef};
\draw   (-2.5,0.3)--(-2.5,-0.5) node[midway, xshift=-3.5mm]{Nudr};
\draw   (-1,0.3)--(-1,-0.5) node[midway, xshift=-3.5mm]{Nudm};
\draw   (0.5,0.3)--(0.5,-0.5) node[midway, xshift=-3mm]{Naf};
\draw   (-3,-1.3)--(-3,-0.5) node[midway, xshift=-3.5mm]{Namf};
\draw   (-1.5,-1.3)--(-1.5,-0.5) node[midway, xshift=-3.5mm]{Nlmf};
\draw   (0.15,-1.3)--(0.15,-0.5) node[midway, xshift=-5mm]{Nngmlc};

\draw [fill=white] (-4,0.3) circle (2pt);
\draw [fill=white] (-2.5,0.3) circle (2pt);
\draw [fill=white] (-1,0.3) circle (2pt);
\draw [fill=white] (0.5,0.3) circle (2pt);
\draw [fill=white] (-3,-1.3) circle (2pt);
\draw [fill=white] (-1.5,-1.3) circle (2pt);
\draw [fill=white] (0.15,-1.3) circle (2pt);

\draw[dashed, rounded corners] (-5, -2.15) rectangle (0.9, 0.8) {};
\node[] at (1, 1) {5GCN};

\end{tikzpicture}
}
\hfill
\subfloat[NG-RAN architecture]
{
 \label{fig:RAN}
\begin{tikzpicture}
    \tikzstyle{every node}=[font=\footnotesize]
    \node[rectangle, draw, fill=morelightgrey] (UE) at (-2,-2) {UE}; 
    \node[rectangle, draw, fill=morelightgrey] (AMF) at (2,-2) {AMF}; 
    \node[rectangle, draw, fill=morelightgrey] (LMF) at (2,0.2) {LMF}; 
    \node[rectangle, draw,  minimum width = 1.9 cm,  minimum height = 4 cm , dashed ]   at (0,-0.75) {} ; 
    \node at (0,1) {NG-RAN};
    \node[rectangle, draw, minimum width = 1.65 cm, minimum height = 1 cm ]   at (0,0.25) {}; 
    \node at (0.3,0.25) {ng-eNB}; 
    \node[rectangle, draw,minimum width = 1.65 cm,minimum height = 1 cm ]   at (0,-2) {} ; 
    \node   at (0.3,-2) {gNB}; 
    \node[rectangle, draw,minimum width = 0.5 cm,minimum height = 0.4 cm , dotted]   at (-0.5,0) {\scriptsize TP}; 
    \node[rectangle, draw,minimum width = 0.5 cm,minimum height = 0.4 cm , dotted]   at (-0.5,0.5) {\scriptsize TP}; 
    \node[rectangle, draw,minimum width = 0.5 cm,minimum height = 0.4 cm , dotted]   at (-0.4,-1.75) {\scriptsize TRP}; 
    \node[rectangle, draw,minimum width = 0.5 cm,minimum height = 0.4 cm , dotted]   at (-0.4,-2.25) {\scriptsize TRP}; 

\draw (UE)--(-0.82,-2) node[midway, yshift=1mm]{\tiny NR-Uu};
\draw (UE)--(-0.82,0.25) node[midway, yshift=2mm , rotate=62]{\tiny LTE-Uu};

\draw  (AMF)--(0.82,-2) node[midway, yshift=1mm]{\tiny NG-C};
\draw  (AMF)--(0.82,0.25) node[midway, xshift=1mm, rotate=-62]{\tiny NG-C};
\draw  (AMF)--(LMF) node[midway, xshift=2mm]{\tiny NLs};

\end{tikzpicture}
}
\hfill
\subfloat[gNB architecture]
{
\label{fig:BS}
\begin{tikzpicture}
    \tikzstyle{every node}=[font=\footnotesize]
        \node[rectangle, draw,minimum width = 1.65 cm,minimum height = 1 cm ] (gNBCU)  at (0,0.6) {gNB-CU}; 
        \node[rectangle, draw,minimum width = 1.65 cm,minimum height = 1.3 cm ] (gNBDU1)   at (-1,-1.2) {};
        \node[rectangle, draw,minimum width = 1.65 cm,minimum height = 1.3 cm ] (gNBDU2)  at (1,-1.2) {};        
        \node  at (-1,-0.75) {gNB-DU};
        \node  at (1,-0.75) {gNB-DU};

        \node[] (NRU1)  at (-1,-2.4) {NR-Uu};
        \node[] (NRU2) at (1,-2.4) {NR-Uu};
        
        \node[rectangle, draw,minimum width = 1.35 cm,minimum height = 0.8 cm ]   at (1,-1.4) {};        
        \node[rectangle, draw,minimum width = 1.35 cm,minimum height = 0.8 cm ]   at (-1,-1.4) {};        

        \node[rectangle, draw,minimum width = 0.5 cm,minimum height = 0.3 cm , dotted]   at (-1.3,-1.25) {\scriptsize TP}; 
        \node[rectangle, draw,minimum width = 0.5 cm,minimum height = 0.3 cm , dotted]   at (-0.7,-1.25) {\scriptsize RP}; 
        \node[rectangle, draw,minimum width = 0.5 cm,minimum height = 0.3 cm , dotted]   at (0.7,-1.25) {\scriptsize TP}; 
        \node[rectangle, draw,minimum width = 0.5 cm,minimum height = 0.3 cm , dotted]   at (1.3,-1.25) {\scriptsize RP}; 

        \node[minimum width = 0.5 cm,minimum height = 0.3 cm]   at (1,-1.65) {\scriptsize TRP};
        \node[minimum width = 0.5 cm,minimum height = 0.3 cm]   at (-1,-1.65) {\scriptsize TRP};
        
        \draw  (gNBCU)--(gNBDU1) node[rectangle, draw=white, fill=white, midway, rotate=60]{\tiny F1};
        \draw  (gNBCU)--(gNBDU2) node[rectangle, draw=white, fill=white, midway, rotate=-60]{\tiny F1};

        \draw  (NRU1)--(gNBDU1);
        \draw  (NRU2)--(gNBDU2);

        \draw  (-0.5,1.1)--(-0.5,1.8) node[rectangle, draw=white, fill=white, midway]{\scriptsize NG};
        \draw  (0.5,1.1)--(0.5,1.8) node[rectangle, draw=white, fill=white, midway]{\scriptsize Xn};
        
        \draw[rounded corners] (-2.1, -2) rectangle (2.1, 1.2) {};
        \node[] at (-1.5, 0.7) {\scriptsize gNB};
\end{tikzpicture}

}

        \caption{Main architectures of 5G positioning~\cite{fischer}. (a) SBI representation of the NR positioning architecture, (b) NG-RAN architecture, (c) gNB architecture.}
\end{figure*}

\section{5G Positioning Technology (Rel-16)}
\label{sec:5G Positioning Technology (Rel-16)}
\textcolor{black}{In this section, we discuss various aspects of \ac{5G}  positioning. We start with the description of the \ac{5G} positioning architecture (Section~\ref{sec:5G positioning architecture}),
then we detail the \ac{5G} frame structure (Section~\ref{sec:5G Frame Structure}) highlighting its impact on the positioning accuracy compared to \ac{LTE} (Section~\ref{sec:Positioning Precision: LTE vs NR}).
In Section~\ref{sec:5G Signals for Positioning}, we describe the different signals for \ac{5G} positioning, both for \ac{UL} and \ac{DL}; the associated positioning methods are in Section~\ref{sec:5G positioning methods}. Lastly, we explain how to extract positioning measurements from the \ac{5G} signals  (Section~\ref{sec:5G measurement extraction}). 
}

\subsection{5G positioning architectures}
\label{sec:5G positioning architecture}

The general architecture of a \ac{5G}  network is shown in Fig.~\ref{fig:architecture}. 
Main components are the \ac{5GCN} and the \ac{RAN}~\cite{fischer}.
The \ac{5GCN} is built on a \ac{SBA}, which guarantees the network functionalities using a set of \acp{NF}. Functions can interact with each other using the \ac{SBI}. The main \acp{NF} are the \ac{LMF} and the \ac{AMF}. The \ac{LMF}  is in charge of all the procedures regarding the UE localization, such as selection of the positioning method,  resource scheduling, and overall coordination, and it is responsible for broadcasting the assistance data to the \acp{UE}. The \ac{AMF}, instead, supports location services, including emergency calls and initiating a localization request for a \ac{UE}. Generally, it can be considered an intermediary node between the \ac{LMF} and the RAN or the UE.

The \ac{RAN} is involved in the handling of the positioning procedures, and it has the duty of transferring messages between the \ac{UE} and the \ac{AMF} or \ac{LMF}, such as positioning messages or broadcast assistance data. The \ac{RAN}, or \ac{NG-RAN}, is formed by an ng-eNB for \ac{LTE} access and a \ac{BS} for NR access, as shown in Fig.~\ref{fig:RAN}.

Differently from the monolithic building block of the \ac{4G} \ac{RAN} architecture, i.e., \ac{eNB}, the architecture of \ac{5G} \ac{BS} can be split into a \ac{gNB-CU} and one or more \acp{gNB-DU}, as shown in Fig.~\ref{fig:BS}. 
The \ac{gNB} can transmit a signal in \ac{DL} or measure a signal in \ac{UL}, enabling the implementation of the various positioning methods. This twofold feature is possible thanks to the \ac{TRP}, which acts as a \ac{TP}, a \ac{RP}, or both.

\subsection{5G frame structure}
\label{sec:5G Frame Structure}
The physical layer of \ac{5G}  is characterized by a frame of duration of 10~ms, as for LTE. However, the frame structure differs in the two protocols. In LTE, the frame is divided into 10 sub-frames of 1~ms, each being composed of 2 slots of 7 \ac{OFDM} symbols in time and occupying 12 subcarriers in the frequency domain. In 5G, each frame is divided into 10 sub-frames of 1 ms duration, and each sub-frame is divided into slots, containing $N_{\text{symb}}^{\text{slot}}=14$ \ac{OFDM} symbols each. The number of slots is variable and depends on the \ac{SCS}, which is univocally defined by the numerology, indicated with $\mu$.  Table~\ref{tab:scs} reports the numerology $\mu$, the number of slots for each sub-frame  $N_{\text{slot}} = 2^{\mu}$, the \ac{SCS} $\Delta f=15 \cdot N_{\text{slot}} $ (in kHz), the \ac{FR}, the maximum bandwidth (in MHz), the average symbol duration  $T_\text{symb} = \frac{1}{\Delta f}$~$\mu$s, and the cyclic prefix length $T_\text{cp}$. Moreover, we associated each numerology with a theoretical ranging accuracy computed as $\approx c/ \text{BW}$.

\begin{table*}[!tb]
    \centering
    \caption{Supported 5G numerologies and main parameters~\cite{ts138101-1,ts138101-2, ts138300}}
    \label{tab:scs}
    \renewcommand{\arraystretch}{1.1} 
    \begin{tabular}{c c c c c c c c c c}
        \toprule
         $\mu$ & $\Delta f$ $[\text{kHz}]$ & FR & BW $[\text{MHz}]$ & $N_{\text{slot}}^{}$ & $T_\text{symb}^{}$ $[\mu \text{s}]$ & $T_\text{cp}^{}$ $[\mu \text{s}]$ & \acapo{Ranging accuracy \\ $\approx c/ \text{BW}$ $[\text{m}]$} & Data & Synch.\\
        \midrule
           &    &   &    &   &      &   &  &  \\[-9pt]
         0 & 15 & 1 & 50 & 1 & 66.7 & 4.69 & 6.00 & \cmark & \cmark \\
         1 & 30 & 1 & 100 & 2 & 33.3 & 2.34 & 3.00 & \cmark & \cmark \\
         2 & 60 & 1/2 & 200 & 4 & 16.7 & 1.17& 1.50 & \cmark & $\xmark$ \\
         3 & 120 & 2 & 400 & 8 & 8.33 & 0.57 & 0.75 & \cmark & \cmark \\
         4 & 240 & 2 & 400 & 16 & 4.17 & 0.29 & -  & $\xmark$ & \cmark \\
         5 & 480 & 2 & 1600 & 32 & 2.08 & 0.15& 0.19 & \cmark & \cmark \\
         6 & 960 & 2 & 2000 & 64 &  1.04 & 0.07 & 0.15 & \cmark & \cmark \\
         \bottomrule
    \end{tabular}
\end{table*}

In LTE, the numerology was limited to $\mu=0$. The \ac{3GPP} Rel-15 extended it up to numerology $\mu=4$~\cite{tr138912}, and the latest \ac{3GPP} Rel-17 has further enhanced the numerology up to $\mu=6$~\cite{ts138101-2}.
While the maximum supported channel bandwidth for LTE is 20~MHz, in \ac{5G} it is 100~MHz for \ac{FR}1~\cite{8316765}, 400~MHz for \ac{FR}2 in Rel-16 and 2~GHz for \ac{FR}2 in Rel-17~\cite{ts138101-2}. Note that numerology $\mu=4$ is not intended to support data transmission~\cite{ts138300}, but only synchronization. 
On the contrary, numerology $\mu=2$ only supports data transmission and not synchronization.

\begin{figure}[t]
    \centering

\begin{tikzpicture}

\foreach \x in {-10,-9,-8,-7,-6,-5,-4,-3,-2,-1,0,1,...,9}
{
\foreach \y in {-10,-9,-8,-7,-6,-5,-4,-3,-2,-1,0,1,...,9}
{
    \draw [lightgrey] (0+0.3*\x, 0+0.3*\y) rectangle (0.3+0.3*\x, 0.3+0.3*\y);
}
}

\foreach \x in {-7,-6,-5,-4,-3,-2,-1,0,1,...,6}
{
\foreach \y in {-10,...,-5,-4,-3,-2,-1,0,1,...,9}
{
    \filldraw [lightgrey, fill=bubblegum, opacity=0.3] (0+0.3*\x, 0+0.3*\y) rectangle (0.3+0.3*\x, 0.3+0.3*\y);
}
}

\foreach \x in {-7,-6,-5,-4,-3,-2,-1,0,1,...,6}
{
\foreach \y in {-6,-5,-4,-3,-2,-1,0,1,...,5}
{
    \draw (0+0.3*\x, 0+0.3*\y) rectangle (0.3+0.3*\x, 0.3+0.3*\y);
}
}

\draw[-{Stealth}] (-3,-3) -- (4,-3) node [midway, yshift = -3mm] {\small Symbols [Time]};
\draw[-{Stealth}] (-3,-3) -- (-3,4) node [midway, xshift = -3mm , rotate=90] {\small Sub-carrier [Frequency]};

\draw[stealth-stealth] (-2.1,-2.05) -- (2.1,-2.05) node [midway, yshift = -3mm] {$N_{\text{symb}}^{\text{slot}}$};
\draw[stealth-stealth] (-2.3,-1.8) -- (-2.3,1.8) node [midway, xshift = -3.5mm] {\small $N^{\text{RB}}_{\text{SC}}$};
\draw[stealth-stealth] (2.3,1.5) -- (2.3,1.2) node [midway, xshift = 3mm] {\small $\Delta f$};

\draw[dashed] (-2.1,-3) -- (-2.1,3) ;
\draw[dashed] (2.1,-3) -- (2.1,3) ;

\draw [fill=blue!30!] (1.2, 0.9) rectangle (1.5, 1.2);
\draw (2.3+1, 3.5-2) rectangle (2.3+0.3+1, 3.5+0.3-2) node [midway, xshift=5mm]{\small RE};
\filldraw [fill= blue!30!] (2.3+1, 3.5-2) rectangle (2.3+0.3+1, 3.5+0.3-2);
\draw [magenta] (2.3+1, 3.5-1.5) rectangle (2.3+0.3+1, 3.5+0.3-1.5) node [black, midway, xshift=5mm] {\small RB};
\draw [dashed] (2.3+1, 3.5-1) rectangle (2.3+0.3+1, 3.5+0.3-1) node [midway, xshift=5mm]{\small Slot};
\filldraw [fill=bubblegum,opacity=0.3] (2.3+1, 3.5-1) rectangle (2.3+0.3+1, 3.5+0.3-1);

\draw[magenta, rounded corners] (-2.2, -1.9) rectangle (2.2,1.9) {};

\end{tikzpicture}
    
    \caption{Representation of 5G resource grid in time and frequency domains, with highlights on the RB, RE, and slot.} 
    \label{fig:grid}
\end{figure}
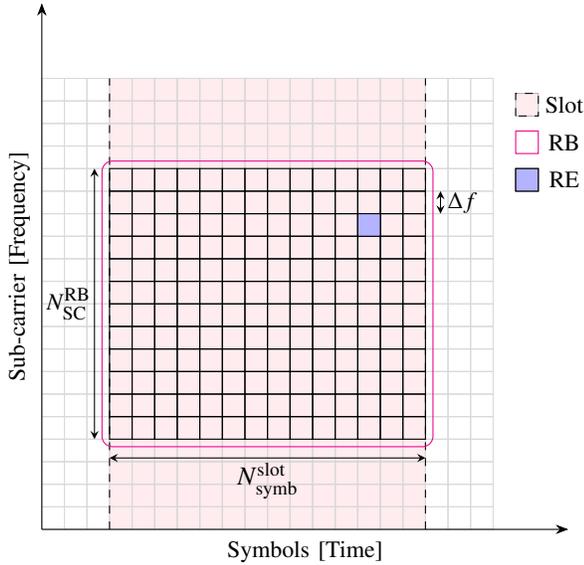

Fig.~\ref{fig:grid} defines the resource grid in the time and frequency domain. A \ac{RB} is a set of $N^{\text{RB}}_{\text{SC}} = 12$ sub-carriers inside a slot of 14 \ac{OFDM} symbols. A \ac{RE} is the smallest unit in the resource grid, constituted by a single symbol in time and a single sub-carrier in frequency.
Gathering all the parameters,  the signal bandwidth is computed as
\begin{equation}
    BW = N^{\text{RB}}_{} \cdot \Delta f \cdot N^{\text{RB}}_{\text{SC}},
    \label{eq:BW}
\end{equation}
where $N^{\text{RB}}_{}$ is the number of utilized \acp{RB}, and the data rate (in Mbps) is~\cite{ts138306}:
\begin{align}
    & DR = \\ \nonumber
    &10^{-6} \cdot \sum^J_{j=1} \left( v_{j, \text{layers}}^{} \cdot Q_{j, m}^{} \cdot f_j^{} \cdot R^{}_{\text{\text{max}}} \cdot \frac{12\cdot N^{\text{RB}}_{} }{T^{}_\text{symb}} \cdot (1-OH_j^{})\right),
    \label{eq:DR}
\end{align}
where $J$ is the number of aggregated component carriers in a band, $R^{}_{\text{\text{max}}} = \frac{948}{1024}$, $v_{j, \text{layers}}^{}$ is the maximum number of supported layers (8 in DL, 4 in UL), $Q_{j,m}^{}$ is the maximum supported modulation order, $f_j^{} \in \{1, 0.8, 0.75, 0.4\}$ is a scaling factor, 
$T^{}_\text{symb}$ is the average \ac{OFDM} symbol duration in a subframe for numerology $\mu$~\cite{ts138101-1, ts138101-2}, and $OH_j^{}$ is the overhead which can take the following values:
\begin{itemize}
\item $OH_j^{}=0.14$, for \ac{FR}1 in \ac{DL},
\item $OH_j^{}=0.18$, for \ac{FR}2 in \ac{DL},
\item $OH_j^{}=0.08$, for \ac{FR}1 in \ac{UL},
\item $OH_j^{}=0.10$, for \ac{FR}2 in \ac{UL}.
\end{itemize}

\subsection{Time-domain accuracy: LTE vs NR}
\label{sec:Positioning Precision: LTE vs NR}

With the addition of \ac{FR}2 bands, larger signal bandwidths and higher data rates are available. Larger signal bandwidth is the key to unlocking high-accuracy positioning, as the resolution in delay estimation, which is roughly equal to the inverse of the bandwidth (i.e., the sampling time), improves and enhances the capability to resolve multipath. 

To highlight the improvement brought by \ac{5G} NR with respect to \ac{LTE}, we analyze the following: the temporal resolution of the different numerologies and the corresponding ranging accuracy.
The minimum sampling time is:
\begin{equation}
    T_s^{} = \frac{1}{ \Delta f_{\text{max}}^{} \cdot N_f^{}},
    \label{eq:ts}
\end{equation}
with $N_f^{}$ as the number of Fourier points, which provides 
a granularity in the ranging domain  $\Delta r =T_s^{} \cdot c$.
For LTE  (numerology $\mu=0$), we get the following delay and range resolution:
\begin{align}
        T_s^{\text{LTE}} &= \frac{1}{15.000 \cdot 2048} \approx 32.55 \,\text{ns},
        \\
        \Delta r^{\text{LTE}} &= T_s^{\text{LTE}} \cdot c \approx 10 \, \text{m};
\end{align}
while for \ac{5G} Rel-16 ($\mu=3$) it is:
\begin{align}
    T_s^{\text{5G Rel-16}} &= \frac{1}{120.000 \cdot 4096} \approx 2.03 \,\text{ns},
    \\
    \Delta r^{\text{5G Rel-16}} &= T_s^{\text{5G Rel-16}} \cdot c \approx 60.8 \,\text{cm}.
\end{align}
Instead, taking into consideration the highest numerology introduced by Rel-17 ($\mu=6$), we obtain:
\begin{align}
    T_s^{\text{5G Rel-17}} &= \frac{1}{960.000 \cdot 4096} \approx 0.25 \,\text{ns},
    \\
    \Delta r^{\text{5G Rel-17}} &= T_s^{\text{5G Rel-17}} \cdot c \approx 7.6 \,\text{cm}.
\end{align}
The finer granularity of \ac{5G}  \ac{NR} compared to LTE highlights the huge potential in accurate positioning of \ac{5G} at \acp{mmWave}~\cite{GarMaiPhi2020}. 
On the other hand,  the coverage of a \ac{BS} transmitting in \ac{FR}2 is highly reduced, leading to a densification of \ac{BS} installations. This is not necessarily a drawback. Indeed, while adding more \acp{BS} will cost more from the cellular operators' point of view, it also allows greater frequency reuse. Moreover, smaller cell size might provide satisfactory positioning performance even using the basic \ac{CID} method, which can be used for non-critical applications such as geo-marketing.

\subsection {5G positioning signals}

\label{sec:5G Signals for Positioning}
In Rel-16, the \ac{3GPP} standard updates and redefines two reference signals in order to overcome the positioning problems of previous  releases~\cite{ts138211}. 
Former signals, such as \ac{CSI-RS} and \ac{SS} (which composes the \acp{SSB}), were not designed specifically for positioning because of the following limitations.
A first major limitation is their inability to solve the hearability issue arising from interference by neighboring cells~\cite{yap_accuracy_2002}. This is crucial for positioning, as the \ac{UE} must receive signals simultaneously from multiple \acp{BS} to perform multi lateration/angulation.
On the other hand, signals from nearby cells shadow weak signals coming from far-away cells, making their detection difficult at the \ac{UE}. Lastly, \acp{CSI-RS} and \acp{SS} have weak correlation properties due to low density of \acp{RE} and their pattern. Therefore, they might not spread well across all of the sub-carriers in the frequency-domain. 
For these reasons, the \ac{PRS} for \ac{DL} transmission and the \ac{SRS} for \ac{UL} transmission have been introduced in Rel-16 with the aim of allowing precise positioning by the \ac{5G}  cellular network. 

In the following, we describe the features of \ac{SSB}, \ac{CSI-RS}, \ac{PRS}, and \ac{SRS}, whose main differences affecting positioning are summarized in Table~\ref{tab:pos_signals}. The number of beams for \ac{SRS} and \ac{PRS} are associated with the number of \ac{RE} in a slot.
\begin{table}[!tb]
\caption{Comparison of positioning signals in 3GPP Rel-16} 
\label{tab:pos_signals}
\renewcommand{\arraystretch}{1.1} 
\linespread{0.5}\selectfont\centering
    \centering

    \begin{tabular}{l c c c c }
        \toprule
        \textbf{Signal} & \textbf{Max BW [MHz]} & \textbf{Number of beams} & \textbf{\acapo{Designed for \\Positioning}}  \\ \midrule 
        
        & & & \\[-4pt]
        
        SSB& 60 &4, 8, 64& $\xmark$\\[3pt]
        CSI-RS& 400 & 2-8 & $\xmark$\\[3pt]
        PRS& 400 & 2-12 & \cmark\\[3pt]
        SRS& 400 & 1-12 & \cmark\\
        
        \bottomrule    
    \end{tabular}
\end{table}

\input{Figure_tex/SSB_composition.tex}
\begin{table*}[!t]
\caption{SSB pattern specifications \cite{ssb_standard,ssb_standard_patter}}
\label{tab:SSB_patterns}
\renewcommand{\arraystretch}{1.1} 
\linespread{0.5}\selectfont\centering
    \centering

    \begin{tabular}{l c c c c }
\toprule
\textbf{SCS} & \textbf{{ Starting OFDM symbol}} & {{ $f_c^{} \leq 3$ GHz}} & {{ $3 < f_c^{} \leq 6$ GHz}} & {{ $f_c^{} > 6$ GHz}} \\ \midrule 

& & & & \\[-4pt]
        
        {  Case A: 15 kHz} & $\{2, 8\}$ + 14 $n$ & {  $n = 0,1$ (4 SSBs)} & \shortstack{  $n = 0,1,2,3$ (8 SSBs)} & NA\\[3pt] 

        {  Case B: 30 kHz} & $\{4, 8, 16, 20\}$ + 28 $n$ & $n = 0$ (4 SSBs)& {  $n = 0,1$(8 SSBs)} & NA\\[3pt] 

        {  Case C: 30 kHz} & $\{2, 8\}$ + 14 $n$ & {  $n = 0,1$ (4 SSBs)} & {  $n = 0,1,2,3$ (8 SSBs)} & NA\\[3pt] 

        {  Case D: 120 kHz} & $\{4,8,16,20\}$ + 28 $n$ & NA & NA & {  $n = \{i\}_{i=0}^{18}$ (64 SSBs)} \\[3pt] 

          Case E:  240 kHz &    $\{8,12,16,20$   $32,36,40,44\}$ + 56 $n$ & NA & NA &   $n = \{i\}_{i=0}^{8}$ (64 SSBs) \\
        
        \bottomrule
        
    
    \end{tabular}
\end{table*}


\input{Figure_tex/SSB_pattern.tex}

\subsubsection{SSB}
The \ac{SSB} consist of the \ac{SS}, \ac{PBCH}, and \ac{DMRS}. \acp{SSB} are periodically transmitted in broadcast by a \ac{TRP} within spatially contained bursts (\ac{SS} burst set) in a beam sweeping pattern (i.e., each \ac{SSB} over a specific spatial beam).
The main objectives of the \ac{SSB}, also known as  SS/PBCH block, are the following.
To have an active \ac{5G} connection,  an \ac{UE} has to perform a cell-search procedure to identify, locate, and synchronize with a \ac{TRP}. 
The cell-search during the initial access is conducted through the use of \ac{PSS} and \ac{SSS}, which constitute the \ac{SS}.
Additionally, the \ac{UE} uses \ac{DL} signals such as the \ac{PDSCH} and \ac{PBCH} to obtain the necessary system parameters for the connection. The \ac{UE} also detects the \ac{DMRS}, which acts as a reference signal for decoding the \ac{PDSCH}  and \ac{PBCH}. 
Each \ac{SSB} is sent over a different spatial direction at different timing by the \ac{TRP}, and the \ac{UE} measures the signal strength of each \ac{SSB}. Based on the measuring results, the \ac{UE} can determine and report to the \ac{TRP} the index of the strongest (in terms of power) \ac{SSB}.

The structure of the \ac{SSB} is reported in Fig.~\ref{fig:SSB_composition}. It is constituted by 20 \acp{RB} and 4 \ac{OFDM} symbols in the frequency and time domains, respectively. Depending on the adopted carrier frequency $f_c^{}$, different numbers of consecutive \acp{SSB} ($N^{\text{SSB}}$) compose an \ac{SS} burst set. Intuitively, the higher the carrier frequency, the narrower and more directive the beam will be. 
For frequency below 3~GHz, $N^{\text{SSB}} = 4$; for frequency between 3 and 6~GHz $N^{\text{SSB}} = 8$; and for frequency between 6 and 52.6~GHz $N^{\text{SSB}} = 64$. Depending on the \ac{SCS} and carrier frequency, the starting \ac{OFDM} symbol of the \ac{SSB} varies according to a specific pattern, as described by \ac{3GPP} specification in~\cite{ssb_standard,ssb_standard_patter}.
Patterns are categorized as Case A, B, C, D, and~E, and they mainly differ according to the \ac{SCS} and carrier frequency $f_c$ as indicated in Table~\ref{tab:SSB_patterns}.  Fig.~\ref{fig:SSB_pattern} depicts every \ac{SSB} pattern and demonstrates how \acp{TRP} operating at higher frequencies (such as millimeter waves) employ more beams overall. A \ac{TRP}'s ability to comprehensively scan the spatial domain using more directed beams is indicated by a higher  $N^{\text{SSB}}$.


\begin{table*}[ht]
    \centering
    \caption{Resource element offsets of PRS for all the comb patterns \cite{ts138211, fischer}}
    \label{tab:prsoffset}
    \begin{tabular}{ c c c c c }
        \toprule
        \diagbox{$K_{\text{size}}^{}$}{$N_{\text{symb}}^{\text{slot}}$} & 2 & 4 & 6 & 12 \\
        \midrule
        2 & $\{0, 1\}$ & $\{0, 1, 0, 1\}$ & $\{0, 1, 0, 1, 0, 1\}$ & $\{0, 1, 0, 1, 0, 1, 0, 1, 0, 1, 0, 1\}$  \\
        4 & \texttt{-} & $\{0, 2, 1, 3\}$ & \texttt{-} & $\{0, 2, 1, 3, 0, 2, 1, 3, 0, 2, 1, 3\}$ \\
        6 & \texttt{-} & \texttt{-} & $\{0, 3, 1, 4, 2, 5\}$ & $\{0, 3, 1, 4, 2, 5, 0, 3, 1, 4, 2, 5\}$\\
        12 & \texttt{-} & \texttt{-} & \texttt{-} & $\{0, 6, 3, 9, 1, 7, 4, 10, 2, 8, 5, 11\}$\\
        \bottomrule
    \end{tabular}
\end{table*}
\begin{table*}[ht]
\centering
\caption{Resource element offsets of SRS for all the possible combinations. \cite{ts138211, fischer}}
\label{tab:srsoffset}
\begin{tabular}{ c c c c c c}
\toprule
\diagbox{$K_{\text{size}}^{}$}{$N_{\text{symb}}^{\text{slot}}$} & 1 & 2 & 4 & 8 & 12 \\
\midrule
2 & $\{0\}$ & $\{0, 1\}$ & $\{0, 1, 0, 1\}$ & \texttt{-} & \texttt{-}  \\
4 & \texttt{-} & $\{0, 2\}$ & $\{0, 2, 1, 3\}$ & $\{0, 2, 1, 3, 0, 2, 1, 3\}$ & $\{0, 2, 1, 3, 0, 2, 1, 3, 0, 2, 1, 3\}$ \\
8 & \texttt{-} & \texttt{-} & $\{0, 4, 2, 6\}$ & $\{0, 4, 2, 6, 1, 5, 3, 7\}$ & $\{0, 4, 2, 6, 1, 5, 3, 7, 0, 4, 2, 6\}$ \\
\bottomrule
\end{tabular}

\end{table*}

\subsubsection{CSI-RS}
\ac{CSI-RS} were introduced in Rel-10 with the aim of acquiring the channel state information. In order to support up to eight layers of spatial multiplexing, the configuration of \acp{CSI-RS} can be defined accordingly with the same number of signals for a \ac{TRP}. 
In time-domain, the \ac{CSI-RS} periodicity can be configured such that there can be from 2 to 8 \acp{CSI-RS} in every frame. For a given periodicity, it is also possible to configure the subframe offset. The \ac{CSI-RS} is transmitted in every RB in the frequency-domain. In this way, \ac{CSI-RS} can cover the entire cell bandwidth. The \acp{RE} actually used depend on the defined \ac{CSI-RS} configuration.
In addition to conventional \ac{CSI-RS}, also known as \ac{NZP-CSI-RS}, it is possible to configure \ac{ZP-CSI-RS} with the same structure~\cite{DAHLMAN2014161}. 

\begin{figure*}[t]
\subfloat[Comb-2]{
\begin{tikzpicture}

\foreach \x in {-7,-6,-5,-4,-3,-2,-1,0,1,...,6}
{
\foreach \y in {-6,-6,-5,-4,-3,-2,-1,0,1,...,5}
{
    \draw [lightgrey] (0+0.25*\x, 0+0.25*\y) rectangle (0.25+0.25*\x, 0.25+0.25*\y);
}
}
\foreach \x in {1,3,...,13}
{
\node [rotate=0] at(-1.62+0.25*\x,-1.7) {\scriptsize \x};
}
\foreach \y in {2,4,...,12}
{
\node [rotate=90] at(-2,-1.64+0.25*\y) {\scriptsize \y};
}
\node [rotate=0] at(0,-2.1) {\small OFDM symbol};
\node [rotate=90] at(-2.3,0) {\small Sub-carrier };

\foreach \y in {1,3,...,11}
{
\draw [fill=lightgrey] (-0.25*4, -1.75+0.25*\y) rectangle (-0.25*3, -1.5+0.25*\y);
}
\foreach \y in {2,4,...,12}
{
\draw [fill=lightgrey] (-0.25*3, -1.75+0.25*\y) rectangle (-0.25*2, -1.5+0.25*\y);
}

\end{tikzpicture}
}
\subfloat[Comb-4]{
\begin{tikzpicture}

\foreach \x in {-7,-6,-5,-4,-3,-2,-1,0,1,...,6}
{
\foreach \y in {-6,-6,-5,-4,-3,-2,-1,0,1,...,5}
{
    \draw [lightgrey] (0+0.25*\x, 0+0.25*\y) rectangle (0.25+0.25*\x, 0.25+0.25*\y);
}
}
\foreach \x in {1,3,...,13}
{
\node [rotate=0] at(-1.62+0.25*\x,-1.7) {\scriptsize \x};
}
\foreach \y in {2,4,...,12}
{
\node [rotate=90] at(-2,-1.64+0.25*\y) {\scriptsize \y};
}
\node [rotate=0] at(0,-2.1) {\small OFDM symbol};
\node [rotate=90] at(-2.3,0) {\small Sub-carrier };

\foreach \y in {1,5,9}
{
\draw [fill=lightgrey] (-1.25+0.25*2, -1.75+0.25*\y) rectangle (-1.25+0.25*1, -1.5+0.25*\y);
}
\foreach \y in {3,7,11}
{
\draw [fill=lightgrey] (-1.25+0.25*3, -1.75+0.25*\y) rectangle (-1.25+0.25*2, -1.5+0.25*\y);
}
\foreach \y in {2,6,10}
{
\draw [fill=lightgrey] (-1.25+0.25*3, -1.75+0.25*\y) rectangle (-1.25+0.25*4, -1.5+0.25*\y);
}
\foreach \y in {4,8,12}
{
\draw [fill=lightgrey] (-1.25+0.25*4, -1.75+0.25*\y) rectangle (-1.25+0.25*5, -1.5+0.25*\y);
}

\end{tikzpicture}
}
\subfloat[Comb-6]{
\begin{tikzpicture}

\foreach \x in {-7,-6,-5,-4,-3,-2,-1,0,1,...,6}
{
\foreach \y in {-6,-6,-5,-4,-3,-2,-1,0,1,...,5}
{
    \draw [lightgrey] (0+0.25*\x, 0+0.25*\y) rectangle (0.25+0.25*\x, 0.25+0.25*\y);
}
}
\foreach \x in {1,3,...,13}
{
\node [rotate=0] at(-1.62+0.25*\x,-1.7) {\scriptsize \x};
}
\foreach \y in {2,4,...,12}
{
\node [rotate=90] at(-2,-1.64+0.25*\y) {\scriptsize \y};
}
\node [rotate=0] at(0,-2.1) {\small OFDM symbol};
\node [rotate=90] at(-2.3,0) {\small Sub-carrier };

\foreach \y in {1,7}
{
\draw [fill=lightgrey] (-1.25+0.25*2, -1.75+0.25*\y) rectangle (-1.25+0.25*1, -1.5+0.25*\y);
}
\foreach \y in {4,10}
{
\draw [fill=lightgrey] (-1.25+0.25*3, -1.75+0.25*\y) rectangle (-1.25+0.25*2, -1.5+0.25*\y);
}
\foreach \y in {2,8}
{
\draw [fill=lightgrey] (-1.25+0.25*4, -1.75+0.25*\y) rectangle (-1.25+0.25*3, -1.5+0.25*\y);
}
\foreach \y in {5,11}
{
\draw [fill=lightgrey] (-1.25+0.25*5, -1.75+0.25*\y) rectangle (-1.25+0.25*4, -1.5+0.25*\y);
}
\foreach \y in {3,9}
{
\draw [fill=lightgrey] (-1.25+0.25*6, -1.75+0.25*\y) rectangle (-1.25+0.25*5, -1.5+0.25*\y);
}
\foreach \y in {6,12}
{
\draw [fill=lightgrey] (-1.25+0.25*7, -1.75+0.25*\y) rectangle (-1.25+0.25*6, -1.5+0.25*\y);
}
\end{tikzpicture}
}
\subfloat[Comb-12]{
\begin{tikzpicture}

\foreach \x in {-7,-6,-5,-4,-3,-2,-1,0,1,...,6}
{
\foreach \y in {-6,-6,-5,-4,-3,-2,-1,0,1,...,5}
{
    \draw [lightgrey] (0+0.25*\x, 0+0.25*\y) rectangle (0.25+0.25*\x, 0.25+0.25*\y);
}
}
\foreach \x in {1,3,...,13}
{
\node [rotate=0] at(-1.62+0.25*\x,-1.7) {\scriptsize \x};
}
\foreach \y in {2,4,...,12}
{
\node [rotate=90] at(-2,-1.64+0.25*\y) {\scriptsize \y};
}
\node [rotate=0] at(0,-2.1) {\small OFDM symbol};
\node [rotate=90] at(-2.3,0) {\small Sub-carrier };

\draw [fill=lightgrey] (-1.25+0.25*-1, -1.75+0.25*1) rectangle (-1.25+0.25*0, -1.5+0.25*1);
\draw [fill=lightgrey] (-1.25+0.25*1, -1.75+0.25*7) rectangle (-1.25+0.25*0, -1.5+0.25*7);
\draw [fill=lightgrey] (-1.25+0.25*2, -1.75+0.25*4) rectangle (-1.25+0.25*1, -1.5+0.25*4);
\draw [fill=lightgrey] (-1.25+0.25*3, -1.75+0.25*10) rectangle (-1.25+0.25*2, -1.5+0.25*10);
\draw [fill=lightgrey] (-1.25+0.25*4, -1.75+0.25*2) rectangle (-1.25+0.25*3, -1.5+0.25*2);
\draw [fill=lightgrey] (-1.25+0.25*5, -1.75+0.25*8) rectangle (-1.25+0.25*4, -1.5+0.25*8);
\draw [fill=lightgrey] (-1.25+0.25*6, -1.75+0.25*5) rectangle (-1.25+0.25*5, -1.5+0.25*5);
\draw [fill=lightgrey] (-1.25+0.25*7, -1.75+0.25*11) rectangle (-1.25+0.25*6, -1.5+0.25*11);
\draw [fill=lightgrey] (-1.25+0.25*8, -1.75+0.25*3) rectangle (-1.25+0.25*7, -1.5+0.25*3);
\draw [fill=lightgrey] (-1.25+0.25*9, -1.75+0.25*9) rectangle (-1.25+0.25*8, -1.5+0.25*9);
\draw [fill=lightgrey] (-1.25+0.25*10, -1.75+0.25*6) rectangle (-1.25+0.25*9, -1.5+0.25*6);
\draw [fill=lightgrey] (-1.25+0.25*11, -1.75+0.25*12) rectangle (-1.25+0.25*10, -1.5+0.25*12);
\end{tikzpicture}
}
\caption{Representation of four different PRS time/frequency comb patterns, as described by 3GPP Rel-16 \cite{ts138213, fischer}.}
    \label{fig:prscomb}
\end{figure*}
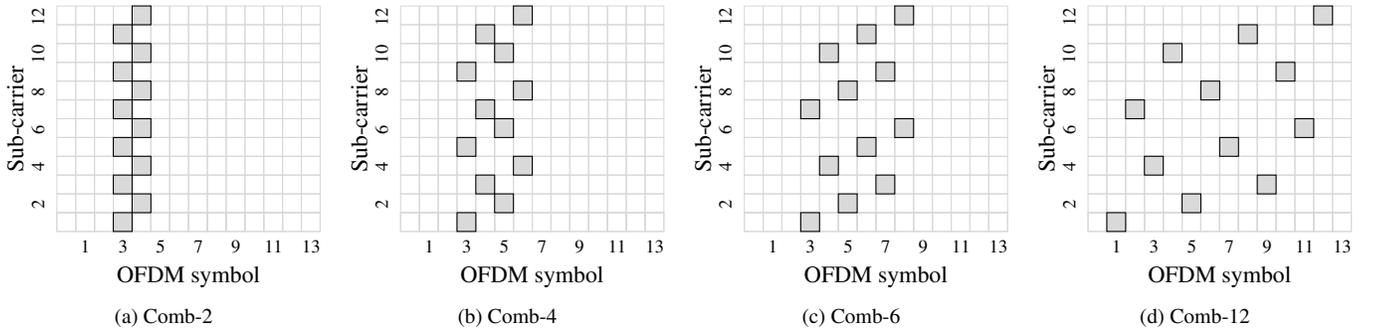

\subsubsection{PRS}
\ac{PRS}, also known as \ac{DL}-\ac{PRS}, is similar to the homonym \ac{LTE} \ac{DL} signal and it is specifically designed to allow the \ac{UE} receiving signals from multiple \acp{BS}. 
A key feature of \acp{PRS} is the improved hearability thanks to the muting concept: multiple \acp{BS} can transmit the \ac{PRS} in a coordinated way by literally muting less relevant \ac{PRS} transmissions to avoid interferences. Furthermore, the staggered pattern of the \ac{PRS} \acp{RE} results into better correlation properties that facilitate the peak detection.
The so-called \textit{comb pattern} structures are shown in Fig.~\ref{fig:prscomb}. With a comb-$N$ pattern ($N \in \{2, 4, 6, 12\}$), $N$ different \acp{TRP} can be frequency multiplexed within the same time slot, assigning different frequency offsets.
Different combinations are possible, assigning a comb size and the number of \ac{OFDM} symbols. Table~\ref{tab:prsoffset} reports the \ac{RE} offsets in the frequency domain given all the combination pairs formed by the comb size ($K_{\text{size}}$) and the number of symbols ($N_{\text{symb}}^{\text{slot}} \in \{2, 4, 6, 12\}$).
Each \ac{PRS} can be further customized by assigning different periodicity ($T^{\text{PRS}}_{\text{per}}$), slot offset ($T^{\text{PRS}}_{\text{offset}}$), \ac{RB} offset ($T^{\text{PRS}}_{\text{offset, RB}}$), and \ac{RE} offset ($T^{\text{PRS}}_{\text{offset, RE}}$) values to fulfill different service requirements (e.g., latency-sensitive applications should opt for frequent \ac{PRS} transmissions, while energy-saving devices would require a low periodicity) and deal with multiple \acp{PRS}. According to \ac{3GPP} \ac{TS} 28.211~\cite[Section 7.4.1.7.4]{ts138211}, $T^{\text{PRS}}_{\text{per}} \in 2^\mu \cdot \{4, 5, 8, 10, 16, 20, 32, 40, 64, 80, 160, 320, 640, 1280, 2560, \linebreak 5120, 10240 \}$ slots and $T^{\text{PRS}}_{\text{off}} \in \{ 0, 1, \dots, T^{\text{PRS}}_{\text{per}}-1 \}$ slots.

\subsubsection{SRS}

\ac{SRS}, often referred to as \ac{UL}-\ac{SRS} to differentiate it with respect to the Rel-15 version, is the \ac{UL} equivalent of \ac{PRS} and it is updated in Rel-16 for positioning purposes. Similar to its \ac{DL} counterpart, the \acp{RE} are arranged in a comb pattern. The comb size set $K_{\text{size}}^{}=\{2, 4\}$ of Rel-15 is extended in Rel-16 to $\{2, 4, 8\}$, while the number of symbols consecutively available are $N^{\text{slot}}_{\text{symb}}=\{1, 2, 4, 8, 12\}$, in contrast to the precedent version which disposed only of $\{1, 2, 4\}$ within the last six symbols of a slot.  All the available combinations with the number of symbols are listed in Table~\ref{tab:srsoffset}. Since \ac{SRS} derives from the same-named signal of Rel-15, it inherits some parameters, such as resource type and periodicity.
The \ac{SRS} resource type can be periodic, semi-persistent, and aperiodic. The periodicity $T^{\text{SRS}}_{\text{per}}$ is available for semi-persistent and periodic \ac{SRS}. In addition to the periodicities $T^{\text{SRS}}_{\text{per}} \in \{1, 2, 4, 5, 8, 10, 16, 20, 32, 40, 64, 80, 160, 320, 640, 1280, 2560\}$ slots available in Rel-15, Rel-16 SRS can also handle $T^{\text{SRS}}_{\text{per}} \in \{5120, 10240, 20480, 40960, 81920\}$~slots. $T^{\text{SRS}}_{\text{per}}=20480$~slots is applicable for $\Delta f=\{30, 60, 120\}$~kHz only; $T^{\text{SRS}}_{\text{per}}=40960$~slots is applicable for $\Delta f=\{60, 120\}$~kHz only; and $T^{\text{SRS}}_{\text{per}}=81920$~slots is exclusive for $\Delta f=120$~kHz.

Rel-16 \ac{SRS} also inherits the bandwidth configuration parameters $B_{\text{SRS}}^{}$, and $C_{\text{SRS}}^{}$, where $B_{\text{SRS}}^{} \in \{0,1,2,3\}$ is the column index of the higher-layer parameter of the frequency hopping (\ac{3GPP} \ac{TS} 38.211~\cite[Table 6.4.1.4.3-1]{ts138211}) if configured, otherwise $B_{\text{SRS}}^{}=0$. The row of the table is selected according to the index  $C_{\text{SRS}}^{} \in \{0, \dots, 63\}$. These values control the bandwidth allocated to the \ac{SRS}. The number of \acp{RB} is given by the specific value denoted as $m_\text{SRS}$ in the table mentioned above.
The frequency hopping of \ac{SRS} is configured by the parameter $b_{\text{hop}}^{} \in \{0,1,2,3\}$. With $b_\text{hop}^{} \geq B_{\text{SRS}}^{}=0$, the frequency hopping is disabled. In Rel-16, frequency hopping is not supported; however, part of its parameters are bandwidth indications, which are still applicable. At last, $n_{\text{RRC}}^{} \in \{0, \dots, 67\}$ is an additional circular frequency-domain offset of \ac{SRS}, as a multiple of 4 \acp{RB}. These properties determine the actual frequency-domain location of the \ac{SRS}.

\subsection {5G positioning methods}
\label{sec:5G positioning methods}
In this section, we detail the main \ac{5G}  positioning methods relying on the delay and angular measurements described in Section~\ref{sec:System Model}. In particular, the outlined methods are: \ac{DL}-\ac{TDOA}, \ac{DL}-\ac{AOD}, \ac{UL}-\ac{AOA} and multi-\ac{RTT}.

\subsubsection{DL-TDOA}
\label{sec:dltdoa}
\ac{DL}-\ac{TDOA} is similar to \ac{OTDOA} in LTE, as they are both based on \ac{TOA} measurements of \ac{DL} signals from multiple \acp{TRP}. The \ac{TDOA} is computed as the difference between two \ac{TOA} measurements.
Considering two \acp{BS} $i$ and $i'$, with $i$ being the reference \ac{BS}, the following three quantities are associated to the \ac{DL}-\ac{TDOA}:
\begin{itemize}
    \item \ac{RSTD}: $t_{\text{rx},i'}^{}-t_{\text{rx},i}^{}$, where $t_{\text{rx},i}^{}$ and $t_{\text{rx},i'}^{}$ are  the reception time instants  of  signals from \acp{BS} $i$ and $i'$, respectively. The \ac{RSTD} defines the time interval observed by the UE between the reception of \ac{DL} reference signals from two different \acp{BS};
    \item \ac{RTD}: $t_{\text{tx},i'}^{}-t_{\text{tx},i}^{}$, where $t_{\text{tx},i}^{}$ and $t_{\text{tx},i'}^{}$ are the transmit time instants of signal from \ac{BS} $i$ and $i'$, respectively. The \ac{RTD} denotes the synchronization between two \acp{BS}, i.e., if two \acp{RTD} are perfectly synchronized, the \ac{RTD} is 0;
    \item \ac{GTD}: $({d_{i'}^{}-d_i^{}})\cdot{c^{-1}}$, where $d_i^{}$ and $d_{i'}^{}$ are respectively the lengths of the propagation paths between the UE and the \acp{BS} $i$ and $i'$, respectively. It represents the ideal hyperbolic line of position.
\end{itemize}
In a noiseless scenario, the following relationship holds~\cite{fischer}:
\begin{equation}
    \text{GTD} = \text{RSTD} - \text{RTD}.
\end{equation}
In simulation analyses, perfect synchronization between \acp{BS} is typically assumed, i.e.,  all \acp{BS} transmit exactly in the allocated time slots, and no clock offset contributes to the measurement error.
On  the other hand, in real operating conditions with the currently deployed \ac{5G}  network, synchronization errors leads to major bias in ranging measurements, up to hundreds of meters~\cite{Koivisto_j17,sci_rep_5G}. This is a primary limitation of \ac{5G}  precise positioning at present (more details are provided in Section~\ref{sec:limitations}). As a matter of fact, current \ac{5G}  networks implement a master-and-slave-based \ac{PTP}~\cite{4579760} protocol which only achieves a synchronization that is accurate up to $\pm1.5$ \textmu s, as recommended by the \ac{ITU}~\cite{20178271}. This converts to a distance error of about $\pm450$~m, hugely limiting the positioning performance.

\subsubsection{DL-AOD}
\label{sec:dlaod}
\ac{DL}-{AOD} positioning can be obtained thanks to the computation of \ac{DL} \ac{RSRP} measurements of beams by the \ac{UE}. The \acp{BS} may transmit signals in a beam-sweeping manner that can be measured by the \ac{UE}. The more the beam is directed to the \ac{UE} and not impaired by obstacles, the higher the \ac{RSRP}. The resulting vector of all \ac{RSRP} measurements (one for each beam)  could be considered as a  \ac{RF} fingerprint and used to perform positioning by a pattern-matching  approach~\cite{rascoskoi2018}. 

Another solution, which is also the one adopted in this work, is the \textit{beam management} procedure~\cite{GioPolZor:J19}.
This procedure is used to acquire and maintain a link pair between the UE and a \ac{BS}.
\ac{3GPP} \ac{TR} 38.802~\cite[section 6.1.6.1]{tr138802}, defines the beam management as the combination of the following three procedures:
\begin{itemize}
    \item[P1)] This procedure focuses on the initial acquisition based on \ac{SSB} and it employs  analog beamforming. During the initial acquisition, beam sweeping takes place at both transmit and receive ends to select the best beam pair based on the \ac{RSRP} measurement. In general, the selected beams are wide and may not be optimally paired for data transmission and reception.

    \item[P2)] This procedure, which is referred to as \textit{beam refinement}, focuses on transmit-end beam refinement, where beam sweeping is performed at the transmit side while keeping the receive beam fixed. The procedure is based on \ac{NZP-CSI-RS} for \ac{DL} transmit-end beam refinement and \ac{SRS}  for UL transmit-end beam refinement. P2 makes use of digital beamforming.
    
    \item[P3)] This procedure focuses on receive-end beam adjustment, where the beam sweeping happens at the receiving end given the current transmit beam. This process aims to find the best receive beam. For this procedure, a set of reference signal resources are transmitted with the same transmit beam, and the UE or \ac{BS} receives the signal using different beams from different directions covering an angular range. Finally, the best receive beam is selected based on the RSRP measurements on all receive beams.
\end{itemize}

The technical report defining beam management refers to Rel-14, where \ac{NZP-CSI-RS} is mentioned for the P2 procedure in \ac{DL}. However, in Rel-16, \ac{NZP-CSI-RS} is no longer used for positioning purposes.
In the analyses and results presented in this tutorial, we consider the P2 procedure in \ac{DL} based on \ac{PRS}.
Moreover, we are interested only in the first two phases of the procedure to obtain the \ac{AOD}. The P3 procedure could be used for \ac{AOA} estimation only in the case of a large antenna array available to the UE side. However, most likely scenarios include a UE device with one or very few antennas due to size, battery, and weight constraints (e.g., a smartphone). For this reason, estimating the \ac{AOA} at the UE side is very challenging at present.

After the initial beam establishment, obtaining a unicast data transmission with high directivity requires a beam much finer than the SSB beam. Therefore, a set of \ac{PRS} resources are configured and transmitted over different directions by using finer beams within the angular range of the beam from the initial acquisition process. Then, the UE measures all these beams by capturing the signals with a fixed receive beam. The best transmit beam is selected using \ac{PRS}-\ac{RSRP} measurements (defined in \ac{3GPP} \ac{TS} 38.215~\cite[Section 5.1.28]{ts138215}) on all the transmit beams, which allow to determine the best \ac{AOD}. Lastly, the \ac{AOA}  measurements needed for positioning with \ac{NLS} are derived from the \acp{AOD}.
\begin{figure}[t]
    \centering
    \begin{tikzpicture}
        \node[]at(0,0){\includegraphics[width=\linewidth]{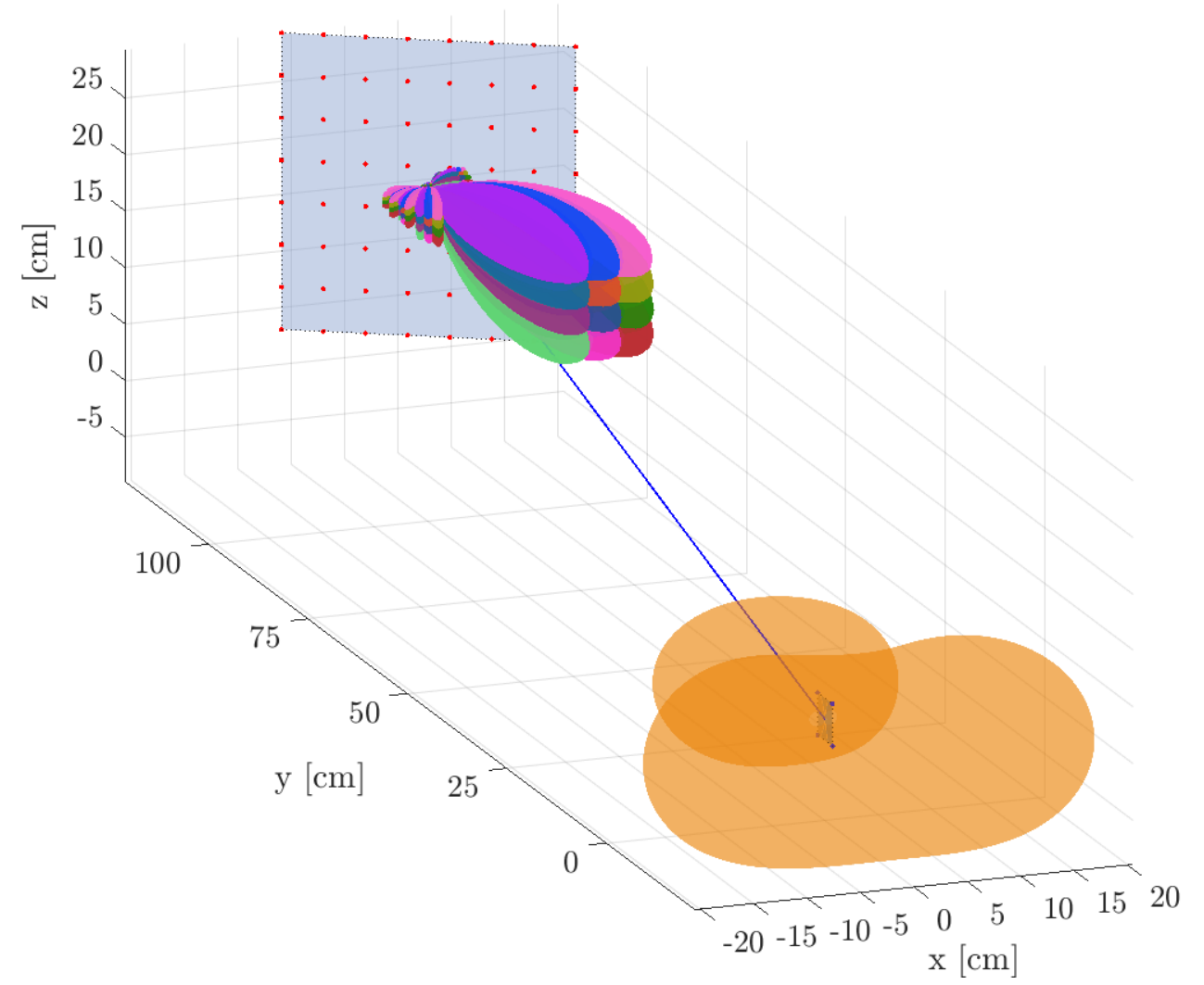}};
        \node[]at(-1.2,3.6){\small Tx Site};
        \node[]at(2.7,-0.8){\small Rx Site};
    \end{tikzpicture}
    
    \caption{Beam refinement phase within the beam management procedure for DL-AOD estimation with $N^\text{PRS}=12$. In the example, 12 different Tx spatial PRS beams are formed over different angles in a confined angular domain. The beam with the highest RSRP is chosen by the \ac{UE}. The blue line indicates the direct path; the best beam is the light green one. 
    }
    
    \label{fig:dlaod}
\end{figure}
\input{Figure_tex/PRS_resource_set}
Fig.~\ref{fig:dlaod} illustrates the beam refinement with an example. The orange beam is selected during P1 at the UE end, while all the colored beams refer to the \ac{PRS}  resources sent in \ac{DL} by the \ac{BS}. The straight blue line identifies the direct path that links UE and \acp{BS}, and it shows clearly that the \ac{PRS}  with the highest \ac{RSRP} will be the one with index 1 (light green) because is the one with more directivity to the UE.

The number of finer beams depends on the number of \ac{PRS} resources employed. Since in our work, all the \acp{PRS} are delivered in a single slot, the maximum number of beams is 12. In Fig.~\ref{fig:p2prs}, we show an RB with the set of \ac{PRS} in use, which is an example of comb 12 with 12 \ac{OFDM} symbols and 12 resources.
A critical aspect of beam selection is related to the duration of the beam searching procedure, which reduces the data rate of the link, especially if exhaustive searches are carried out. For this reason, literature works have proposed to speed up the searching procedure by exploring in-band signalling~\cite{garcia_transmitter_2018, sim_deep_2020} or the repeatability of the wireless environment to learn the 
geo-referenced optimal beams~\cite{Mizmizi2021,Brambilla_beam_align_2020}.

\subsubsection{UL-AOA}
\label{sec:ulaoa}
\ac{UL}-\ac{AOA} is a network-based positioning method where the \ac{BS} exploits the signals transmitted by the UE, i.e., the \ac{SRS}, to determine the \ac{AOA} both in zenith and azimuth directions. As for the \ac{DL}-{AOD}, a directional antenna is required to calculate the \ac{AOA}. This is somehow a usual assumption given that \ac{5G}  \ac{NR} supports multi-antenna transmission and reception. According to the standard, there are several methods for determining the \ac{AOA}.

Classical \ac{AOA}  estimation is performed with conventional beamforming, as described by procedure P3 in Section~\ref{sec:dlaod}. These methods do not make any assumptions about how the incoming signal and noise should be modeled. They require electrically pointing beams in every direction (or a predetermined selection of directions) and looking for power output peaks. The beamforming is achieved by applying a Fourier-based spectrum analysis to the spatio-temporal received samples. However, with these methods, the beamwidth of the array limits the angular resolution, necessitating a large number of antenna components to attain high precision.

Other more advanced techniques are high-resolution subspace-based methods like \ac{MUSIC}~\cite{1143830} and \ac{ESPRIT}~\cite{32276}. This family of methods is better suited for lower frequencies, i.e., \ac{FR}1, where digital beamformers are more widely accessible. They process the eigenstructure of the incident signal by computing spatial covariance matrices using digital samples from each antenna element output. Due to the array aperture's modest size at lower frequencies, the spatial resolution is only moderate, i.e., beams are relatively broad. As a result, contrary to conventional beamforming, high-resolution approaches are particularly useful at lower frequencies because they may reduce the angular resolution to values smaller than the array's beamwidth without requiring the array aperture to be expanded.
With the former technique, we are able to extract the \ac{AOA} measurement, i.e., the angle between the \ac{UE} and a \ac{BS}, while with the latter type of technique, we analyze the received signal. 

\subsubsection{Multi-RTT}
\label{sec:mrtt}
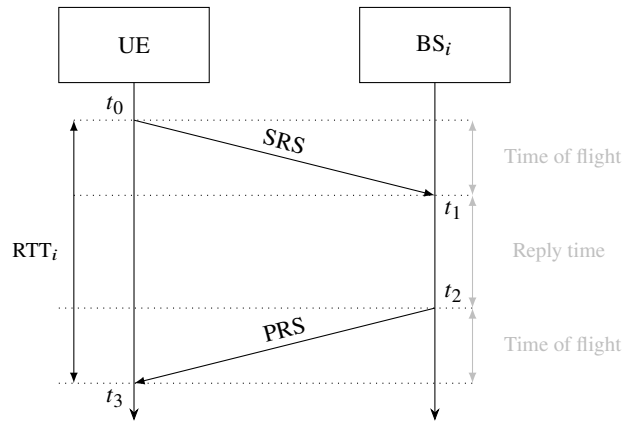
\begin{figure}[t]
    \centering

\begin{tikzpicture}
        \tikzstyle{every node}=[font=\footnotesize]

         \draw[-{Stealth}] (-2,0) -- (-2,-5); 
         \draw[-{Stealth}] (2,0) -- (2,-5); 

         \draw[-{latex}] (-2,-1) -- (2,-2); 
         \draw[-{latex}] (2,-3.5) -- (-2,-4.5); 

         \draw[dotted] (-2.8,-1) -- (2.5,-1); 
         \draw[dotted] (-2.8,-2) -- (2.5,-2); 
         \draw[dotted] (-3,-3.5) -- (2.5,-3.5); 
         \draw[dotted] (-3,-4.5) -- (2.5,-4.5); 
         
        \draw[{latex}-{latex}] (-2.8,-4.5) -- (-2.8,-1)  node [midway, xshift = -5mm] {$\text{RTT}_i$}; 
        \draw[lightgray,{latex}-{latex}] (2.5,-2) -- (2.5,-1) node [midway, xshift = 12mm, text=lightgray] {Time of flight}; 
        \draw[lightgray,{latex}-{latex}] (2.5,-2) -- (2.5,-3.5) node [midway, xshift = 11.5mm, text=lightgray] {Reply time}; 
        \draw[lightgray,{latex}-{latex}] (2.5,-3.5) -- (2.5,-4.5) node [midway, xshift = 12mm, text=lightgray] {Time of flight};

         \draw (-2,0) node [rectangle, minimum width=2cm, minimum height=1cm, draw, fill=white] {\small \shortstack{UE}};
         \draw (2,0) node [rectangle, minimum width=2cm, minimum height=1cm, draw, fill=white] {\small \shortstack{\ac{BS}$_i$}};

         \draw (-2.25,-0.8) node [] {\small $t_0$};
         \draw (2.25,-2.2) node [] {\small $t_1$};
         \draw (2.25,-3.3) node [] {\small $t_2$};
         \draw (-2.25,-4.7) node [] {\small $t_3$};

         \draw (0,-3.78) node [rotate=15] {\small PRS};
        \draw (0,-1.28) node [rotate=-15] {\small SRS};
    \end{tikzpicture}
    
    \caption{TOF estimation via multi-RTT procedure in 5G using UL and DL measurements. The procedure starts with the UE sending an SRS to the BS, which responds with a PRS. The overall RTT is computed at UE side, knowing the reply time of BS.}
    \label{fig:mrttproc}
\end{figure}

\ac{DL}-\ac{TDOA} requires precise synchronization among the \acp{BS}, which is not obvious in a real scenario. \ac{RTT} does not require any synchronizations, even if a coarse time synchronization is desirable to increase hearability from multiple \acp{BS}. The synchronization accuracy needed for \ac{TDOA} is in nanoseconds, while for \ac{RTT}, it is enough to be in microseconds~\cite{fischer}. For this reason, an \ac{RTT}  measurement would be a more suitable choice for the currently deployed networks. Similar to \ac{TDOA}, the basic measurement is \ac{TOA}, one in UL based on \ac{SRS}  and one in \ac{DL}  based on \ac{PRS}, as shown in Fig.~\ref{fig:mrttproc}. The two-time differences used to compute the \ac{RTT} value are referred to as the same clock: $t_3-t_0$ is referred to as the UE clock, while $t_2-t_1$ is referred to as the \ac{BS} one. Thanks to this, synchronization is not needed anymore. However, in multi-\ac{RTT}, several \acp{BS} are involved simultaneously, and, with a microsecond level synchronization, it is possible to send back the signals in different time slots or in the same time slot with different frequency offsets. 
With a static \ac{UE}, it is possible to send the signal of each \ac{BS} in different time slots. In the case of mobile positioning, this choice would lead to higher measurement errors.
Generally, all the measurements need to be concurrently made to mitigate the errors. 

\subsection {Extraction of 5G positioning measurements}
\label{sec:5G measurement extraction}




In this section, we provide examples of how it is possible to address \ac{NLOS} detection (Section~\ref{sec:NLOS_detection}), and we describe the selected procedure used to extract the positioning measurements from the \ac{5G} signals, considering both \ac{DL} (Section~\ref{sec:meas_extraction_DL}) and \ac{UL} (Section~\ref{sec:meas_extraction_UL}).

\subsubsection{NLOS detection}
\label{sec:NLOS_detection}
The identification of \ac{NLOS} propagation condition refers to algorithms able to detect whether the radio signal has been received from reflected paths rather than the direct link.
This allows to properly account for possible excess delay in ranging measurement at the tracking algorithm, thus improving the final localization performance~\cite{Cao_j21,Nicoli_j08,Morelli_j07}. \textcolor{black}{Furthermore, knowing the environmental map of the area (also known as \ac{HD} map) could give insights about possible reflectors or virtual anchors; such a-priori information can be exploited together with multipath and \ac{NLOS} measurements into positioning algorithms~\cite{HDmap}. The same concept can be applied to \ac{RIS}, where the reflectors are actively identified and exploited for positioning and velocity estimation~\cite{RIS_NLOS}.
Rel-17 includes the capability to indicate whether the received signal is received over a direct or reflected path. However, the standard currently lacks detailed technical specifications regarding its implementation.}

\textcolor{black}{In the literature, several \ac{NLOS} detection and mitigation techniques have been developed in the past. We here report some of them, including statistical methods and \ac{ML} solutions~\cite{Díez23}. 
The oldest prior art is well-summarized in~\cite{GuvCho2009}, which includes relaxed constrained localization, identify and discard, and weighted \ac{LS}-based techniques. 
\textcolor{black}{Constrained localization is based on quadratic programming techniques, where the constraints can be relaxed to include \ac{NLOS} measurements. Identification and discard consists of considering sub-groups of \acp{BS} to discern the \ac{LOS} and \ac{NLOS} ones. Lastly, from \ac{LS}-based techniques, the residual error in output from the algorithm can be used to detect \ac{NLOS} measurements.}
Regarding more recent works, instead, the authors in~\cite{Marano_j10} designed non-parametric techniques utilizing \ac{LS}-\ac{SVM} to discriminate \ac{LOS} from \ac{NLOS} conditions \textcolor{black}{(classification)} and mitigate the biases of \ac{NLOS} range estimates \textcolor{black}{(regression)}. \textcolor{black}{The selected features are mainly the power and the maximum amplitude of the received signal and the mean excess delay. Different mitigation strategies are proposed based on \acp{BS} \ac{NLOS} probability and the number of \acp{BS} in \ac{LOS}, outperforming previous state-of-the-art techniques.}
In~\cite{changhui20}, \ac{DNN} methods were employed, combining \ac{CNN} and \ac{LSTM} networks to solve the classification problem. \textcolor{black}{The results demonstrate a classification accuracy above 80\%.}
In~\cite{Barbieri_j21}, a Bayesian filter that jointly tracks the time-varying visibility conditions and the  \ac{UE} motion has been proposed, and it is demonstrated to efficiently handle \ac{NLOS} in harsh industrial environments \textcolor{black}{with an accuracy of $\approx$50~cm in 95\% of the cases.}
In~\cite{Linsalata22}, the environmental conditions are predicted by exploiting the information of vehicle onboard sensors; the so-called dynamic \ac{LOS}-map is used to improve the \ac{V2X} performance by selecting optimal relays. 
In~\cite{tedeschini_latent_2023}, a semi-supervised anomaly detection technique was used to identify \ac{NLOS} conditions by means of an \ac{AE} structure applied to the full \ac{CIR}.
A neural-enhanced sum-product algorithm using an ad-hoc factor graph has been designed in~\cite{Witrisal2024}, \textcolor{black}{employing a channel estimation and detection algorithm for the measurements and an \ac{AE} for features extraction. The method therein demonstrates highly robust positioning and tracking capability while attaining the \ac{PCRB} even when the training data is confined to local regions.}
\textcolor{black}{In~\cite{Poorter2024}, an automatic optimization for transfer learning has been recently proposed for \ac{NLOS} error detection and correction for feature and \ac{CIR} data. With the \ac{CIR}-based approach, the results reveal 93\% of \ac{NLOS} detection capability and positioning accuracy of $\approx$10~cm, unlocking a high-precision positioning for \ac{UWB} systems.}
\textcolor{black}{Lastly, \cite{Qin_NLOS} describes various statistical and optimization techniques for \ac{NLOS} error estimation. While the most promising methods in the literature rely on integrating \ac{RSS} measurements, the authors propose a novel distance-dependent uncertainty model for dynamic \ac{NLOS} environments. This model shows promising results, achieving an error of less than 1~m without requiring prior information.}
}

\begin{figure*}[!ht]
    \centering
    
\begin{tikzpicture}
            \tikzstyle{every node}=[font=\footnotesize]
            \draw (-5,-4) node (A) [rectangle, rounded corners, minimum width=2cm, minimum height=1cm, draw, fill=blue!20, text opacity=1] {\small \shortstack{PRS, PDSCH\\ Generation}};
            \draw (0,-4) node (B1) [rectangle, rounded corners, minimum width=2cm, minimum height=1cm, draw, fill=blue!20, text opacity=1] {\small \shortstack{OFDM \\Modulation}};
            \draw (2.7,-4) node (C1) [rectangle, rounded corners, minimum width=2cm, minimum height=1cm, draw, fill=white!20, text opacity=1] {\small \shortstack{Propagation  \\ Channel}};
            \draw (11,-4) node (D1) [rectangle, rounded corners, minimum width=2cm, minimum height=1cm, draw, fill=orange!20, text opacity=1] {\small \shortstack{Timing \\ Estimation}};

            \draw (-2.5,-2) node (B2) [rectangle, rounded corners, minimum width=2cm, minimum height=1cm, draw, fill=blue!20, text opacity=1] {\small \shortstack{Tx Beam \\Sweeping}};
            \draw (0,-2) node (C2) [rectangle, rounded corners, minimum width=2cm, minimum height=1cm, draw, fill=blue!20, text opacity=1] {\small \shortstack{OFDM \\Modulation}};
            \draw (2.7,-2) node (D2) [rectangle, rounded corners, minimum width=2cm, minimum height=1cm, draw, fill=white!20, text opacity=1] {\small \shortstack{Propagation \\ Channel}};
            \draw (5.5,-2) node (E2) [rectangle, rounded corners, minimum width=2cm, minimum height=1cm, draw, fill=orange!20, text opacity=1] {\small
            \shortstack{OFDM \\Demodulation}};
            \draw (11,-2) node (F2) [rectangle, rounded corners, minimum width=2cm, minimum height=1cm, draw, fill=orange!20, text opacity=1] {\small \shortstack{Angle \\Estimation}};

            \draw (-5,0) node (A3) [rectangle, rounded corners, minimum width=2cm, minimum height=1cm, draw, fill=blue!20, text opacity=1] {\small \shortstack{SSB, DMRS\\ Generation}};
            \draw (-2.5,0) node (B3) [rectangle, rounded corners, minimum width=2cm, minimum height=1cm, draw, fill=blue!20, text opacity=1] {\small \shortstack{Tx Beam \\ Sweeping}};
            \draw (0,0) node (C3) [rectangle, rounded corners, minimum width=2cm, minimum height=1cm, draw, fill=blue!20, text opacity=1] {\small \shortstack{OFDM \\Modulation}};
            \draw (2.7,0) node (D3) [rectangle, rounded corners, minimum width=2cm, minimum height=1cm, draw, fill=white!20, text opacity=1] {\small \shortstack{Propagation \\ Channel}};
            \draw (5.5,0) node (E3) [rectangle, rounded corners, minimum width=2cm, minimum height=1cm, draw, fill=orange!20, text opacity=1] {\small \shortstack{OFDM \\Demodulation}};
            \draw (8.3,0) node (F3) [rectangle, rounded corners, minimum width=2cm, minimum height=1cm, draw, fill=orange!20, text opacity=1] {\small \shortstack{Rx Beam \\Sweeping}};
            \draw (11,0) node (G3) [rectangle, rounded corners, minimum width=2cm, minimum height=1cm, draw, fill=orange!20, text opacity=1] {\small \shortstack{Beam Pair \\Determination}};

            \draw[-{Stealth}] (A.east) -- (B1);
            \draw[-{Stealth}] (B1) -- (C1);
            \draw[-{Stealth}] (C1) -- (D1);

            \draw[-{Stealth}] (A) -- (-5,-2) -- (B2);
            \draw[-{Stealth}] (B2) -- (C2);
            \draw[-{Stealth}] (C2) -- (D2);
            \draw[-{Stealth}] (D2) -- (E2);
            \draw[-{Stealth}] (E2) -- (F2);
            
            \draw[-{Stealth}] (A3) -- (B3);
            \draw[-{Stealth}] (B3) -- (C3);
            \draw[-{Stealth}] (C3) -- (D3);
            \draw[-{Stealth}] (D3) -- (E3);
            \draw[-{Stealth}] (E3) -- (F3);
            \draw[-{Stealth}] (F3) -- (G3);
            
        \end{tikzpicture}
        \caption{DL block diagram for location measurement extraction. Top row represents the beam pair selection in DL-AOD estimation; whereas the bottom one reports the angle refinement and TOF extraction. BS, propagation channel and UE are indicated with blue, white, and orange colors, respectively. The OFDM Demodulation block includes the channel estimation.}
    \label{fig:blockdiagram_1}
\end{figure*}
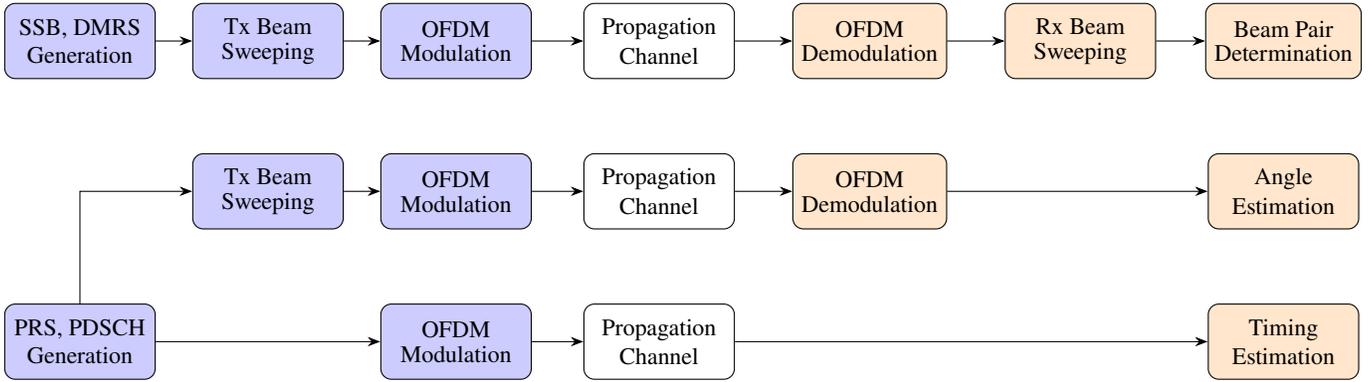
\begin{figure*}[!ht]
    \centering
    
       \begin{tikzpicture}
            \tikzstyle{every node}=[font=\footnotesize]
            \draw (-5,-1) node (A) [rectangle, rounded corners, minimum width=2cm, minimum height=1cm, draw, fill=orange!20, text opacity=1] {\small
            \shortstack{SRS, PUSCH\\ Generation}};
            \draw (-2,-1) node (B) [rectangle, rounded corners, minimum width=2cm, minimum height=1cm, draw, fill=orange!20, text opacity=1] {\small \shortstack{OFDM \\Modulation}};
            \draw (1.5,-1) node (C) [rectangle, rounded corners, minimum width=2cm, minimum height=1cm, draw, fill=white!20, text opacity=1] {\small
            \shortstack{Propagation Channel}};
            \draw (10,0) node (D1) [rectangle, rounded corners, minimum width=2cm, minimum height=1cm, draw, fill=blue!20, text opacity=1] {\small
            \shortstack{Timing Estimation}};

            \draw (6,-2) node (D2) [rectangle, rounded corners, minimum width=2cm, minimum height=1cm, draw, fill=blue!20, text opacity=1] {\small
            \shortstack{OFDM \\Demodulation}};
            \draw (10,-2) node (E2) [rectangle, rounded corners, minimum width=2cm, minimum height=1cm, draw, fill=blue!20, text opacity=1] {\small
            \shortstack{Angle Estimation}};

            \draw[-{Stealth}] (A.east) -- (B.west);
            \draw[-{Stealth}] (B) -- (C);
            \draw[-{Stealth}] (C.east) -- (4,-1) -- (4,0) -- (D1.west);

            \draw[-{Stealth}] (C.east) -- (4,-1) -- (4,-2) -- (D2.west);
            \draw[-{Stealth}] (D2) -- (E2);

        \end{tikzpicture}
\caption{UL block diagram for location measurement extraction. BS,  propagation channel, and UE are indicated with blue, white, and orange colors, respectively.} 
    \label{fig:blockdiagram_2}
\end{figure*}
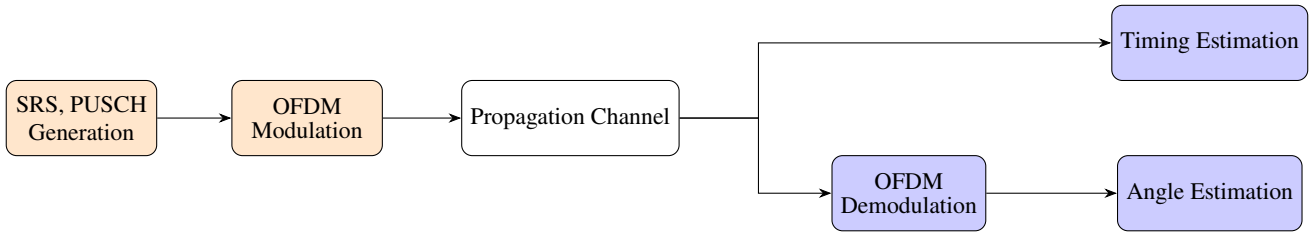

\subsubsection{Downlink}
\label{sec:meas_extraction_DL}

For \ac{DL} positioning, we proceed according to the block diagram illustrated in Fig.~\ref{fig:blockdiagram_1}, where the blocks pertaining to the \ac{BS} are colored in blue, while the \ac{UE} is in orange. Two types of signals are used: \acp{SSB} and \acp{PRS}. 
\acp{SSB} are generated to perform the procedure P1, while \acp{PRS} are used for the procedure P2 (see Section~\ref{sec:dlaod}) and the timing estimation.
After \ac{SSB} and \ac{DMRS} generation, both the \ac{Tx} \ac{BS} and \ac{Rx} \ac{UE} perform beam sweeping over all the configured angular domain. Typical conditions include an omnidirectional \ac{UE} and a tri-sector \ac{BS}, although many other configurations are possible.
Signals are generated according to an \ac{OFDM} modulation, and after channel propagation, they are demodulated, and the channel is estimated. 
The beam determination is then performed at the \ac{Rx} \ac{UE} side by selecting the beam pair with the highest received power.  

\textcolor{black}{Recalling the channel matrix $\mathbfcal{H}_{\tau}^{}$ from~\eqref{eq:channel}, by 
defining the beam codebooks comprising $L$ and $V$ candidate beamforming
vectors at \ac{Rx} and \ac{Tx} sides respectively as 
$\mathcal{W}_{\mathrm{rx}}^{} = \{\boldsymbol{\omega}_{\mathrm{rx}, 1}^{} , \ldots , \boldsymbol{\omega}_{\mathrm{rx}, L}^{} \} \in \mathbb{C}^{N_{\mathrm{rx}} \times L}$
and
$\mathcal{W}_{\mathrm{tx}}^{} = \{\boldsymbol{\omega}_{\mathrm{tx}, 1}^{} , \ldots , \boldsymbol{\omega}_{\mathrm{tx}, V}^{} \} \in \mathbb{C}^{N_{\mathrm{tx}} \times V}$, the selection of the optimal beam pair follows an optimization problem defined as
\begin{align}
    \underset{\ell , \nu}{\mathrm{arg \,max}}^{} &\, 
    \sum_{\tau}
    \left| \boldsymbol{\omega}_{\mathrm{rx}, \ell}^{\mathrm{H}} \, \mathbfcal{H}_{\tau}^{} \, \boldsymbol{\omega}_{\mathrm{tx}, \nu}^{} \right|^2
    \label{eq:beam_pair_selection}
    \\
    \nonumber
    &\text{s.t.} \,\,\boldsymbol{\omega}_{\mathrm{rx}, \ell}^{} \in \mathcal{W}_{\mathrm{rx}}^{}, \, \text{with} \, \ell=1, \ldots, L,
    \\
    \nonumber
     &\text{s.t.} \,\, \boldsymbol{\omega}_{\mathrm{tx}, \nu}^{} \in \mathcal{W}_{\mathrm{tx}}^{}, \, \text{with} \, \nu=1, \ldots, V.
\end{align}
Since each vector in the codebooks corresponds to a specific pair of azimuth and elevation angles, 
the solution to \eqref{eq:beam_pair_selection} determines optimal pair of \ac{AOA} and \ac{AOD}.
}

Up to 8 \acp{SSB} can be transmitted in a frame at FR1, a value that raises to 64 for FR2, and they can be steered across the entire \ac{BS} sector in azimuth ($A_{\phi}^{}$) and elevation ($A_{\psi}^{}$). Given the number of steering vectors in azimuth $N^{\text{SSB}}_\phi$ and the number of steering vectors in elevation $N^{\text{SSB}}_{\psi}$, we can define the \ac{SSB} resolution for azimuth and elevation, respectively, as 
$\phi^{\text{SSB}}_{\text{RES}} = A_{\phi}^{}/N^{\text{SSB}}_{\phi}$ and $\psi^{\text{SSB}}_{\text{RES}} = A_{\psi}^{}/N^{\text{SSB}}_{\psi}$.
Then, \ac{PRS} and \ac{PDSCH} are generated. For the \ac{AOD} estimation, $N^{\text{PRS}}$ narrow beams are shot within the spatial domain selected in the \ac{SSB} reporting (see Fig.~\ref{fig:dlaod}). Since $N^{\text{PRS}}=N^{\text{slot}}_{\text{symb}}$ and they can be steered in azimuth and elevation, we define the number of steering vectors in azimuth as $N^{\text{PRS}}_\phi$ and the number of steering vectors in elevation as $N^{\text{PRS}}_\psi$. Therefore, we can depict the \ac{PRS} resolution for azimuth and elevation respectively as $\phi^{\text{PRS}}_{\text{RES}} = \phi^{\text{SSB}}_{\text{RES}}/N^{\text{PRS}}_\phi$ and $\psi^{\text{PRS}}_{\text{RES}} = \psi^{\text{SSB}}_{\text{RES}}/N^{\text{PRS}}_\psi$. 

Since \ac{OFDM} signals are employed, it is worth delving into a more comprehensive exploration of the techniques for effectively managing them~\cite{Babich_2017, Cerutti_j20, Simeone_2004}. 
Before transmitting the signal across the wireless channel, the discrete signal can be oversampled during the \ac{IFFT} process, followed by the addition of a cyclic prefix.
%
After the propagation, a first coarse synchronization is performed, usually detecting the \ac{PSS} of the \ac{SSB} in the time-domain~\cite{time_synch}. 
Then, before the FFT, the cyclic prefix is removed.
Moreover, in the context of multi-link communications, it becomes essential to differentiate between various \acp{BS} based on their respective Cell-IDs and the corresponding $T^{\text{PRS}}_{\text{offset, RE}}$.
For timing estimation, one \ac{PRS} is modulated, and the \ac{TOA} is estimated at the \ac{UE} side by computing a cross-correlation between the received waveform and the replica of the transmitted waveform at the \ac{Rx}.
\textcolor{black}{Recalling the \ac{Tx} and \ac{Rx} signal $\boldsymbol{y}_t^{} \in \mathbb{C}^{N_{\mathrm{tx}} \times 1}$ and $\boldsymbol{z}_t^{} \in \mathbb{C}^{N_{\mathrm{rx}} \times 1}$ from \eqref{eq:rx_signal}, which are sampled with sample time $T_s^{} = 1 /(\Delta f\cdot N_f)$, we define 
the cross-correlation $r_t^{}$ as
\begin{equation}
    r_t^{} = 
    \sum^{N_{\mathrm{rx}}}_{n_{\mathrm{rx}}=1}
    \sum^{N_{\mathrm{tx}}}_{n_{\mathrm{tx}}=1}
    \sum^{N_{s}-1}_{n_{s}=0}
    z^{}_{n_{\mathrm{rx}},n_{s}}\cdot y^{*}_{n_{\mathrm{tx}},n_{{s}}-t},
\end{equation}
where $N_s^{}$ is the number of samples.}
Then, the highest peak of the cross-correlation can be used to detect the \ac{TOA}, even if the use of advanced techniques for first peak detection is advisable to ensure more accurate results~\cite{giorgetti_time--arrival_2013, kumari_characterization_2015, wang_performance_2020, zehra_proactive_2022}.
This is particularly pertinent in scenarios with significant multipath effects, as the primary peak associated with the first path may be weaker, with the strongest peak potentially originating from a signal reflection. The \ac{TOA} can be later employed for \ac{TDOA} or \ac{RTT} estimate.


\subsubsection{Uplink}
\label{sec:meas_extraction_UL}

For \ac{UL} positioning, we proceed according to the block diagram illustrated in Fig.~\ref{fig:blockdiagram_2}
In \ac{UL} positioning, only \ac{SRS} signals are employed. For both time and angle estimation, the first three steps are the same as for \ac{DL}, i.e., \ac{SRS} and \ac{PUSCH} generation, \ac{OFDM} modulation, and channel propagation. Afterward, \ac{TOA} estimation follows the same rules described in Section~\ref{sec:meas_extraction_DL}. Instead, for angles, we demodulate the signal, and then a high-resolution \ac{MUSIC} algorithm is used (see Section~\ref{sec:ulaoa}).
\ac{MUSIC} algorithm enables an accurate estimate of \ac{AOA} of signals in cases when the \ac{Rx} is equipped with \ac{MIMO} technology. The process of applying the \ac{MUSIC} algorithm in the \ac{UL} scenario can be described as follows.

After \ac{OFDM} demodulation and noise-filtering, the sample covariance matrix of the data is computed. By taking into account the time correlation between different antenna-element readings, the covariance matrix allows for an effective separation between signal and noise. Indeed, subsequently, the covariance matrix is decomposed into its eigenvectors and eigenvalues, where eigenvectors corresponding to the largest eigenvalues form the signal subspace, while those corresponding to smaller eigenvalues form the noise subspace. Lastly, the algorithm searches over a specified grid of \acp{AOA}, identifying the arrival vectors whose projection into the noise subspace is minimal. This information is used to estimate the \ac{AOA}.


\input{Figure_tex/Fig_Ray_Tracing}

\section{Simulation Experiments}
\label{sec:Simulation Analysis}
In this section, we provide a thorough analysis of the performance of \ac{5G}  positioning assessed over multiple scenarios and with different system configurations.
We start by defining the adopted performance metrics in Section~\ref{sec:Performance Metrics}, then we present the simulation environments in Section~\ref{sec:Simulation Environment}, and the system settings in Section~\ref{sec:Simulation Parameters}.
The simulations consider the use of \ac{PRS}, \ac{SRS}, and \ac{SSB} as defined in Section~\ref{sec:5G Signals for Positioning}.
\textcolor{black}{Lastly, numerical results are reported in
Section~\ref{sec:numerical_results}.}

\subsection{Performance metrics}
\label{sec:Performance Metrics}

We analyze the positioning performance in terms of the accuracy of the location estimate, i.e., in terms of the 2D location estimate error  $\Delta \boldsymbol{u} = \hat{\boldsymbol{u}} - \boldsymbol{u}$, whose l2 norm $\Delta{u}=\lVert\Delta \boldsymbol{u} \lVert$ represents the distance between the true and the estimated \ac{UE} locations.  

We consider several accuracy metrics (averaged over the UE positions and Monte Carlo iterations), including the bias vector $\boldsymbol{b}=\text{E}[\Delta\boldsymbol{u}]$, with  $b=\lVert\boldsymbol{b}\lVert$  representing the distance between the mean location fix and the true location, the \ac{RMSE}  (also known as root mean square distance) defined as $\text{RMSE}=\sqrt{\text{E}[\Delta{u}^2]}$, and the \ac{MAE} defined as $\text{MAE}=\text{E}[\Delta{u}]$ (mean distance between the location fix and the true location). In addition to the mentioned average metrics, we also consider the \ac{CDF} and the \ac{PDF} of $\Delta{u}$.
\textcolor{black}{We also report the \ac{PEB} value computed from the \ac{CRB}, recalling that 
$
    \text{RMSE} \geq 
    \sqrt{ \text{tr} ( \boldsymbol{J}(\boldsymbol{u})^{-1} )}
$
and
$
    \text{PEB} = 
    \sqrt{ \text{tr} ( [ \boldsymbol{J}(\boldsymbol{u})^{-1} ]^{}_{1:2, 1:2})} 
$
\cite{Shahmansoori_j19}.}

The location accuracy is known to depend on two main factors: the statistics of the measurement errors  $n_{i}$ in \eqref{eq:measurement_single}
and the geometric arrangement of the \acp{BS} with respect to the \ac{UE}, referred to as geometric factor \cite{kaplan_understanding_2006}. In our analyses, we investigate both of them by analyzing the measurement statistics 
and the variation of the location error ellipse over the space.

\begin{figure}[!tb]  
\subfloat[Satellite view (Map data: $\copyright$2023 Google Earth) \label{Fig_outdoor_screenshot}]{
\begin{tikzpicture}
        \node[]at(0,0){\includegraphics[width=1\columnwidth,
        trim= 5pt 10pt 0 0, clip]{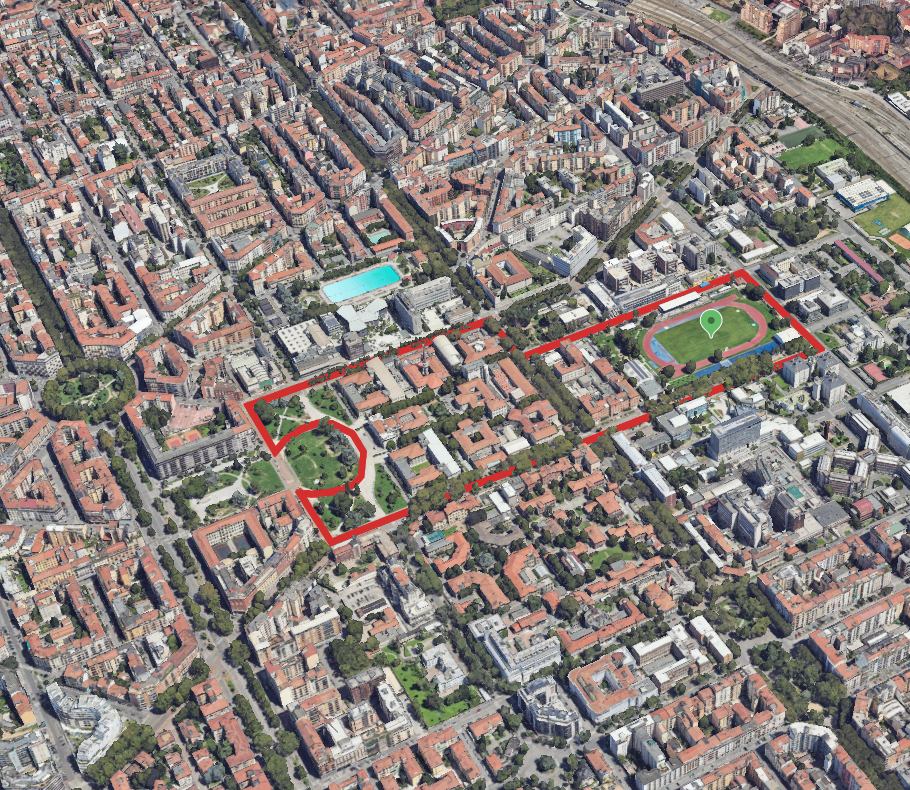}};
\end{tikzpicture}
}

\subfloat[Simulated deployment \label{fig:outdoor_coverage}]{
    \begin{tikzpicture}
        \node[]at(0,0){\includegraphics[width=1\columnwidth]{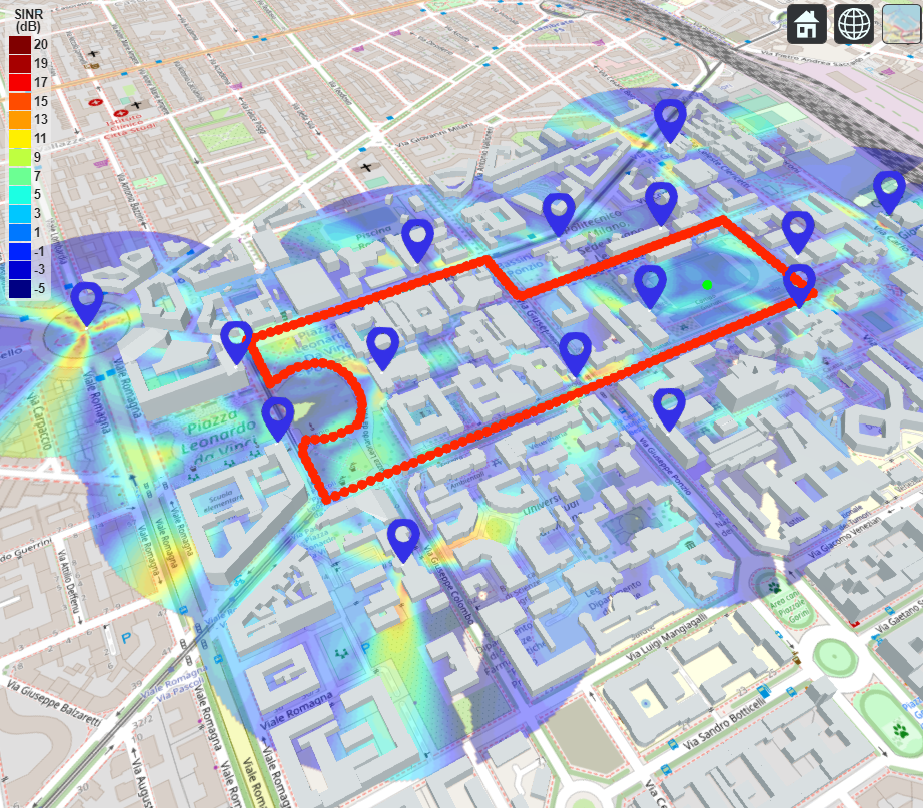}};
        \draw (3,-3.1) node [rectangle, minimum width=2.3cm, minimum height=1.2cm, draw, fill=white]{};
        \node [anchor=west] at (2.3,-2.75){\small BS};
        \node [anchor=west] at (2.3,-3.1) {\small Static UE};
        \node [anchor=west] at (2.3,-3.475) {\small Mobile UE};

        \node[]at(2.1,-2.75){\includegraphics[width=0.29cm]{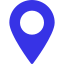}};
        \node[]at(2.1,-3.12){\includegraphics[width=0.29cm]{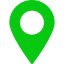}};
        \node[]at(2.1,-3.46){\includegraphics[width=0.15cm]{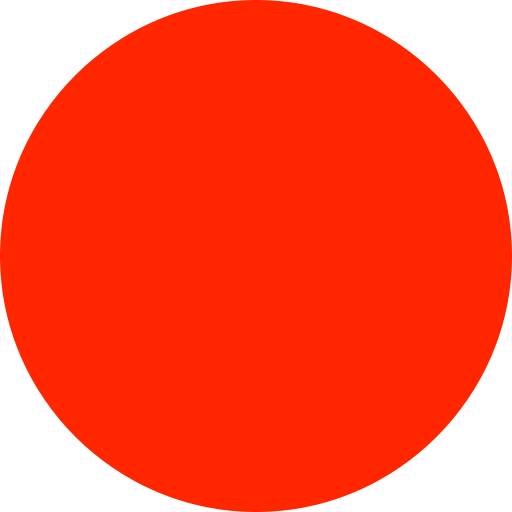}};

        \node[]at(2.35,1.275){\includegraphics[width=0.5cm]{photo/green.png}};

    \end{tikzpicture}
    }
    \caption{
    \textcolor{black}{Outdoor urban scenario in Politecnico di Milano, Leonardo campus. 
    (a) 3D satellite view of the area (Map data: $\copyright$2023 Google Earth).
    (b) \acp{BS} deployment, coverage, and \ac{UE} locations.
    Blue markers indicate the positions of 5G BSs, green marker the UE position used for static experiments, while red circles the UE trajectory for mobile simulations.}
    } \label{fig:outdoor_scenario}
        
\end{figure}

\subsection{Simulation environment}
\label{sec:Simulation Environment}

The \ac{RT} tool provided by Matlab$^{\circledR}$~\cite{MATLAB} is used to perform the \ac{5G} positioning simulations. It allows to faithfully model the \ac{PRS} and \ac{SRS} signals according to \ac{3GPP} Rel-16 and propagate them over a \ac{3D} environment accounting for the presence of buildings and associated multipath effects. The propagation model can be designed with an arbitrary number of reflections, depending on the context. 
The \ac{3D} environment is modeled with the \textit{Site Viewer} feature,
which, combined with \ac{RT}, allows to recreate realistic scenarios for performance analyses. An example of a simulation environment in Matlab$^{\circledR}$ is shown in  Fig.~\ref{fig:raytracer}, 
\textcolor{black}{where a \ac{UE} (green marker) is placed in the middle of a courtyard and it is surrounded by three \acp{BS}  (blue markers).}
The drawn rays represent the signal propagation paths computed by the \ac{RT} for each \ac{BS}, colored according to the path loss value and showing both LOS and NLOS conditions.

\begin{figure}[!tb]
    \centering

    \subfloat[Indoor environment \label{fig:made_adv}]{
         \begin{tikzpicture}
            \node[]at(0,0){\includegraphics[width=\columnwidth, trim= 0 8cm 0 0, clip]{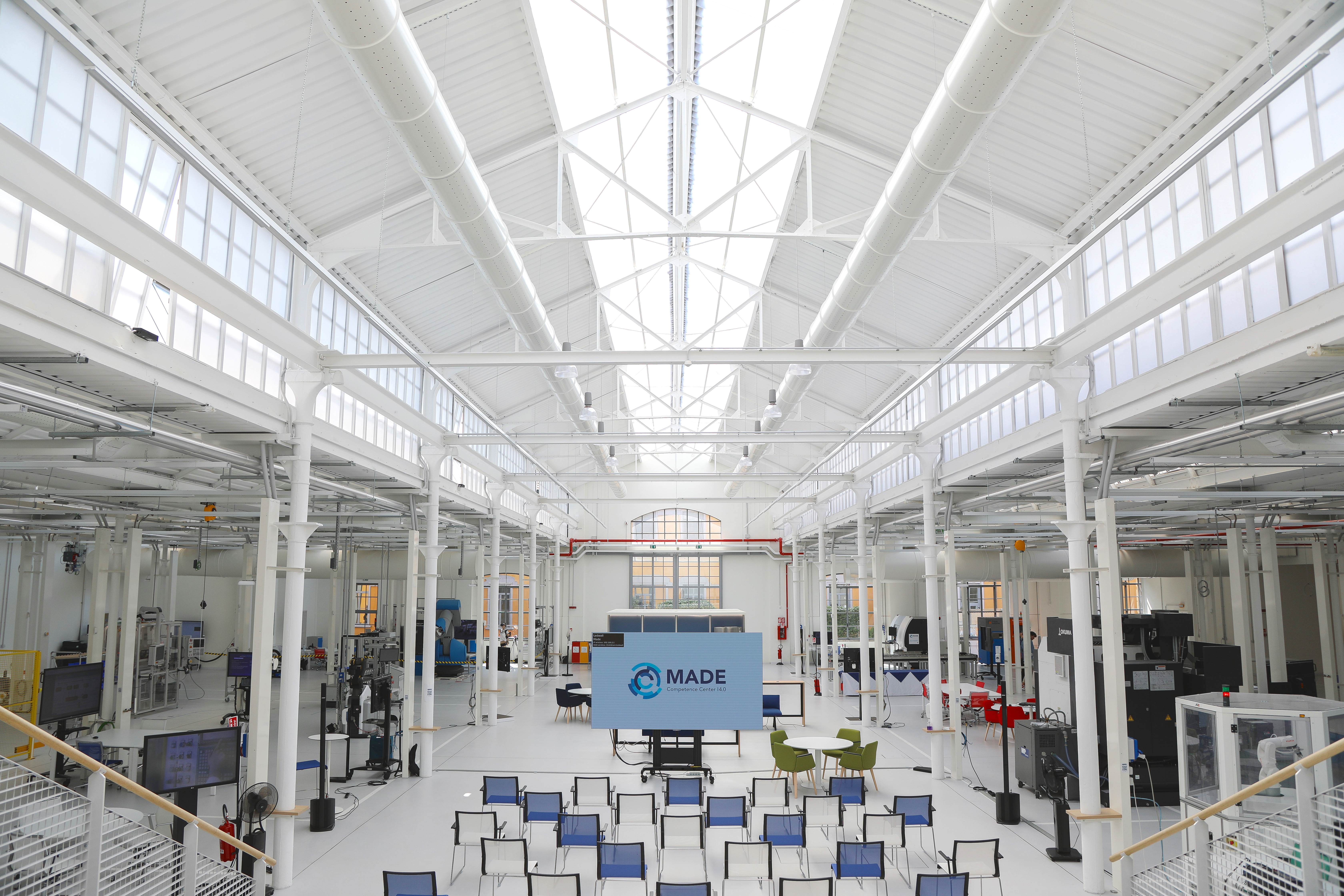}};
            \draw [draw=blue,dotted,ultra thick] (-0.7,-1.8) rectangle (-2.3,-0.2);
             \node[text=blue]at(-1.5,0.1){Industrial};

            \node[text=red, anchor=east]at(2.3,-2.2){Office};
            \draw[draw=red,-Latex,line width=1.5pt] (2.3,-2.2)--(3.8,-1.5);
        \end{tikzpicture}
        }
        \\
   \subfloat[3D rendering \label{fig:made_rendering}]{
        \begin{tikzpicture}
            \node[]at(0,0){\includegraphics[width=\columnwidth,
        trim= 0 0 0 0, clip]{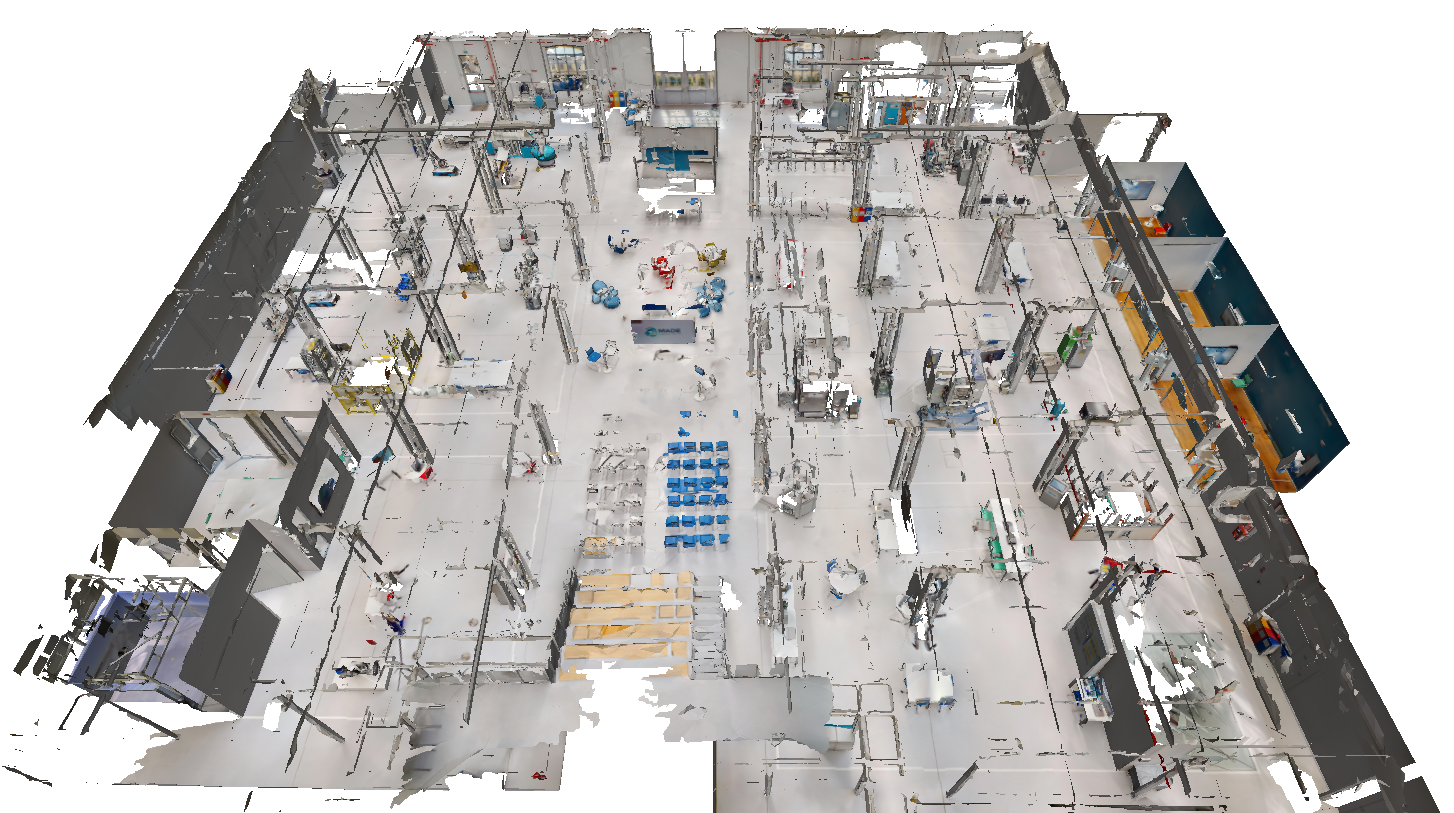}};
            \draw [draw=red,dotted,ultra thick] (3.5,-0.2) rectangle (2.2,1.2);
            \node[text=red]at(2.87,1.5){Office};
            
            \draw [draw=blue,dotted,ultra thick] (-0.4,0.5) rectangle (-2.1,2.2);
             \node[text=blue]at(-1.27,2.5){Industrial};
        \end{tikzpicture}  
    }
    \caption{\textcolor{black}{MADE Competence Center, Politecnico di Milano, Bovisa Durando campus, Milan. (a) Picture of the indoor scenario, (b) 3D rendering from lidar acquisition.}}
        \label{fig:made}
\end{figure}


\begin{figure}[!tb]
  \centering
   \subfloat[Office area\label{fig:indoor_office}]{
         \centering
         \begin{tikzpicture}
              \node[]at(0,0){\includegraphics[width=0.95\columnwidth]{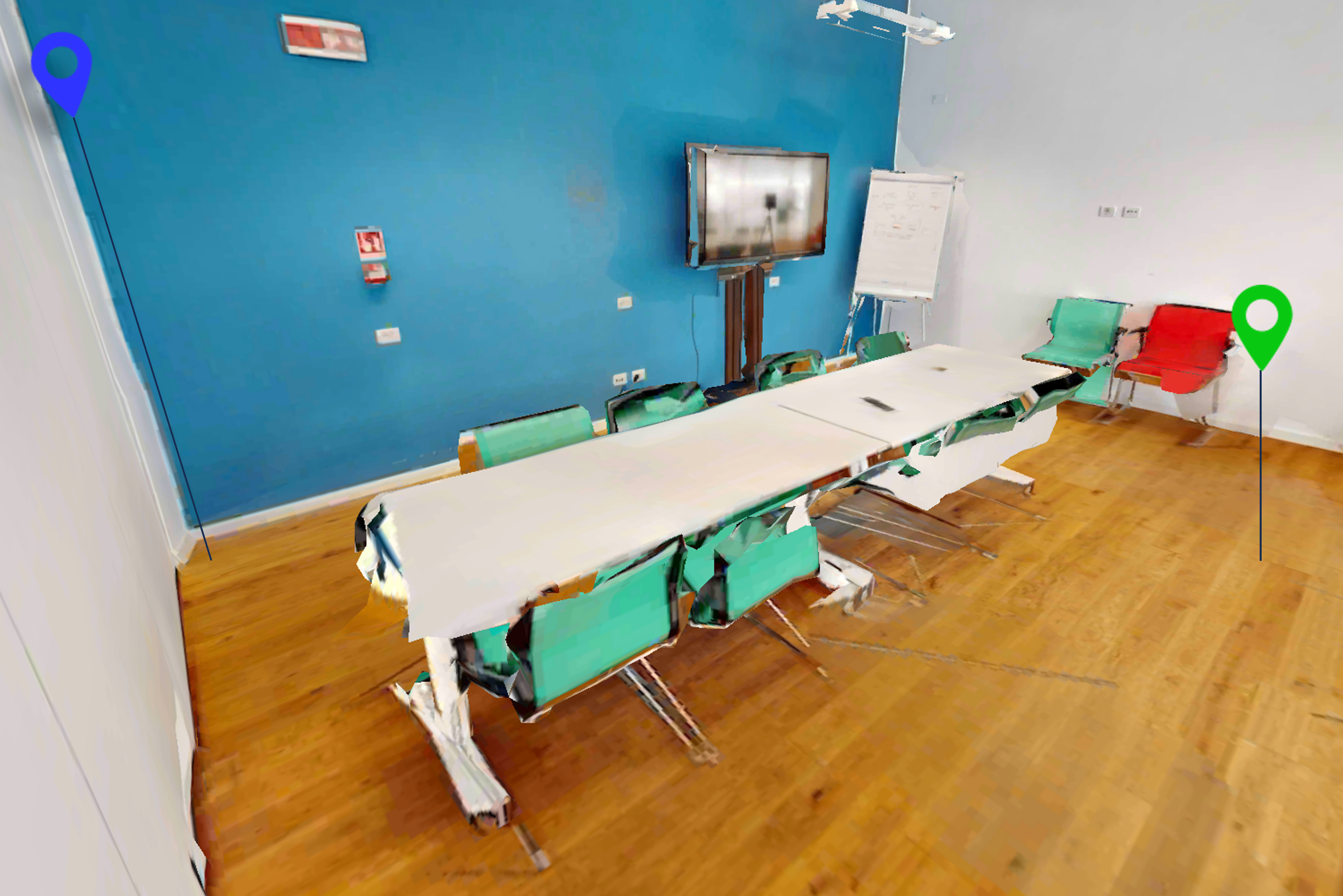}};
              \node[]at(3.65,1.3){UE};
              \node[color=white]at(-3.15,2.35){BS};
         \end{tikzpicture}
         }
         \\
    \subfloat[Industrial area  \label{fig:indoor_industrial}]{    
    \begin{tikzpicture}
        \node[]at(0,0){\includegraphics[width=0.95\columnwidth]{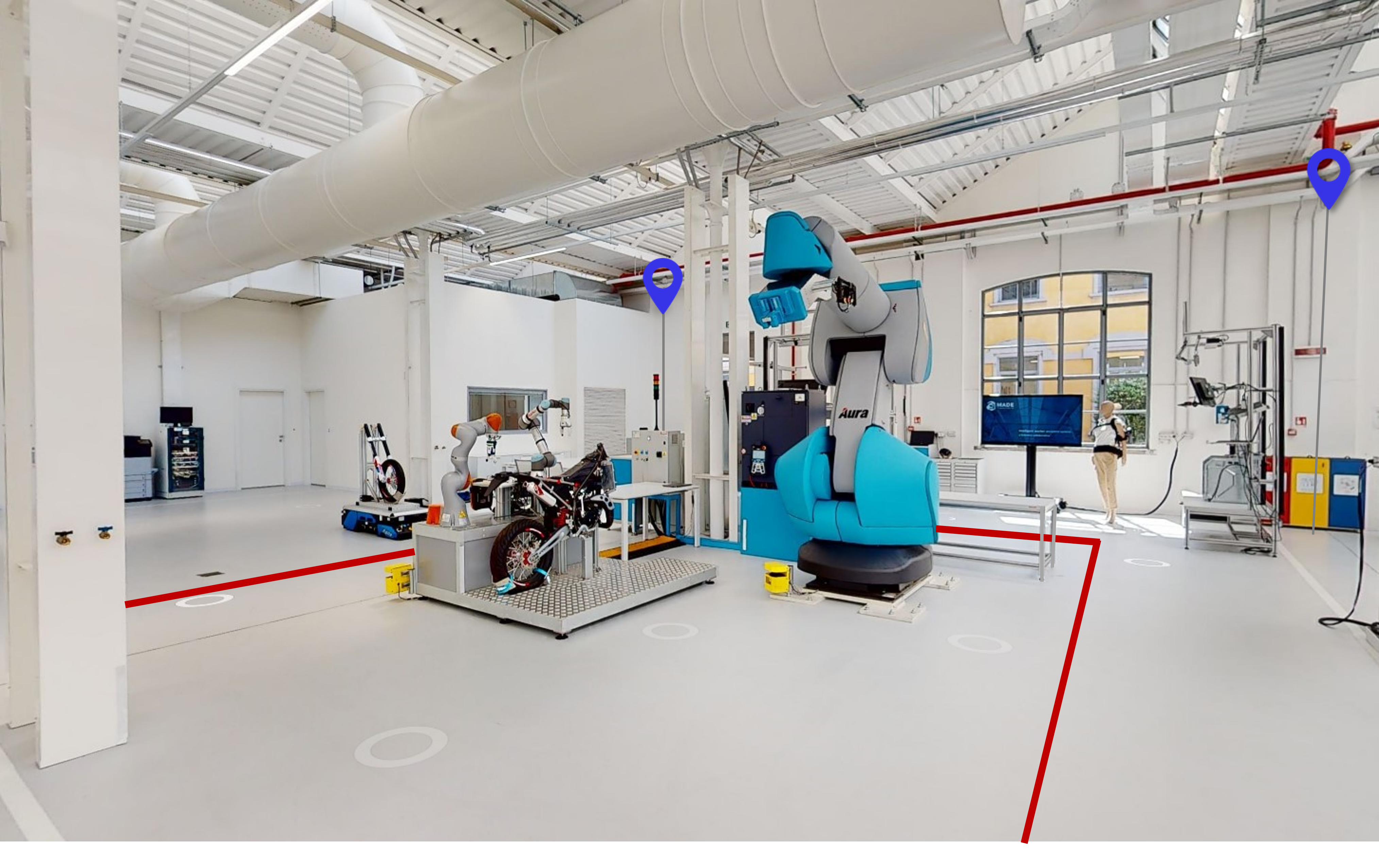}};
        \draw (3.4,-1.75) node [rectangle, minimum width=1.4cm, minimum height=1cm, draw, fill=white] {};
        \draw [draw=darkred] (2.8,-1.555)--(3.2,-1.555);
        \draw (3.6,-1.55) node [] {\small UE};
        \node[]at(3,-1.9){\includegraphics[width=0.29cm]{photo/blue.png}};
        \draw (3.6,-1.9) node [] {\small BS};

        \draw[-Latex,darkred, line width=0.8pt] (-2.24,-0.84)--(-2.26,-0.845);
        \draw[-Latex,darkred, line width=1pt] (2.145,-2)--(2.17,-1.9);
        \draw[-Latex,darkred] (3.1,-1.555)--(3.12,-1.555);
    \end{tikzpicture}
    }
    \caption{\textcolor{black}{Indoor scenario in Politecnico di Milano, Bovisa Durando campus, MADE Competence Center.  (a) Indoor office area rendering, (b) Indoor industrial area, highlighted in Fig.~\ref{fig:made} in red and blue, respectively.}}
        \label{fig:indoor_scenarios}
\end{figure}


We perform \ac{5G}  positioning simulations in both outdoor \textcolor{black}{(Fig.~\ref{fig:outdoor_scenario}) and indoor (Fig.~\ref{fig:indoor_scenarios}) }environments, with static and dynamic \ac{UE} conditions.
In particular, we consider an outdoor urban area around the Politecnico di Milano \textit{Leonardo campus} \textcolor{black}{(see the satellite view in Fig.~\ref{Fig_outdoor_screenshot})}, representative of an urban mobility use case,  and an indoor environment within the Politecnico di Milano \textit{Bovisa Durando campus} , representative of an industrial use case \textcolor{black}{(see the photo in Fig.~\ref{fig:made_adv})}, inside the MADE Competence Center, a laboratory facility on Industry 4.0 that simulates a digital factory and hosts a wide range of industrial machinery. For the former, \textit{OpenStreetMap} files containing the geographical information about buildings have been imported in Matlab$^{\circledR}$; for the latter, we imported a \ac{3D} lidar scanning of the MADE Competence Center. 

\input{Figure_tex/BS_visibility_map}

The outdoor scenario consists of a 1~km$^2$ outdoor urban area, in which we deployed 15 \ac{5G}  sites \textcolor{black}{(see Fig.~\ref{fig:outdoor_coverage})}, each composed of 3 antenna panels oriented at 0°, 120°, and -120° with respect to East, at a height of 4~m from the support point. 
Despite the fact that this deployment does not match the current installation of mobile operators in the area, as they do not guarantee enough density and multi-\ac{BS} visibility for cellular positioning, it is selected as a trade-off between the needs of guaranteeing enough \acp{BS} visibility and limiting the overall number of \acp{BS}. 
More efficient deployments can be designed using optimization algorithms~\cite{yang_5g_2022}, while higher performances can be achieved by further increasing the \ac{BS} density. 
The visibility map for the considered deployment over the simulated UE trajectory is shown in Fig.~\ref{fig:BS_visibility}.
Note that for \ac{mmWave} urban scenarios, the 3GPP standard recommends a dense deployment similar to the proposed one with a distance of 200~m between each \ac{BS}~\cite{tr138913}, as confirmed by further coverage studies in the literature~\cite{AhaFar2021}.  

The \ac{3D} rendering resulting from the lidar acquisition of the indoor scenario is reported in Fig.~\ref{fig:made_rendering}, where the two considered sub-areas representative of an office area and a factory area are highlighted. A more detailed visualization of such areas is shown in Fig.~\ref{fig:indoor_office} and  Fig.~\ref{fig:indoor_industrial}, respectively.
For the office room, we placed a single tri-sectorial cell, while in the industrial area, we deployed 4 \acp{BS} in the four edges near the columns, pointing towards the center.





\subsection{Simulation parameters}
\label{sec:Simulation Parameters}
For the simulation settings, we refer to two scenarios described in \ac{3GPP} \ac{TR} 38.857~\cite[Table 6-1]{tr138857}. The scenario for \ac{FR}1 specification considers $\mu=1$ ($\Delta f=30$ kHz and $BW=100$ MHz) with a carrier frequency $f_c^{}=3.5$ GHz. Instead, for \ac{FR}2, the scenario has a numerology $\mu=3$ ($\Delta f=120$ kHz and $BW=400$ MHz) with a carrier frequency $f_c^{}=28$ GHz.


The simulated radio devices employ a \ac{URA}, defined by the tuple ($M_g^{}$, $N_g^{}$, $M_a^{}$, $N_a^{}$, $P$), where $M_g^{}$ is the number of panels in the vertical plane, $N_g^{}$ the number of panels in the horizontal plane, $M_a^{}$ the number of antenna elements in the vertical plane, $N_a^{}$ the number of antenna elements in the horizontal plane, and $P$ the polarization of the antenna panel ($P \in \{0, 1\}$)~\cite{tr138901}. 
In the considered experiments, the UE has an antenna array defined by the tuple (1, 1, 2, 2, 1), while \acp{BS} 
default configuration is (1, 1, 4, 4, 1) for ranging measurements and (1, 1, 8, 8, 1) for angles. 
Each \ac{BS} is 3GPP standard compliant~\cite{tr138901} and is configured with 33~dBm of transmission power for the outdoor scenario and 23~dBm in indoor \cite{tr138802, tr138104}.
The use of \ac{MIMO} systems allows the implementation of the \ac{MUSIC} for an accurate estimate of \acp{AOA}, which is more effective at the \ac{BS} side rather than at the \ac{UE} as the number of antennas is higher.

The channel is modeled according to the standard using a \ac{CDL} impulse response for NLOS profiles, which can be defined up to a maximum bandwidth of 2~GHz~\cite{tr138901}. The \ac{CDL} model adopted for the simulations is the customized one, where channel parameters can be adapted to the \ac{RT}~\cite{raytracing} multipath configuration. The number of path reflections is set to two with the \ac{SBR} method.

The noise power spectral density ($N_0^{}$) is modeled as follows:
\begin{equation}
    N_0^{} = k_{\text{B}}^{} \cdot BW \cdot T_e^{},
\end{equation}
with $k_{\text{B}}^{}$ as the Boltzmann constant [JK$^{-1}$], $BW$ the bandwidth [Hz], and $T_e^{}=T_{\text{ant}}^{} + 290(NF-1)$ the noise temperature [K], where $T_{\text{ant}}^{}$ is the temperature [K], and $NF$ is the linearized noise figure, both referring to the receive antenna. For DL measurements, $NF=9$ dB in FR1 and $NF=10$~dB in FR2, while for UL measurements, $NF=5$ dB in FR1 and $NF=7$~dB in FR2. Instead, $T_{\text{ant}}^{} = 298 $~K (25°~C)~\cite{tr138857}.

The \acp{PRS} are defined for ranging measurements with $T^{\text{PRS}}_{\text{offset, RE}} = 0$, and starting symbol index $l_0^{} = 0$; $K_{\text{size}}^{} = 12$ and $N^{\text{slot}}_{\text{symb}} = 12$ without muting; $T^{\text{PRS}}_{\text{rep}} = 1$ slot and $T^{\text{PRS}}_{\text{per}}=10240$ slots. Each \ac{BS} sends a \ac{PRS} with $T_{\text{offset}}^{\text{PRS}} = 2$ slots with respect to the other \acp{BS} in order to avoid overlaps~\cite{ts138211}.
For the beam refinement procedure, we need to use more \acp{RE} since each \ac{RE} corresponds to a beam. Therefore, with a comb-12 pattern, we are able to create a maximum of 12 beams all at once beamformed in frequency. 
Alternatively, it might be feasible to increase the number of beams while reducing the number of \acp{RE} through the implementation of time-based beamforming.
To accomplish this task, our settings consider $T^{\text{PRS}}_{\text{per}}=10240$ slots, while $T_{\text{offset}}^{\text{PRS}}$ and the \ac{RE} offset $T_{\text{offset, RE}}^{\text{PRS}}$ are $1\times12$ arrays, the former has the same value repeated (as before each \ac{BS} has an offset of 2 slots with respect to the others), and the latter has incremental values between 0 and 11. All the other values are unchanged. 

The \acp{SRS}, instead, need to be configured for \ac{3GPP} Rel-16 positioning, with $N^{\text{slot}}_{\text{symb}} = 8$ and $K_{\text{size}}^{} = 8$, starting frequency index $f_0^{} = 0$,  starting symbol index $l_0^{} = 0$, and $n_\text{RRC}^{} = 0$, which is an additional offset from $l_0$ specified in blocks of 4 \acp{RB}. For the bandwidth configuration, we set the values $B_\text{SRS}^{} = 0$ and $C_\text{SRS}^{}=63$ to unlock the maximum bandwidth (i.e., $m_\text{SRS}^{} = 272$), and  $b_\text{hop}^{} = 0$ to disable the frequency hopping. We also enable the periodic resource type with period and repetition as $T^{\text{SRS}}_{\text{per}} = 10240$ and $T^{\text{SRS}}_{\text{rep}}=2$ slots~\cite{ts138213}.
For the data transmission, we define the \ac{PDSCH} and \ac{PUSCH}, assuming to have a single transmission layer.

Regarding the algorithm implementations, the \ac{NLS} is implemented by setting the step-size scaling parameter $\eta=0.01$, a maximum of $1000$ iterations, and a stopping condition of $\lVert\boldsymbol{\hat{u}}_{k}^{} - \boldsymbol{\hat{u}}_{k-1}^{}\rVert < 10^{-4}$~m. While the \ac{NLS} is generally used for static \ac{UE} positioning,  the \ac{EKF} is preferable to estimate mobile \ac{UE}. The mobility model is a random walk, and the driving process covariance matrix is defined as $\boldsymbol{Q}_{t}^{} = \text{diag}\left(\sigma_x^2,\sigma_y^2,\sigma_z^2\right)$, where the diagonal entries denote the uncorrelated standard deviations along the three axes, respectively.

\subsection{Numerical results}
\label{sec:numerical_results}
In the following, we evaluate the accuracy performance of \ac{5G} positioning in the selected outdoor and indoor environments, with various configurations of system parameters.
The code used for the simulation in the outdoor scenario is publicly available\footnote{Link to the public code repository: \hyperlink{https://github.com/Ita97/A-tutorial-on-5G-positioning}{https://github.com/Ita97/A-tutorial-on-5G-positioning}}.

\subsubsection{Outdoor environment}
\label{sec:numerical_results_outdoor}

For the outdoor case, we first present a statistical analysis of the location-related measurements extracted from the \ac{5G} radio signals. We then consider a static positioning use case (green pin in Fig.~\ref{fig:outdoor_scenario})
where we assess the effect of the numerology, the type of measurements, and the \ac{BS} antenna array configuration using as positioning algorithm the \ac{NLS} with Gauss-Newton implementation (see Section~\ref{sec:Positioning algorithms}). 
Finally, we discuss a dynamic use-case with the \ac{UE} moving along the red trajectory in Fig.~\ref{fig:outdoor_scenario}, where we assess the tracking performance of \ac{EKF} localization (see Section~\ref{sec:EKF_tracking}) using different types and numbers of measurements.

\begin{figure}[!t]
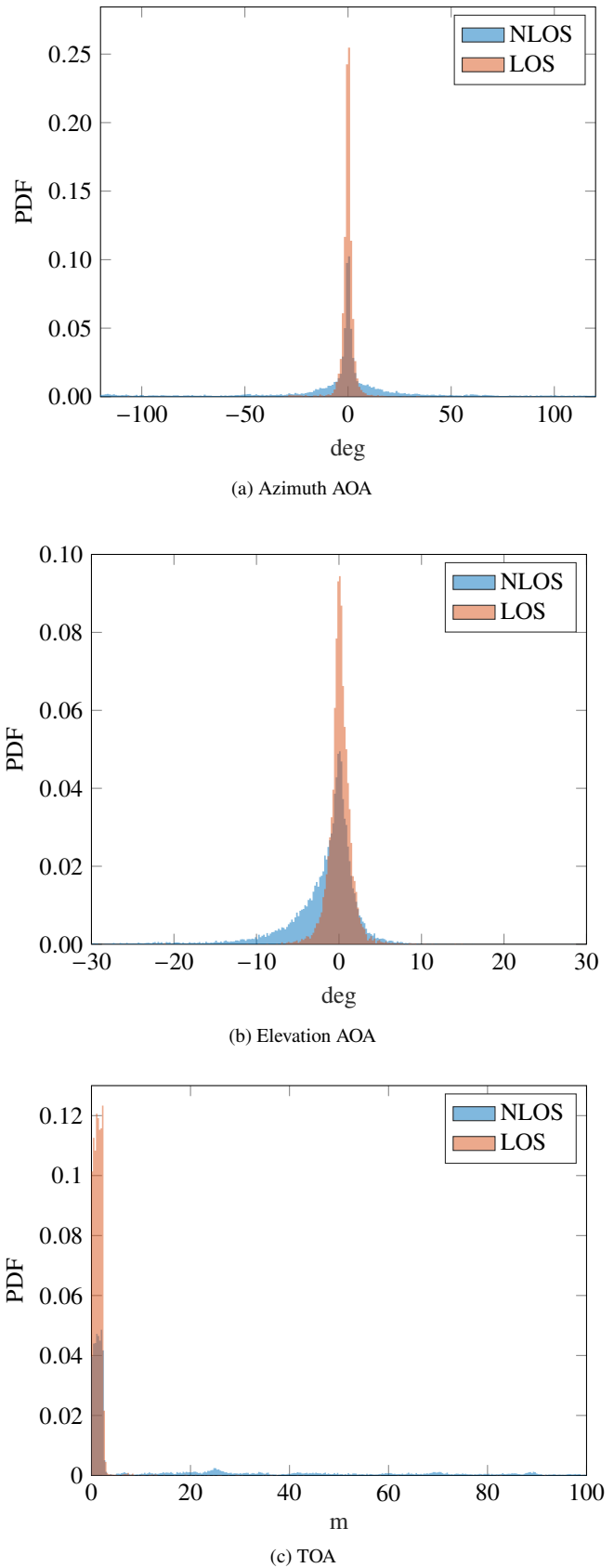

    \centering
    \subfloat[Azimuth AOA\label{fig:azimuth_error}]{
        \input{Figure_tex/pdf_azimuth_error_deg}  
    }\\
    \vspace{5pt}
    \subfloat[Elevation AOA\label{fig:elevation_error}]{
        \input{Figure_tex/pdf_elevation_error_deg}
    }\\
    \vspace{5pt}
    \subfloat[TOA \label{fig:ranging_error}]{
        \input{Figure_tex/pdf_toa_error}
    }

    \caption{Outdoor urban scenario - analysis of the measurement accuracy: measurement errors in \ac{LOS} (orange) and \ac{NLOS} (blue) conditions for $\mu=1$. (a) azimuth AOA; (b) elevation AOA; (c) TOA.}
    \label{fig:meas_analysis}
\end{figure} 
\paragraph{5G measurement accuracy} 
Before assessing the performance of 5G positioning, it is worth analyzing the statistics of the location measurements extracted from the received 5G radio signals. They will then be used for multi-lateration/angulation. We recall that signal propagation from the \ac{Tx} to the \ac{Rx} is simulated using the Matlab$^{\circledR}$ \ac{RT} tool.  

We report in Fig.~\ref{fig:meas_analysis} the \ac{PDF} of the measurement error in \eqref{eq:measurement_single}, i.e., $p(n_{i}^{})$, that is observed by collecting the location parameters along the red trajectory of the dynamic scenario in Fig.~\ref{fig:outdoor_scenario}. We analyze the measurement errors obtained with the numerology $\mu=1$ on the azimuth AOA (Fig.~\ref{fig:azimuth_error}), elevation AOA (Fig.~\ref{fig:elevation_error}), and TOA (Fig.~\ref{fig:ranging_error}), distinguishing between \ac{LOS} and \ac{NLOS} conditions. 
Regarding the azimuth AOA, we observe a symmetric distribution of the errors centered around 0 deg, with larger support for the NLOS case. The symmetry, on the other hand, is not observed on the elevation angle in NLOS conditions, as most of the errors are negatively biased in elevation due to the terrain reflections,
whereas ranging inaccuracies are mostly positive since the \ac{TOF} is usually the first peak in the cross-correlation. Therefore, in the case of peaks generated by multipath or NLOS measurements, the range estimate is higher than the real distance.

\begin{figure}[!t]
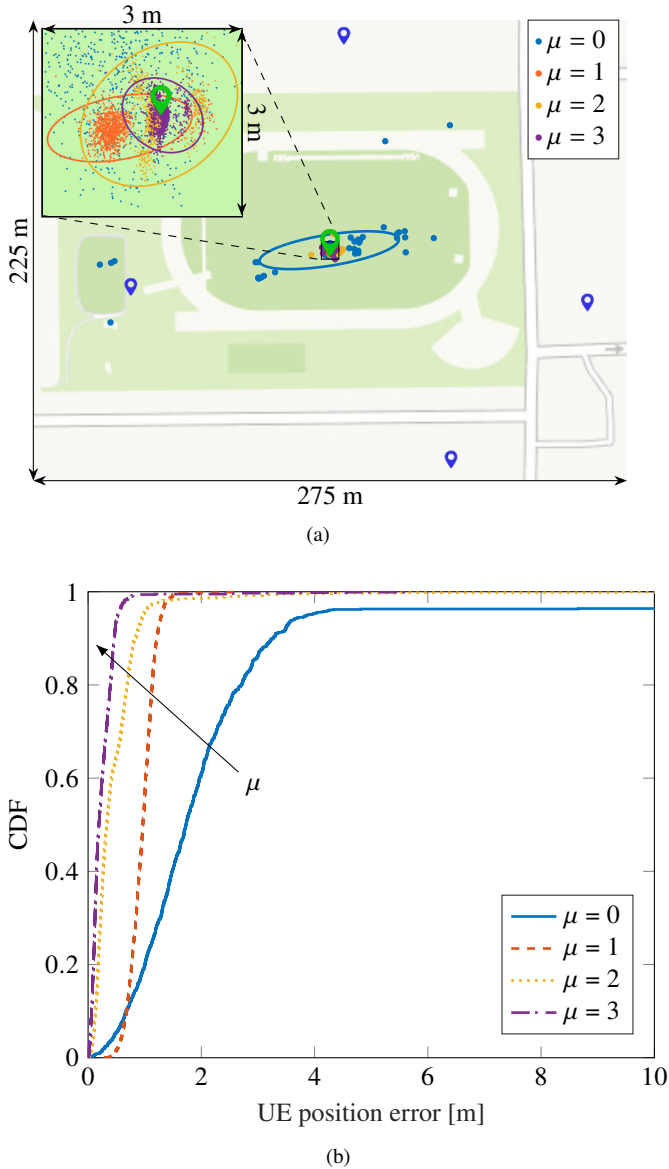


    \subfloat[\label{fig:tdoa_cep}]{
    \input{Figure_tex/cep_tdoa}
    }

    \subfloat[\label{fig:measerrcdf}]{
    \input{Figure_tex/cdf_tdoa_pos_err}
    }
 \caption{Static outdoor UE positioning in Politecnico di Milano, Leonardo campus area - effect of numerology using DL-TDOA measurements. (a) scatterplot of the position estimates and associated error ellipses  for  $\mu \in \{0,3\}$. (b) CDF of UE position error with DL-TDOA  measurements.}
    \label{fig:tdoa_static}
\end{figure}
\paragraph{Impact of the numerology} 
As first assessment of 5G positioning, we evaluate the impact of the numerology $\mu \in \{0,3\}$ (i.e., both FR1 and FR2) in static conditions, using
 \ac{DL}-\ac{TDOA} measurements.
The static positioning outdoor scenario is characterized by an open area  (i.e., a running track) surrounded by four \acp{BS}. This emulates a condition where no obstacles are present, resulting in a nearly ideal \ac{LOS} environment for positioning.

As a first example, Fig.~\ref{fig:tdoa_cep} shows the scatter plot of the location fixes obtained by the \ac{NLS} algorithm and the associated error ellipses for all the considered numerology, i.e., $\mu=0$ in blue, $\mu=1$ in orange, $\mu=2$ in yellow and $\mu=3$ in purple. 
A first takeaway is related to the non-recommended use of the lowest numerology for positioning tasks, as such configuration leads to large positioning errors, even in ideal \ac{LOS} conditions.
A more detailed comparison of the positioning performances given in terms of \ac{CDF} of the \ac{UE} position error in
Fig.~\ref{fig:measerrcdf}.  

A quantitative summary of performance metrics is reported in Table~\ref{tab:numerologycomparison}, in terms of measurement accuracy $\sigma_{\text{TDOA}}^{}$, \ac{2D} \ac{RMSE}, \ac{MAE} and bias. 
Analyzing the values in the table for $\mu=3$ and $\mu=0$, we quantify an improvement of $97.3$\% on the \ac{2D} \ac{RMSE}.

\begin{table}[!tb]
    \centering
    \caption{Summary of results for static UE outdoor positioning with DL-TDOA measurements using different numerologies}
    \label{tab:numerologycomparison}
    \begin{tabular}{c c c c c}
    \toprule
        $\mu$ & 0 & 1 & 2 & 3 \\
        \midrule
        $\sigma_{\text{TDOA}}^{}$ [m] & 5.99 & 0.98 & 0.58 & 0.30 \\
        2D RMSE [m] & 14.7 & 0.98 & 0.76 & 0.40 \\
        2D MAE [m] & 3.72 & 0.96 & 0.47 & 0.25 \\
        2D bias [m] & 1.86 & 0.81 & 0.09 & 0.09 \\
        \textcolor{black}{PEB [m]} & 4.13 & 0.68 & 0.4 &  0.21 \\
        \bottomrule
    \end{tabular}
\end{table}

\begin{table}[!tb]
    \centering
    \caption{Static outdoor positioning - summary results for different positioning methodologies in FR1 ($\mu=1$)}
    \label{tab:methodcomparison}
    \begin{tabular}{c c c c c}
    \toprule
         & DL-TDOA & multi-RTT & UL-AOA & DL-AOD \\
        \midrule
        $\sigma_{\text{n}}^{}$ [m] & 0.98 & 0.59 & - & - \\
        $\sigma_{\text{AOA, az}}^{}$ [deg] & - & - & 2.64 & 4.01 \\
        $\sigma_{\text{AOA, el}}^{}$ [deg] & - & - & 1.55 & 0.57 \\
        2D RMSE [m] & 0.98 & 0.89 & 8.60 & 10.57 \\
        2D MAE [m] & 0.96 & 0.84 & 3.55 & 10.55 \\
        2D bias [m] & 0.81 & 0.81 & 0.39 & 10.55 \\
         \textcolor{black}{PEB [m]} & 0.68 & 0.53 & 5.30 & 8.05 \\
        \bottomrule
    \end{tabular}
\end{table}

\paragraph{Impact of measurement type} 
\label{sec:measurement_impact}
We extend the analysis on static \ac{UE} positioning by focusing on numerology $\mu=1$ and evaluating the effect of the measurement type on the positioning performance. This comparison includes \ac{DL}-\ac{TDOA}, multi-\ac{RTT}, \ac{UL}-\ac{AOA} and \ac{DL}-\ac{AOD} methodologies.

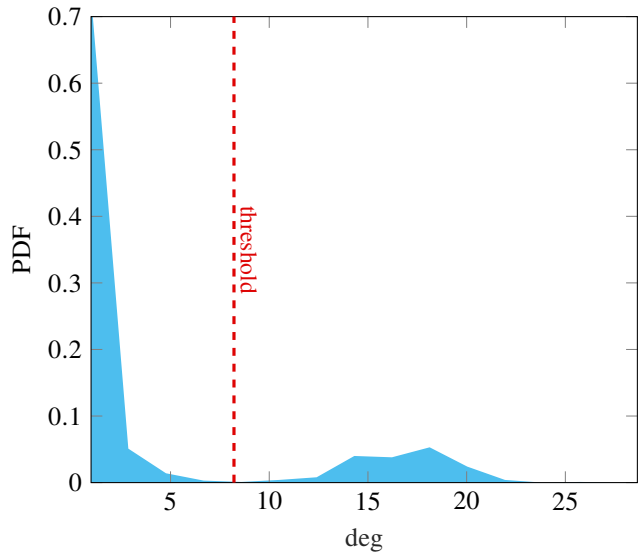
\begin{figure}
    \centering
%
%
\definecolor{mycolor1}{rgb}{0.30196,0.74510,0.93333}%
\definecolor{mycolor2}{rgb}{0.87059,0.01569,0.01569}%
\begin{tikzpicture}

\begin{axis}[%
width=0.82\linewidth,
height=0.7\linewidth,
at={(0\linewidth,0\linewidth)},
scale only axis,
area style,
stack plots=y,
xmin=0.9740282517224,
xmax=28.6478897565412,
xlabel style={font=\color{white!15!black}},
xlabel={deg},
ymin=0,
ymax=0.7,
ylabel style={font=\color{white!15!black}},
ylabel={PDF}, ylabel near ticks,
]
\addplot[fill=mycolor1, draw=none] table[row sep=crcr]{%
0.953974728892821	0.739\\
2.86192418667846	0.051\\
4.7698736444641	0.014\\
6.67782310224975	0.003\\
8.58577256003539	0.001\\
10.493722017821	0.004\\
12.4016714756067	0.008\\
14.3096209333923	0.04\\
16.217570391178	0.038\\
18.1255198489636	0.053\\
20.0334693067492	0.024\\
21.9414187645349	0.004\\
23.8493682223205	0\\
25.7573176801062	0.001\\
27.6652671378918	0\\
29.5732165956774	0.001\\
31.4811660534631	0.001\\
33.3891155112487	0\\
35.2970649690344	0\\
37.20501442682	0\\
39.1129638846057	0\\
41.0209133423913	0\\
42.9288628001769	0\\
44.8368122579626	0\\
46.7447617157482	0\\
48.6527111735339	0\\
50.5606606313195	0\\
52.4686100891051	0.001\\
54.3765595468908	0\\
56.2845090046764	0\\
58.1924584624621	0\\
60.1004079202477	0\\
62.0083573780334	0\\
63.916306835819	0\\
65.8242562936046	0\\
67.7322057513903	0\\
69.6401552091759	0\\
71.5481046669616	0\\
73.4560541247472	0.001\\
75.3640035825328	0.001\\
77.2719530403185	0.006\\
79.1799024981041	0.008\\
81.0878519558898	0\\
82.9958014136754	0\\
84.903750871461	0\\
86.8117003292467	0\\
88.7196497870323	0\\
90.627599244818	0\\
92.5355487026036	0\\
94.4434981603893	0.001\\
}
\closedcycle;

\addplot [color=mycolor2, dashed, line width=1.0pt]
  table[row sep=crcr]{%
8.58577256003539	0\\
8.58577256003539	0.8\\
};

\end{axis}

\draw[color=mycolor2, dashed, line width=1.2pt] (1.9,0)--(1.9,6.2) node [midway, xshift = 2mm, rotate=-90, color=mycolor2] {\small threshold};

\end{tikzpicture}%
    \caption{Static outdoor positioning - multipath detection on the residual error. PDF of the mean absolute residual error of NLS estimation using UL-AOA measurements. The red dashed line represents the threshold to discriminate multipath-affected positioning outputs.}
    \label{fig:aoa_meas_res_error}
\end{figure}

\begin{figure}
    \centering
    \input{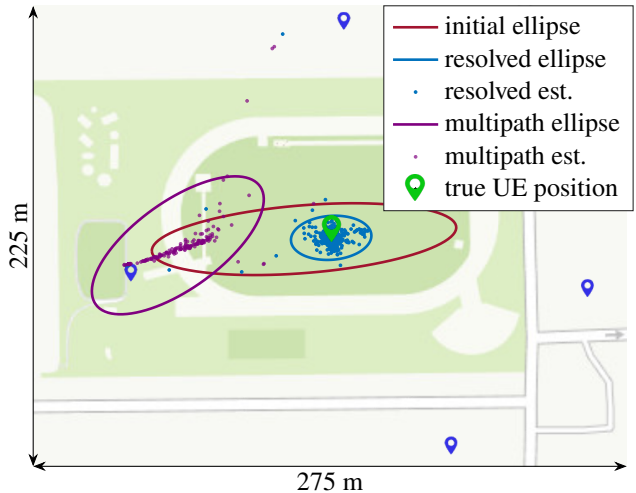}
    \caption{Static outdoor positioning with UL-AOA measurements: position estimates and associated error ellipses.}
    \label{fig:aoa_cep}
\end{figure}

Dealing with angle estimation, note that the \ac{MUSIC} algorithm used in UL estimation is more prone to the multipath effect than the beam management procedure employed for DL-AOD estimate due to the finer beam resolution. 
The critical determination of whether the signal is received via indirect propagation paths holds significant importance in identifying unreliable measurements that should be discarded. To this aim, a strategy could be to inspect the residual error $\Delta\boldsymbol{\rho}$ of the \ac{NLS} algorithm. For the considered static outdoor positioning test, the \ac{PDF} of the mean absolute residual error
is reported in Fig.~\ref{fig:aoa_meas_res_error}, which exhibits a clear bi-modal shape. The second peak (at around 15-20 deg) comes from the contributions of indirect paths; thus, it is possible to identify a threshold (red dashed line) discriminating between UL-AOA from LOS and NLOS paths.
The implication of using such a threshold is highlighted in Fig.~\ref{fig:aoa_cep}, in which we show the position estimated and associated error ellipse with and without discarding  UL-AOA NLOS measurements. In case we do not detect NLOS measurements, i.e., we equally consider all the UL-AOAs, the error ellipse is quite high (red ellipse). On the other hand, by detecting the NLOS measurements and discarding them (shown in purple), the final error ellipse (in blue) is smaller and centered around the true UE position.

Table~\ref{tab:methodcomparison} reports the results of the comparison between the different methods in terms of the standard deviation of measurement error ($\sigma_{\text{TDOA}}^{}$, $\sigma_{\text{RTT}}^{}$, and $\sigma_{\text{AOA}}^{}$), and the following positioning metrics: \ac{2D} \ac{RMSE}, \ac{MAE} and bias. 
Focusing only on angle-based positioning, our observations reveal that the DL-AOD positioning approach, executed via the beam management procedure, yields to high positioning errors despite its reduced susceptibility to multipath interference. Instead, the UL-AOA positioning methodology exhibits a heightened susceptibility to the multipath phenomenon. The removal of NLOS measurements results into a notable enhancement in positioning accuracy. Specifically, the mean of the positioning estimates closely approximates the true \ac{UE} position.
Lastly, we point out that ranging-based methodologies, i.e., DL-TDOA and multi-RTT, yield superior accuracy in terms of \ac{RMSE} and \ac{MAE} compared to their angle-based counterparts, as they are less impacted by the incorrect geometrical information coming from multipath. 
Moreover, the degree of error induced by the angles is highly dependent on the distance and the \ac{BS} array configuration.
Among the ranging-based approaches, multi-RTT measurements demonstrate a higher level of accuracy compared to DL-TDOA. 
This advantage is justified by the fact that, at first, we do not account for synchronization errors at the \ac{UE} side and assume perfect knowledge regarding the reply time. Then, it is also explained by the additive property of the variance of measurement noise on the two communication links involved in a TDOA  computation.



\begin{figure}
    \centering
    \begin{tikzpicture}
        \node[]at(0,0){\includegraphics[width=0.95\linewidth]{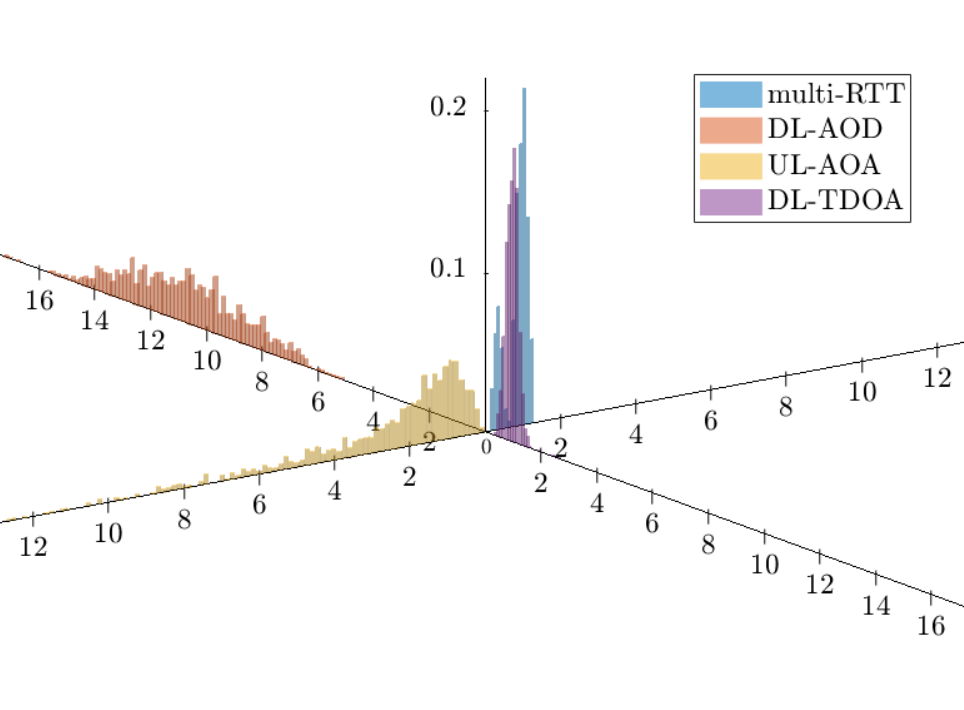}};
            
        \node[]at(2.5,-0.8){\small [m]};
        \node[]at(2.,-2){\small [m]};
        \node[]at(-2.5,-0.4){\small [m]};
        \node[]at(-2.,-1.6){\small [m]};
        \node[]at(0,2.7){\small PDF};
    \end{tikzpicture}
    \caption{Static outdoor positioning -  \acp{PDF} of the positioning error for different types of measurement. 
    }
    \label{fig:pdf_methods}
\end{figure}

A comparison of all the four considered positioning methodologies is given in
Fig.~\ref{fig:pdf_methods} in terms of \acp{PDF} of UE positioning error. 
The colored histograms reveal that ranging-based methodologies have a support of less than  3~m, while angle-based methods exhibit errors exceeding 10~m. However, it is noteworthy that the UL-AOA approach achieves an error peak close to one meter, similar to the performance of \acp{TDOA} and \acp{RTT}. By contrast, the DL-AOD method exhibits a conspicuous bias, evidenced by a peak error of approximately 10~m. 


\begin{table}[!tb]
    \centering
    \caption{Static outdoor positioning - Impact of \acp{BS} array size in UE positioning with \ac{UL}-\ac{AOA} measurements at FR1 ($\mu=1$)}
    \label{tab:angleaccuracy}
    \begin{tabular}{c c c c}
    \toprule
         & $4\times4$ & $8\times8$ & $16\times16$  \\
         \midrule
         & & & \\ [-9pt]
         $\sigma_{\text{AOA, az}}^{}$ [deg] & 2.81 & 2.64 & 0.95 \\
        $\sigma_{\text{AOA, el}}^{}$ [deg]& 1.83 & 1.55 & 0.75 \\
        2D RMSE [m] & 9.03 & 8.60 & 2.36 \\
        2D MAE [m] & 4.85 & 3.55 & 1.73 \\
        2D bias [m] & 1.0 & 0.39 & 0.34 \\
        \textcolor{black}{PEB [m]} & 5.64 & 5.3 & 1.9 \\
        \bottomrule
    \end{tabular}
\end{table}

\paragraph{Impact of BS antenna configuration} 
As a last analysis on static \ac{UE} positioning, we analyze the impact of different configurations of \ac{BS} antennas in \ac{UL}-\ac{AOA} measurements in FR1 ($\mu=1$). Specifically, the communication hardware at \acp{BS} is compared for the following tuples: (1, 1, 4, 4, 1), (1, 1, 8, 8, 1), and (1, 1, 16, 16, 1).
This analysis aims to evaluate the impact of the number of \ac{MIMO} antennas in accurately estimating the \ac{AOA}. 
Table~\ref{tab:angleaccuracy} reports the results of the comparison in terms of azimuth and elevation accuracy ($\sigma_{\text{AOA, az}}^{}$ and $\sigma_{\text{AOA, el}}^{}$, respectively) and resulting UE positioning in terms of \ac{2D} \ac{RMSE}, \ac{MAE} and bias of the estimate. These results are reported after the application of the residual error method (explained in Section~\ref{sec:measurement_impact}) to get rid of positioning estimations biased by the multipath effect.
The use of common array sizes, such as with a panel of $16\times16$ elements, allows the collection of angle measurements that are accurate up to 1 deg in LOS condition. With these settings (4 \acp{BS} in LOS surrounding the UE), the 5G network is capable of localizing the UE with an error of about 2 m using only UL-AOA information.

\input{Figure_tex/trajectory_heatmap}
\input{Figure_tex/trajectory_cdf}

\paragraph{Outdoor mobile scenario}
This analysis aims to assess the tracking performance of a 5G mobile positioning system based on \ac{EKF} in mixed LOS/NLOS conditions with a variable number of visible \acp{BS}.
The UE mobility model is a random walk~\cite{Gustafsson_m05} with a sampling time of 0.7134\,s, according to the \ac{PRS} periodicity $T^\text{PRS}_\text{rep}$.
We consider 5G signals in FR1 with numerology $\mu=1$,  and the use of \ac{DL}-\ac{TDOA}, \ac{UL}-\ac{AOA} and the combination of the two types of measurement. 

The 5G positioning results are first analyzed with the heatmap of the positioning error in
Fig.~\ref{fig:trajectory_heatmap}, complemented with the associated \acp{CDF} in Fig.~\ref{fig:trajectory_cdf} and the summary in Table~\ref{tab:trajectory}.
Looking at the heatmaps in Fig.~\ref{fig:trajectory_heatmap}, higher errors are visible in the bottom left part and in the upper right part of the trajectory, where the visibility is poor, i.e., no LOS \acp{BS} or at most one are present (see Fig.~\ref{fig:BS_visibility}). The areas well covered by many \acp{BS}, such as the top-left and bottom-right portions of the trajectory, guarantee better positioning.
We recall that at least two \acp{BS} are required to have one TDOA measurement; thus, having poor visibility conditions is detrimental to DL-TDOA methodology. On the other hand, AOA-based methods are highly susceptible to multipath, and the method of residuals described in Section~\ref{sec:measurement_impact} cannot be employed within the \ac{EKF}. 
Overall, the joint use of \ac{DL}-\ac{TDOA} and \ac{UL}-\ac{AOA} leads to better positioning, as the tracking algorithm is frequently updating the estimate with measurements, minimizing outage conditions and avoiding to rely on motion model prediction over long time periods. 
Table~\ref{tab:trajectory} depicts the overall accuracy of the trajectory, showing the need for higher \ac{BS} density to attain satisfactory results when solely relying on 5G measurements.

A breakdown of the achieved UE position error according to the number of available DL-TDOA measurements $M$ is reported in Fig.~\ref{fig:accuracyVSmeasurements}. Notice that with only one or two \ac{TDOA} measurements, the results are very poor as the information gain provided by the measurements in the EKF is limited by the weak geometric condition. By increasing the number of simultaneously available measurements, as expected, the positioning accuracy improves. Having a number of measurements higher than 3 guarantees good accuracy ($\approx 1\,$m). This confirms the importance of guaranteeing good visibility and coverage conditions for unlocking precise positioning services.

\input{Figure_tex/error_per_meas_TDOA}

\begin{table}[!tb]
    \centering
    \caption{Outdoor mobile positioning - Summary results according to the type of employed measurements}
    \label{tab:trajectory}
    \begin{tabular}{c c c c}
    \toprule
         & DL-TDOA & UL-AOA & DL-TDOA \& UL-AOA \\
        \midrule
        2D RMSE [m] & 49.11 & 17.48  & 14.57 \\
        2D MAE [m] & 20.06 & 11.16 & 7.42 \\
       \bottomrule
    \end{tabular}
\end{table}

\subsubsection{Indoor environment}
\label{sec:numerical_results_indoor}
For the indoor environment (Fig.~\ref{fig:indoor_scenarios}), we focus on two scenarios: an office with a single \ac{BS} and an industrial area full of metallic objects (e.g., machinery and robots). This selection allows us to assess the 5G capabilities for a perspective consumer application (e.g., smartphone location-based services with FR2 support), as well as to analyze the introduction of 5G positioning into industrial production and manufacturing environments (e.g., by a 5G private network providing positioning services inside a factory).

\paragraph{Office area}
In the office scenario illustrated in Fig.~\ref{fig:indoor_office},
we focus on static \ac{UE} positioning with a single \ac{BS} using the \ac{NLS} algorithm. 
We consider \ac{RTT} and \ac{UL}-\ac{AOA} measurements extracted from \ac{PRS} and \ac{SRS}. An \ac{FR}2 communication link is 
simulated with numerology $\mu=3$ ($BW=400$~MHz).
For the ranging measurements, we adopt a parabolic interpolation~\cite{5431382} to improve the cross-correlation peak detection at the \ac{Rx} side. 
The antenna array is configured with the tuple (1, 1, 8, 8, 1).

In this small environment, we observed a measurement accuracy equal to  $\sigma_{\text{AOA, az}}^{}=\sigma_{\text{AOA, el}}^{}=3.44$~deg, $\sigma_{\text{RTT}}^{}=0.32$~m, 
while the results for UE positioning indicate a
\ac{2D} \ac{RMSE} of 0.66~m and an \ac{MAE} of 0.52~m, with bias of 0.38~m. 

The location fixes provided by the different positioning methods are shown  in
Fig.~\ref{fig:indoor_cep}. 
The presence of multiple clusters manifests the ambiguities generated by multipath on angle estimation. The multipath detection method on residual error (presented in Section~\ref{sec:measurement_impact}) remains constrained when restricted to two measurements. As a matter of fact, the \ac{NLS} will always converge with low $\Delta\rho$.
Nevertheless, opportunities for mitigating this error still exist, especially through the incorporation of supplementary information such as architectural floor plans. Practically, embedding physical constraints on the position estimates will enforce the positioning algorithm to provide outcomes falling within the office area, rejecting estimates falling outside.
An example of such a process is shown in Fig.~\ref{fig:indoor_cep}, where the estimated positions that fall outside the office room are highlighted in pink, while those inside are in blue. The goal of the figure is to point out the improvements that can be obtained by discarding outside estimates in terms of error ellipse: the ellipse is larger in case the room information is not embedded. By incorporating side information on the room map, the achieved positioning has a \ac{2D} \ac{RMSE} of 0.49~m and an \ac{MAE} of 0.41~m, with a bias of 0.31~m. 

\input{Figure_tex/rtt_aoa_cep}

\paragraph{Industrial area}
In the industrial area (Fig.~\ref{fig:indoor_industrial}), we placed 4 tri-sectorial cells in the corners near the columns. 
The simulations refer to a 
worker walking around the area over a U-shaped trajectory. A peculiarity of the scene is the high density of metallic surfaces, which produce strong multipath effects. As for the tracking in Section~\ref{sec:numerical_results_outdoor}, we employed the \ac{EKF}  with a sampling time of 0.7134 s, according to the \ac{PRS} periodicity $T^\text{PRS}_\text{rep}$, and the antenna array is defined by the tuple (1, 1, 4, 4, 1). Also in this case, we adopted numerology $\mu=3$ and the parabolic interpolation for \ac{TOA} peak detection.

The analysis is focused on assessing the tracking ability when using DL-TDOA measurements, comparing the case where the positioning system is able to accurately detect and discard NLOS measurements (green curve) with a solution that uses all \acp{TDOA} regardless of the visibility condition (red curve).

The estimated trajectories are reported in Fig.~\ref{fig:indoor_ekf}, which shows remarkable improvements brought by an
NLOS identification algorithm in discarding unreliable measurements, even in the presence of strong multipath caused by metallic objects and surfaces. 
Fig.~\ref{fig:heatmap-industrial} reports the heatmap of the positioning error, observing that the large positioning errors for the EKF that uses all DL-TDOA measurements are mainly present near the obstacles that prevent direct \acp{BS} visibility. Overall, we achieve a mean accuracy of 1.97~m for the EKF without NLOS mitigation and of 0.28~m for the EKF discarding \ac{NLOS} measurements.

\input{Figure_tex/EKF_TDOA_indoor}
\input{Figure_tex/Fig_heatmap_industrial}

\textcolor{black}{Most of the primary challenges we encountered are addressed in the 3GPP releases following the Rel-16.
In Rel-17, \ac{NLOS} detection will be enhanced by specifying whether each received signal arrives via a direct or reflected path. Additionally, each signal will be characterized by its \ac{TOF} after applying the \ac{TEG} timing correction. The introduction of path-based received power will further refine angle measurements by distinguishing multipath components. In Rel-18, the network is expected to become more intelligent with the integration of AI/ML and advanced positioning techniques, such as \ac{CPP}, unlocking high accuracy positioning even with lower bandwidths and low-powered devices (i.e., \ac{LPHAP}).}

As final remarks on the enabled positioning services described in
Table~\ref{tab:UC5GAA} and Table~\ref{tab:UC5GACIA}, we point out that when relying solely on 5G positioning, without any advanced filtering technique, in outdoor dynamic scenarios only the \textit{vehicle decision assist} \ac{V2X} service with required accuracy of 150~cm can be supported when $\mu=1$. On the other hand, considering the context of indoor industrial use cases, all the services except \textit{goods storage} are feasible.

\section{\textcolor{black}{Lessons Learned and Open Issues}}
\label{sec:lessons_learned_limitations}
\textcolor{black}{In the previous sections, we highlighted the importance of cellular positioning, starting with a historical overview, outlining the major trends of  the research (Section~\ref{sec:State of the art}), providing examples of measurements and algorithms (Section~\ref{sec:System Model} and detailing the latest standard for cellular positioning (Section~\ref{sec:5G Positioning Technology (Rel-16)})
with associated simulations and performance analyses (Section~\ref{sec:Simulation Analysis}). In this section, we discuss the simulation results along with the lessons learned (Section~\ref{sec:lessons_learned}), and we highlight the current limitations of 5G positioning (Section~\ref{sec:limitations}).}

\subsection{\textcolor{black}{Lessons Learned}}
\label{sec:lessons_learned}

\textcolor{black}{In Section~\ref{sec:numerical_results}, we conducted extensive simulation experiments to explore the capability of the 5G technology in providing accurate positioning services.
The objective was to provide quantitative results on the achievable performance for varying 5G numerology,
type of measurements (DL-TDOA, multi-RTT, UL-AOA, or DL-AOD), BS antenna configuration, and BS visibility.
The findings confirm that augmenting the bandwidth and the antenna array aperture enhances the positioning accuracy, as expected. Additionally, the quantity of \acp{BS} in visibility is shown to play a pivotal role in achieving high positioning accuracy. Overall, the fusion of multiple and heterogeneous 5G measurements and the strategic application of tracking filters represent a viable strategy for overcoming the \ac{BS} visibility issue.}

\textcolor{black}{The numerical results suggest that in dynamic outdoor scenarios, a mobile device is not yet capable of using 5G \ac{DL}-\ac{TDOA} to localize itself with a sub-meter accuracy and meet the requirements of the precise positioning services in Table~\ref{tab:UC5GAA}.
For enhancing the positioning performance, it is recommended to use more sophisticated algorithms (e.g., tracking filters \textcolor{black}{and \ac{NLOS} detection techniques}), integrate multiple types of measurements, increase the number of \acp{BS} in visibility, or even combine 5G with additional localization technologies (e.g., \ac{GNSS} or inertial units). On the other hand, in indoor scenarios, 5G \ac{mmWave} positioning is shown to successfully achieve the cm-level accuracy, meeting the stringent requirements of the industrial use cases outlined in Table~\ref{tab:UC5GACIA}. }

\textcolor{black}{Main lessons learned from the above performance analyses are as follows:
\begin{itemize}
    \item \textcolor{black}{\textbf{Channel estimation complexity:}} \ac{CDL} channel estimation requires high computational complexity that grows with the number of antennas, rays, reflections, and diffractions. In this tutorial, we used \ac{MIMO} antenna arrays in all the simulations to ensure high fidelity and realism in the simulated scenario. However, in the case of ranging only, it is possible to reduce the computational complexity by using an equivalent \ac{SISO} channel with a higher \ac{Tx} power that compensates for the \ac{MIMO} beamforming gain.
    \item \textcolor{black}{\textbf{\ac{BS} deployment for joint localization and communication:}} The cellular network design currently relies on satisfying the communication requirements, which differ from a positioning-optimal \acp{BS} placement. The delivery of cellular-based positioning services should account for a trade-off among the coverage, throughput, and geometrical factors for positioning during network planning.
    \item 
    \textcolor{black}{\textbf{\ac{BS} selection:}} The geometric factor of the network deployment highly affects the positioning results, particularly when a mobile \ac{UE} is involved, and the visibility conditions change over time. In these cases, a selection algorithm that automatically identifies the optimal set of \acp{BS} for positioning is recommended. In TDOA-based positioning, the selection should also account for the geometry of the \ac{TDOA} hyperbola, guaranteeing a choice of the reference measurement that avoids ill-conditioned geometrical configurations. 
    \item \textcolor{black}{\textbf{Service continuity:}} While 5G is designed to guarantee high positioning accuracy compared to previous generations of cellular networks, achieving sub-meter and cm-level accuracy consistently across diverse environments is still a challenge. Major impacts are given by the hardware (for accurate \ac{AOA}/\ac{AOD} information), bandwidth (which impacts the ranging accuracy), and propagation conditions (because of multipath and \ac{NLOS} conditions, which are hard to mitigate). More research efforts and industrial commitments are needed to implement an accurate cellular positioning service, ensuring that \acp{KPI} and requirements are respected.
    \item \textcolor{black}{\textbf{Map information:}} Positioning algorithms using only wireless measurements can lead to poor performance  (especially in \ac{NLOS} environments). Assistance data such as environmental maps (both indoor and outdoor) or a-priori information about forbidden areas can be included in an advanced tracking methodology. As an example, the availability of a floorplan of a building can be valuable in discerning whether a location estimate is feasible (or not) or mitigating the error by constraining the position estimates.
    \item \textcolor{black}{\textbf{NLOS impact:}} \ac{UE} positioning in the presence of \ac{NLOS} \acp{BS} is hard even with tracking filters, resulting in high accuracy errors. 
    Ranging measurements from \ac{NLOS} \acp{BS} overestimate the \ac{UE}, while \ac{NLOS} angle measurements misrepresent the spatial direction of the \ac{UE}. A single \ac{NLOS} \ac{TOF} can bring severe degradation if it is used as reference measurement in \ac{TDOA}-based methods. Intuitively, if the direct path is obstructed by a building, the ideal direct path of about 100~m can be confused with a path in \ac{NLOS} of 150 m, resulting in an overall positioning error of about 50~m.
    \ac{NLOS} detection and mitigation techniques are almost a requirement for precise positioning services, especially in urban areas where the density of \ac{BS} deployment cannot guarantee a continuous \ac{LOS} condition in any location.
    \item \textcolor{black}{\textbf{NLOS detection and mitigation:} \ac{NLOS} detection and mitigation techniques proposed in the literature (see Section~\ref{sec:NLOS_detection}) highlight the possibility of either discarding \ac{NLOS} measurements or exploiting them for improving mapping and taking advantage of the multipath environment. In the industrial indoor scenario, we assumed perfect knowledge of the \ac{NLOS} \acp{BS}, taking advantage of the raytracer tool. However, recent work demonstrates high proficiency in \ac{NLOS} detection (between 80\% and 93\%), indicating promising future outcomes. Results with \ac{UWB} technology (currently the wireless technology most similar to 5G, thanks to the large bandwidth) show $\approx$10~cm accuracy, compared with  $\approx$30~cm obtained in our simulations. This is achievable due to the higher bandwidth ($\geq$500~MHz)~\cite{Poorter2024} that is not yet available in 3GPP Rel-16.}
\end{itemize}
}


\textcolor{black}{Most of the primary challenges we encountered are addressed in the 3GPP releases following the Rel-16.
In Rel-17, \ac{NLOS} detection will be enhanced by specifying whether each received signal arrives via a direct or reflected path. Additionally, each signal will be characterized by its \ac{TOF} after applying the \ac{TEG} timing correction. The introduction of path-based received power will further refine angle measurements by distinguishing multipath components. In Rel-18, the network is expected to become more intelligent with the integration of AI/ML and advanced positioning techniques, such as \ac{CPP}, unlocking high accuracy positioning even with lower bandwidths and low-powered devices (i.e., \ac{LPHAP}).}

\textcolor{black}{As final remarks on the enabled positioning services described in
Table~\ref{tab:UC5GAA} and Table~\ref{tab:UC5GACIA}, we point out that when relying solely on 5G positioning, without any advanced filtering technique, in outdoor dynamic scenarios only the \textit{vehicle decision assist} \ac{V2X} service with required accuracy of 150~cm can be supported when $\mu=1$. On the other hand, considering the context of indoor industrial use cases, all the services except \textit{goods storage} are feasible.}

\subsection{\textcolor{black}{Open Issues}}
\label{sec:limitations}

\textcolor{black}{While technical concepts and architectures are well defined from a theoretical point of view, practical implementation in commercial systems is still restrained. 
The discussion in the following sections is thus focused on current impairments that still limit the pervasive adoption of cellular positioning technologies.}

\textcolor{black}{\subsubsection{Antenna position and orientation}
Accurate cellular positioning strictly needs a precise knowledge of the true location of each antenna panel of the \ac{BS} in terms of latitude, longitude, and altitude. At present, the information about the \ac{BS} location is very approximate, e.g., based on \ac{GNSS} surveys, and typically, with no indication of the exact positions of distributed panels, i.e., only one location information is available for each \ac{BS}. Considering that there are sites with non-co-located panels (possible distances of tens of meters between different panels), a lack of this information unavoidably introduces errors in time-based positioning measurements. It follows that precise mapping surveys are needed to build a reliable database of the antenna positions for each \ac{BS}, and this operation can be tedious, time-consuming, and complex due to (not so rare) impervious sites.}

\textcolor{black}{In addition to the antenna position, precise tilting information is also required to guarantee reliable angular information. Manual tilting measurements are subject to errors, and also, in this case, the operations can be risky and complex, even more than measuring the position. Clearly, the antenna supports need to be highly stable to avoid slight rotations over time, i.e., they should be resistant to severe weather conditions.  Furthermore, accurate calibration procedures are requested to guarantee optimal performance of the antenna arrays at \acp{BS}.}

\textcolor{black}{Lastly, exact knowledge of cable length from the antenna to the signal source generator (typically at the baseband unit) and cabling material is required to precisely measure the \ac{TOF}.}

\textcolor{black}{\subsubsection{Synchronization error}
While the recommendation for communication of \ac{ITU} indicates a tolerable synchronization error of $\pm 1.5$~$\mu$s~\cite{20178271}, the requirements for positioning are much stricter.
As a matter of fact, a synchronization error of $\pm 3$~ns results into a positioning error of $\approx$1~m, and the upper bound of $\pm1.5$~$\mu$s, corresponding to $\approx$450~m of ranging error, is clearly incompatible with most of the 5G positioning use case requirements (see Section~\ref{sec:use_cases}), preventing any precise positioning service. 
At present, \ac{5G} networks use \ac{GNSS}-based synchronization or packet-based synchronization with IEEE 1588v2 \ac{PTP}~\cite{Cha:J17}, but these standards cannot provide an accuracy close to 1~ns.
Reaching a near-zero nanosecond error is challenging, but research demonstrates that fiber-based solutions such as the White Rabbit protocol~\cite{WhiteRabbit} can reach synchronization error values of 1~ns or even less~\cite{ZhuFokEri:J19}.
Having a precisely synchronized \ac{5G}  network will ensure a common scanning of the time domain for all \acp{BS}, which would exactly transmit in the allocated time slot, limiting the interference and avoiding introducing degrading effects on time-domain measurements due to clock drifts.}

\textcolor{black}{\subsubsection{BS density}
The foreseen density of \ac{5G} \acp{BS} in urban scenarios is one \ac{BS} every 200~m~\cite{tr138913}. If having such a high number of \acp{BS} increases the investment costs of operators, on the other side, it brings a significant improvement on the cellular positioning use case, boosting the roll-out of commercial services. We demonstrated that it is possible to localize a \ac{UE} with a single \ac{BS} in \ac{LOS}; thus, a high density of \acp{BS} would minimize blind areas and \ac{NLOS} conditions, allowing for a precise cellular positioning service to the users. Clearly, the coverage of a single \ac{BS} would be limited to a few tens of meters, thus demanding the network to perform handover procedures quickly. The advantage of having close \acp{BS} is that it facilitates the indoor/outdoor transition, guaranteeing a seamless positioning service.}

\textcolor{black}{\subsubsection{Hardware availability}
As of today, experimental activities on \ac{5G} positioning are slowed down by a lack of commercial-ready hardware allowing the extraction of physical level parameters. 
As a matter of fact, current practical works mainly adopt modified commercial devices~\cite{ge_experimental_2022,ge_experimental_2023} or ad-hoc hardware~\cite{yammine_experimental_2021,mata_preliminary_2020} which rarely permit the exploitation of raw measurements. 
So far, the only research paper that measures raw 5G TOF is \cite{sci_rep_5G}.
However, the expensive cost of the hardware and non-compact size, together with the not-so-easy accessibility and usability, produce an inevitable slowdown of the research and testing procedures.
The above limitations are valid for both \ac{FR}1 and \ac{FR}2 bands and are further exacerbated for the latter. This lack, which is going to be resolved soon due to the high push from industries, prevents a pervasive assessment of \ac{5G} positioning potentialities at \ac{mmWave} and large bandwidths, which would unleash the rollout of advanced and precise cellular-based location services. The last desired feature is also limited by a restricted deployment of public \ac{mmWave} \acp{BS}.}

\textcolor{black}{\subsubsection{Deployment of private networks}
An additional notable issue pertains to the indoor 5G positioning domain and 
it revolves around the current state of private networks. As of today, it is observed that private networks have not been widely integrated into industrial settings despite the positioning opportunities they hold (see Section~\ref{sec:numerical_results_indoor}). This deficiency in the deployment has prompted industries to seek alternative technologies to fulfill their specific connectivity and positioning requirements. One such alternative that has gained considerable attention is \ac{UWB} technology, particularly in industrial facilities where precise positioning is requested for the automation of workflows~\cite{UWB_logistics}.}





\section{Conclusion and Future Research}
\label{sec:Conclusions}

This tutorial paper on 5G positioning aims to serve as a trusted reference for understanding the potentialities and limitations of the latest cellular localization technology. 
We covered a journey to explore the fundamental concepts, techniques, and challenges associated with \ac{5G} positioning, delving into the technical underpinnings of \ac{5G} networks and how they can enable accurate positioning.
After summarizing the transition from \ac{1G} to \ac{4G}, we detailed the \ac{5G} evolution across the releases of the \ac{3GPP} standard, and we explored the major research trends towards \ac{6G}.
We delved into an explanation of the 5G positioning system and its associated capabilities, as defined by current industry standards, and highlighted how the latest technological enhancements could bring new possibilities for the roll-out of commercial cellular positioning services.

This tutorial is designed to be a valuable resource not only for academic audiences but also for professionals and businesses operating in or considering entry into the market of positioning services. To this extent, we presented results from extensive simulations designed to assess the positioning performance in diverse settings, including outdoor and indoor environments. Several analyses have been conducted to motivate the adoption of 5G technology for industrial positioning, revealing its appeal for indoor applications while simultaneously highlighting the inherent current limitations in outdoor contexts.

The findings revealed the superior accuracy of ranging measurements compared to angle-based methods. Specifically, UL-AOA positioning can be susceptible to the multipath effect, although it is worth noting that the angle accuracy is significantly linked to the dimensions of the antenna array.
Moreover, integrating multipath detection techniques offers the potential to mitigate this influence by eliminating anomalous positioning estimations, yielding refined results.
The simultaneous utilization of angle and ranging measurements proves advantages for achieving precise positioning, particularly in areas characterized by a low density of \acp{BS}. Additionally, we illustrated the methodology for conducting position estimation using a single \ac{BS}, obtaining promising results.
Furthermore, tracking filters demonstrate their efficacy in environments characterized by multipath interference and limited measurement data, such as indoor and urban scenarios.
Compared to urban settings, more reliable outcomes are observed in restricted environments, such as industrial areas. This discrepancy may be attributed to several factors, including the proximity of \acp{BS} to the user, the consistent presence of at least three \acp{BS} in \ac{LOS}, as well as the availability of larger bandwidth (100 vs 400 MHz).


\textcolor{black}{Future research in cellular positioning should focus on enhancing the accuracy and reliability of the positioning service, pushing the boundaries of current capabilities, and providing a cm-level accuracy even in challenging environments. To this extent, the integration with other localization technologies is highly recommended, as well as the use of \ac{AI}-powered techniques.
A transversal aspect covering all the positioning processes is related to data privacy and security, which call for safe measures preserving \ac{UE} location data. The design and implementation of 
secure positioning protocols are mandatory. Their adoption can also be functional for the implementation of dedicated privacy-preserving algorithms, e.g., \ac{FL}.
This implies the involvement of standardization bodies and dedicated efforts contributing to the enhancement of cellular positioning.
The innovation also includes industrial collaboration in offering open-source development platforms facilitating testing and implementation with hardware.}

\textcolor{black}{5G positioning is still in its early stages of development and, most importantly, deployment.
Despite its challenges, positioning in 5G (and the forthcoming 6G) networks holds high potential to revolutionize various industries and applications, especially in autonomous mobility, \acp{UAV}, \acp{NTN}, asset tracking and logistics, \ac{VR}, and metaverse. The use cases in these areas define stringent requirements for positioning, but at the same time, they unlock new possibilities for location-based services. 
Undoubtedly, most of the existing works dealing with 5G positioning consider simulation environments or ad-hoc limited hardware (e.g., \ac{SDR}). The verification of 5G potentialities with real networks should be a high-priority objective of incoming research, validating the impact of \acp{BS} density, propagation conditions, interference, and hardware impairments.}

\textcolor{black}{Advancing 5G positioning requires integrated cooperation of different partners (e.g., universities, industry players, policymakers, and standardization bodies), whose collaboration should drive technological innovation and economic growth.
The definition of clear value propositions and cost-effective deployments, tailored to the specific use cases and industrial needs, is a non-trivial task for enterprises that require economic feasibility of implementation.
From this perspective, agreeing on standardization and regulations that address privacy concerns and guarantee interoperability across several technologies is central for a large-scale adoption in the industry.
Still, companies can deploy private networks and offer communication and positioning services internally, with optimized deployment according to the defined \acp{KPI} and services.}

Given the increasing demand for precise and reliable positioning in various applications, we can envision a promising future for 5G positioning technologies. The progress made in this field, as outlined in this tutorial, underscores the potential for transformative changes in various sectors. We hope that this tutorial serves as a valuable resource for researchers, engineers, and innovators, contributing to the continued evolution and widespread adoption of 5G positioning solutions, ultimately enhancing our daily lives and driving innovation across industries.


\section*{List of Acronyms}
    \begin{acronym}[HBCI]
    \acro{1G}[1G]{first generation}
    \acro{2D}[2D]{two-dimensional}
    \acro{2G}[2G]{second generation}
    \acro{3D}[3D]{three-dimensional}
    \acro{3G}[3G]{third generation}
    \acro{3GPP}[3GPP]{Third Generation Partnership Project}
    \acro{4G}[4G]{fourth generation}
    \acro{5G}[5G]{fifth generation}
    \acro{5GAA}[5GAA]{5G Automotive Association}
    \acro{5GACIA}[5GACIA]{5G Alliance for Connected Industries and Automation}
    \acro{5GCN}[5GCN]{5G core network}
    \acro{6D}[6D]{six-dimensional}
    \acro{6G}[6G]{sixth generation}
    \acro{A-GNSS}[A-GNSS]{assisted-GNSS}
    \acro{A-GPS}[A-GPS]{assisted-GPS}
    \acro{ADCPM}[ADCPM]{angle-delay channel power matrix}
    \acro{AE}[AE]{auto-encoder}
    \acro{AI}[AI]{artificial intelligence}
    \acro{AL}[AL]{Alert limit}
    \acro{AMF}[AMF]{access and mobility management function}
    \acro{AOA}[AOA]{angle of arrival}
    \acro{AOD}[AOD]{angle of departure}
    \acro{AWGN}[AWGN]{additive white Gaussian noise}
    \acro{B5G}[B5G]{beyond 5G}
    \acro{BNN}[BNN]{Bayesian neural network}
    \acro{BS}[BS]{base station}
    \acro{C-ITS}[C-ITS]{cooperative intelligent transport systems}
    \acro{C-V2X}[C-V2X]{cellular V2X}
    \acro{CDF}[CDF]{cumulative density function}
    \acro{CDL}[CDL]{clustered delay line}
    \acro{CEP}[CEP]{circular error probable}
    \acro{CFRM}[CFRM]{channel-frequency response matrix}
    \acro{CID}[CID]{cell-ID}
    \acro{CIR}[CIR]{channel impulse response}
    \acro{CNN}[CNN]{convolutional neural network}
    \acro{CP}[CP]{cooperative positioning}
    \acro{CPP}[CPP]{carrier phase positioning}
    \acro{CRB}[CRB]{Cram\'{e}r-Rao bound}
    \acro{CSI}[CSI]{channel state information}
    \acro{CSI-RS}[CSI-RS]{CSI reference signal}
    \acro{D-MIMO}[D-MIMO]{distributed MIMO}
    \acro{DAS}[DAS]{distributed antenna system}
    \acro{DL}[DL]{downlink}
    \acro{DMRS}[DMRS]{demodulation reference signal}
    \acro{DNN}[DNN]{deep neural network}
    \acro{DRMS}[DRMS]{distance root mean square}
    \acro{DRSS}[DRSS]{difference of received signal strength}
    \acro{DT}[DT]{digital twin}
    \acro{e991}[e991]{enhanced 911}
    \acro{eCID}[eCID]{enhanced cell-ID}
    \acro{EKF}[EKF]{extended Kalman filter}
    \acro{eMBB}[eMBB]{enhanced mobile broadband}
    \acro{eNB}[eNB]{eNodeB}
    \acro{ESPRIT}[ESPRIT]{estimation of signal parameters through rotational invariance technique}
    \acro{E-UTRA}[E-UTRA]{evolved \ac{UTRA}}
    \acro{FCC}[FCC]{Federal Communications Commission}
    \acro{FFT}[FFT]{fast Fourier transform}
    \acro{FIM}[FIM]{Fisher information matrix}
    \acro{FL}[FL]{federated learning}
    \acro{FR}[FR]{frequency range}
    \acro{GAN}[GAN]{generative adversarial network}
    \acro{GDOP}[GDOP]{geometrical dilution of precision}
    \acro{gNB}[gNB]{gNodeB}
    \acro{gNB-CU}[gNB-CU]{gNB central unit}
    \acro{gNB-DU}[gNB-DU]{gNB distributed unit}
    \acro{GNN}[GNN]{graph neural network}
    \acro{GNSS}[GNSS]{global navigation satellite system}
    \acro{GPS}[GPS]{global positioning system}
    \acro{GSM}[GSM]{global system for mobile communications}
    \acro{GTD}[GTD]{geometric time difference}
    \acro{HD}[HD]{high-definition}
    \acro{HL}[HL]{holographic localization}
    \acro{IFFT}[IFFT]{inverse fast Fourier transform}
    \acro{IIOT}[IIoT]{industrial Internet of things}
    \acro{IMM}[IMM]{interactive multiple model}
    \acro{IMU}[IMU]{internal measurement unit}
    \acro{IOO}[IOO]{indoor open office}
    \acro{IOT}[IoT]{Internet of things}
    \acro{ISAC}[ISAC]{integrated sensing and communications}
    \acro{ITU}[ITU]{International Telecommunication Union}
    \acro{KNN}[KNN]{k-nearest neighbors}
    \acro{KPI}[KPI]{key performance indicator}
    \acro{LCS}[LCS]{location service}
    \acro{LIS}[LIS]{large intelligent surface}
    \acro{LMF}[LMF]{location management function}
    \acro{LOS}[LOS]{line of sight}
    \acro{LPHAP}[LPHAP]{low-power high-accuracy positioning}
    \acro{LPP}[LPP]{LTE positioning protocol}
    \acro{LS}[LS]{least square}
    \acro{LSTM}[LSTM]{long short-term memory}
    \acro{LTE}[LTE]{long term evolution}
    \acro{LTE-A}[LTE-A]{LTE advanced}
    \acro{MAE}[MAE]{mean absolute error}
    \acro{MIMO}[MIMO]{multiple-input multiple-output}
    \acro{ML}[ML]{machine learning}
    \acro{MMSE}[MMSE]{minimum mean square error}
    \acro{mMTC}[mMTC]{massive machine-type communication}
    \acro{mmWave}[mmWave]{millimeter wave}
    \acro{MRC}[MRC]{maximum ratio combining}
    \acro{MUSIC}[MUSIC]{multiple signal classification}
    \acro{NF}[NF]{network function}
    \acro{NFC}[NFC]{near-field communication}
    \acro{NG-RAN}[NG-RAN]{next generation RAN}
    \acro{NLOS}[NLOS]{non-line of sight}
    \acro{NLS}[NLS]{non-linear least squares}
    \acro{NR}[NR]{new radio}
    \acro{NTN}[NTN]{non-terrestrial network}
    \acro{NZP-CSI-RS}[NZP-CSI-RS]{non-zero-power CSI-RS}
    \acro{OFDM}[OFDM]{orthogonal frequency-division multiplexing}
    \acro{OFDMA}[OFDMA]{orthogonal frequency-division multiple access}
    \acro{OOD}[OOD]{out of distribution}
    \acro{OTDOA}[OTDOA]{observed TDOA}
    \acro{PBCH}[PBCH]{downlink physical broadcast channel}
    \acro{PCRB}[PCRB]{posterior \ac{CRB}}
    \acro{PDF}[PDF]{probability density function}
    \acro{PDSCH}[PDSCH]{physical downlink shared channel}
    \acro{PEB}[PEB]{position error bound}
    \acro{PF}[PF]{particle filter}
    \acro{PHY}[PHY]{physical}
    \acro{PRS}[PRS]{positioning reference signal}
    \acro{PSS}[PSS]{primary synchronization signal}
    \acro{PTP}[PTP]{precision time protocol}
    \acro{PUSCH}[PUSCH]{physical uplink shared channel}
    \acro{RAN}[RAN]{radio access network}
    \acro{RB}[RB]{resource block}
    \acro{RE}[RE]{resource element}
    \acro{RedCap}[RedCap]{reduced capacity}
    \acro{RF}[RF]{radio frequency}
    \acro{RIM}[RIM]{reconfigurable intelligent meta-surface}
    \acro{RIS}[RIS]{reconfigurable intelligent surface}
    \acro{RMSE}[RMSE]{root mean square error}
    \acro{RP}[RP]{reception point}
    \acro{RSRP}[RSRP]{reference signal received power}
    \acro{RSRPP}[RSRPP]{reference signal received path power}
    \acro{RSS}[RSS]{received signal strength}
    \acro{RSTD}[RSTD]{reference signal time difference}
    \acro{RSU}[RSU]{road-side unit}
    \acro{RT}[RT]{ray tracing}
    \acro{RTD}[RTD]{real-time difference}
    \acro{RTT}[RTT]{round-trip time}
    \acro{RVM}[RVM]{relevance vector machine}
    \acro{Rx}[Rx]{receiver}
    \acro{SBA}[SBA]{service-based architecture}
    \acro{SBI}[SBI]{service-based interface}
    \acro{SBR}[SBR]{shooting and bouncing rays}
    \acro{SC-FDMA}[SC-FDMA]{single-carrier frequency-division multiple access}
    \acro{SCS}[SCS]{sub-carrier spacing}
    \acro{SDR}[SDR]{software-defined receiver}
    \acro{SISO}[SISO]{single-input single-output}
    \acro{SL}[SL]{sidelink}
    \acro{SLAM}[SLAM]{simultaneous localization and mapping}
    \acro{SNR}[SNR]{signal-to-noise ratio}
    \acro{SRS}[SRS]{sounding reference signal}
    \acro{SS}[SS]{synchronization signal}
    \acro{SSB}[SSB]{synchronization signal block}
    \acro{SSS}[SSS]{secondary synchronization signal}
    \acro{SVD}[SVD]{singular value decomposition}
    \acro{SVM}[SVM]{support vector machine}
    \acro{TDOA}[TDOA]{time difference of arrival}
    \acro{TEG}[TEG]{timing error group}
    \acro{THz}[THz]{teraHertz}
    \acro{TIR}[TIR]{Target integrity risk}
    \acro{TIS}[TIS]{transparent intelligent surface}
    \acro{TOA}[TOA]{time of arrival}
    \acro{TOF}[TOF]{time of flight}
    \acro{TP}[TP]{transmission point}
    \acro{TR}[TR]{technical report}
    \acro{TRP}[TRP]{transmission-reception point}
    \acro{TS}[TS]{technical specification}
    \acro{TTA}[TTA]{Time-to-alert}
    \acro{TTFF}[TTFF]{time-to-first-fix}
    \acro{Tx}[Tx]{transmitter}
    \acro{UAV}[UAV]{unmanned aerial vehicles}
    \acro{UE}[UE]{user equipment}
    \acro{UL}[UL]{uplink}
    \acro{UMa}[UMa]{urban macro}
    \acro{UMi}[UMi]{urban micro}
    \acro{UMTS}[UMTS]{universal mobile telecommunications system}
    \acro{URA}[URA]{uniform rectangular array}
    \acro{URLLC}[URLLC]{ultra-reliable low-latency communications}
    \acro{US}[US]{United States}
    \acro{USRP}[USRP]{universal software radio peripheral}
    \acro{UTRA}[UTRA]{universal terrestrial radio access}
    \acro{UWB}[UWB]{ultra-wideband}
    \acro{V2V}[V2V]{vehicle-to-vehicle}
    \acro{V2X}[V2X]{vehicle-to-everything}
    \acro{WNLS}[WNLS]{weighted NLS}
    \acro{VR}[VR]{virtual reality}
    \acro{ZP-CSI-RS}[ZP-CSI-RS]{zero-power CSI-RS}
    \acro{ZF}[ZF]{zero-forcing}
\end{acronym}


\section*{Acknowledgment}
\label{sec:Acknowledgments}

We acknowledge the MADE Competence Center Industry 4.0 for providing the 3D lidar scan of its environment, enabling us to conduct indoor simulations.




%





\bibliographystyle{IEEEtran}
\bibliography{IEEEabrv, Bibliography_new.bib}



\begin{IEEEbiography}[{\includegraphics[width=1in,height=1.25in,clip,keepaspectratio]{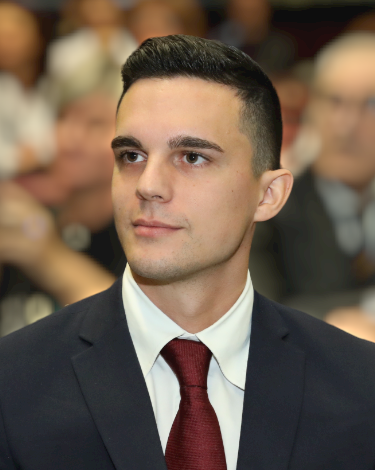}}]{Lorenzo Italiano} (Graduate Student Member, IEEE) obtained the B.Sc. degree in Computer Engineering (2020) from the University of Florence, Italy, where he maturated an interest in Internet of Things and cybersecurity topics. He proceeded with his studies at the Politecnico di Milano, Italy, in Computer Science and Engineering, with a focus on cybersecurity and signal processing, obtaining his M.Sc. degree (cum Laude) in October 2022.
In November 2022, Lorenzo started his Ph.D. career at the Politecnico di Milano. His research interests include 5G positioning, V2X communications, and blockchain technologies. He was a recipient of the Distinguished Paper Award at ICSOC 2022.
\end{IEEEbiography}

\begin{IEEEbiography}[{\includegraphics[width=1in,height=1.25in,clip,keepaspectratio]{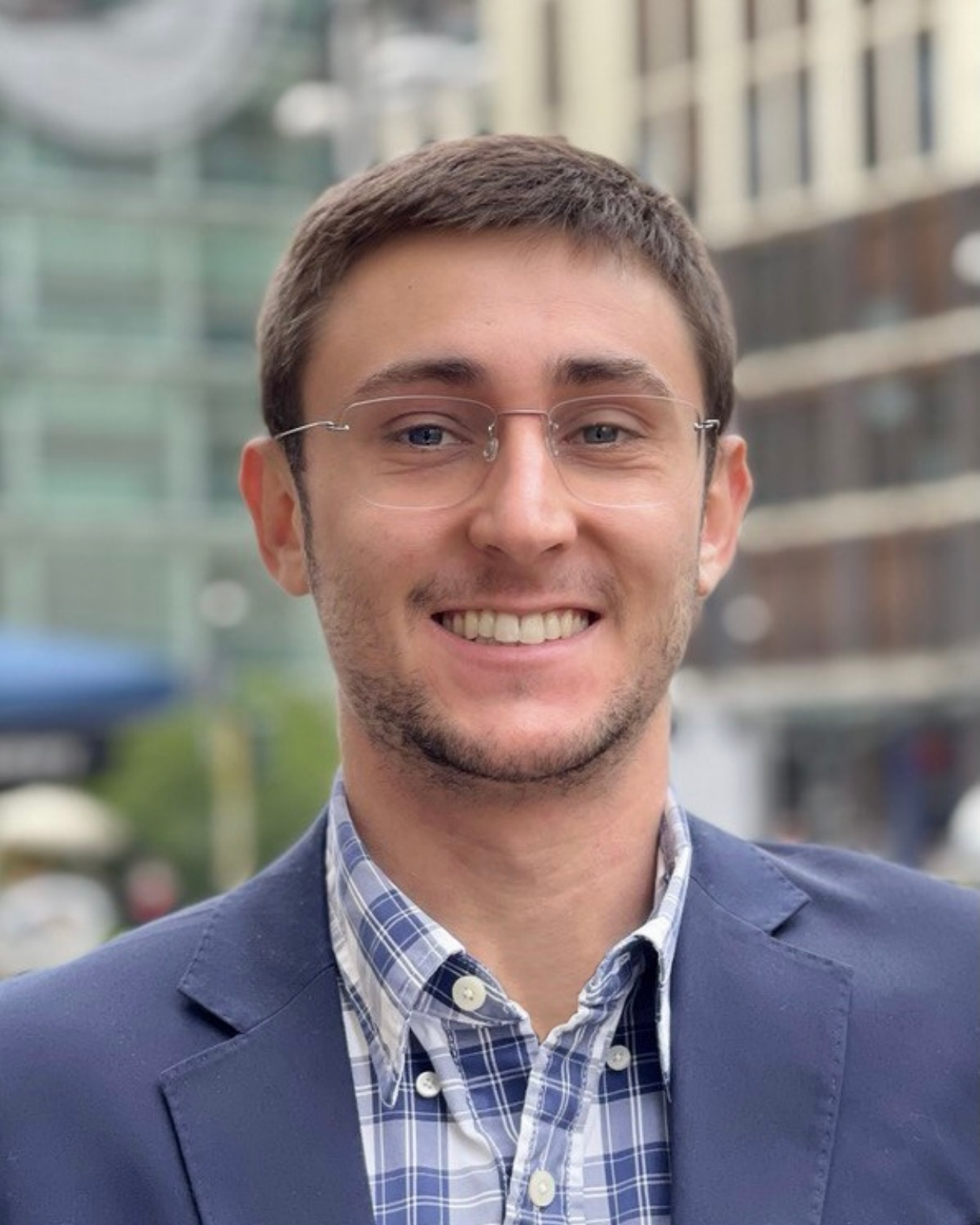}}]{Bernardo Camajori Tedeschini} (Graduate Student Member, IEEE) received the B.Sc. (Hons.) in Computer Science and M.Sc. (Hons.) degrees in Telecommunications Engineering from the Politecnico di Milano, Italy, in 2019 and 2021, respectively. From November 2021, he started as PhD fellow in Information Technology at Dipartimento di Elettronica, Informazione e Bioingegneria (DEIB), Politecnico di Milano.
He is currently a visiting researcher with the Laboratory for Information \& Decision Systems (LIDS) at the Massachusetts Institute of Technology (MIT), Cambridge, MA. His research interests include federated learning, machine learning and localization methods. 
He was a recipient of the Ph.D. grant from the ministry of the Italian government Ministero dell'Istruzione, dell'Università e della Ricerca (MIUR) and the Roberto Rocca Doctoral Fellowship granted by MIT and Politecnico di Milano.
\end{IEEEbiography}

\begin{IEEEbiography}[{\includegraphics[width=1in,height=1.25in,clip,keepaspectratio]{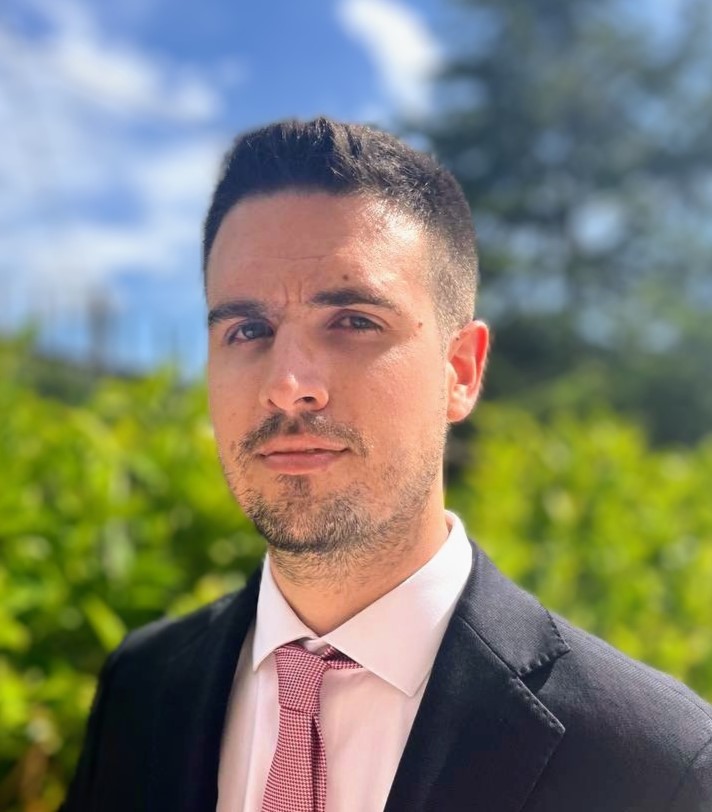}}]{Mattia Brambilla} (Member, IEEE) received the B.Sc. and M.Sc. degrees in telecommunication engineering and the Ph.D. degree (cum laude) in information technology from the Politecnico di Milano, in 2015, 2017, and 2021, respectively. He was a visiting researcher with the NATO Centre for Maritime Research and Experimentation (CMRE), La Spezia, Italy, in 2019. In 2021, he joined the faculty of Dipartimento di Elettronica, Informazione e Bioingegneria (DEIB) at the Politecnico di Milano as Research Fellow. His research interests include signal processing, statistical learning, and data fusion for cooperative localization and communication. He was the recipient of the Best Student Paper Award at the 2018 IEEE Statistical Signal Processing Workshop.
\end{IEEEbiography}

\begin{IEEEbiography}[{\includegraphics[width=1in,height=1.25in,clip,keepaspectratio]{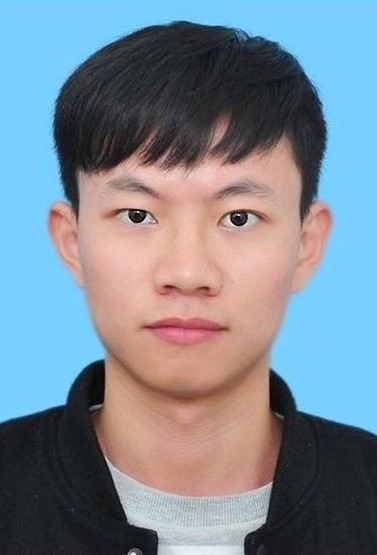}}]{Huiping Huang} (Member, IEEE) is a Marie-Skłodowska Curie Action (MSCA) postdoctoral researcher of the Communication Systems Group at Chalmers University of Technology, Gothenburg, Sweden. He received the Ph.D. degree from the Darmstadt University of Technology, Darmstadt, Germany, in 2023. Prior to that, he received his B.Eng. and M.Sc. degrees from Shenzhen University, Shenzhen, China, in 2015 and 2018, respectively, all in Electrical and Electronic Engineering.
His research interests lie in signal processing and wireless communications, with main focuses on direction-of-arrival estimation, source localization, channel estimation, robust adaptive beamforming, sparse array design, compressed sensing, and optimization theory with applications to radar, sonar, navigation, microphone array processing, and so on.
\end{IEEEbiography}

\begin{IEEEbiography}[{\includegraphics[width=1in,height=1.25in,clip,keepaspectratio]{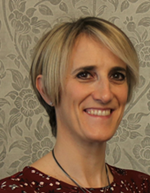}}]{Monica Nicoli} (Senior Member, IEEE) received the M.Sc. (Hons.) and Ph.D. degrees in communication engineering from Politecnico di Milano, Milan, Italy, in 1998 and 2002, respectively. She was a Visiting Researcher with ENI Agip, from 1998 to 1999, and Uppsala University, Sweden, in 2001. In 2002, she joined Politecnico di Milano as a Faculty Member. She is currently an Associate Professor in telecommunications with the Department of Management, Economics and Industrial Engineering. Her research interests include signal processing, machine learning, and wireless communications, with emphasis on smart mobility and Internet of Things (IoT) applications. She was a recipient of the Marisa Bellisario Award, in 1999, and a co-recipient of the best paper awards of the IEEE Symposium on Joint Communications and Sensing, in 2021, the IEEE Statistical Signal Processing Workshop, in 2018, and the IET Intelligent Transport Systems journal, in 2014. She is an Associate Editor of the IEEE Transactions on Intelligent Transportation Systems. 
She has also served as an Associate Editor for the EURASIP Journal on Wireless Communications and Networking, from 2010 to 2017. 
\end{IEEEbiography}

\begin{IEEEbiography}[{\includegraphics[width=1in,height=1.25in,clip,keepaspectratio]{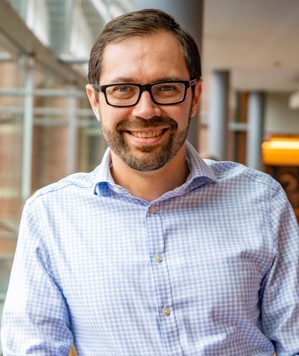}}]{Henk Wymeersch} (Senior Member, IEEE) {(S’01, M’05, SM’19) obtained
the Ph.D. degree in Electrical Engineering/Applied
Sciences in 2005 from Ghent University, Belgium.
He is currently a Professor of Communication Systems with the Department of Electrical Engineering at Chalmers University of Technology, Sweden. He
is also a Distinguished Research Associate with Eindhoven University of Technology. Before joining Chalmers, he was a postdoctoral researcher from
2005 until 2009 with the Laboratory for Information and Decision Systems at the Massachusetts Institute of Technology. Prof. Wymeersch served as Associate Editor for IEEE Communication Letters (2009-2013), IEEE Transactions on Wireless
Communications (since 2013), and IEEE Transactions on Communications (2016-2018). During 2019-2021, he was an IEEE Distinguished Lecturer with the Vehicular Technology Society. His current research interests include the convergence of communication and sensing, in a 5G and Beyond 5G context.}
\end{IEEEbiography}

\vfill


\end{document}